# Beyond Data-Driven: How Physics-Informed Neural Networks are Reshaping Multi-Physics Design and Discovery

Amir H. M. Labeb,[a,#] Basmala Sallam,[b,#] Abdelrahman W. Elsayed,[c,#] Islam I. Abdulaal,[d,#] Amany M. Kamal,[e,#] Omar A. M. Abdelraouf, [f, *]

[a] Department of Physics, Faculty of Science, Qena University (formerly South Valley University), Qena, 83523, Egypt

[b] Physics Department, Faculty of Science, Zagazig University, Zagazig 44519, Egypt

[c] Department of Electronics and Communications Engineering, The American University in Cairo, New Cairo 11835, Egypt

[d] Department of Electronics and Communication Engineering, Alexandria University, Alexandria, 21544, Egypt

[e] Mechanical Power Engineering Department, Faculty of Engineering - Mataria, Capital University (formerly Helwan University), Masaken ElHelmia P.O., Cairo 11718, Egypt.

[f] Institute of Materials Research and Engineering, Agency for Science, Technology, and Research (A*STAR), 2 Fusionopolis Way, #08-03, Innovis, Singapore 138634, Singapore.

#These authors are equally contributed.

*Corresponding author. Email address: Omar_Abdelrahman@a-star.edu.sg

# Abstract

Physics-informed neural networks (PINNs) constitute a rapidly maturing class of scientific machine learning models in which the governing equations of a physical system are embedded directly into the training objective as soft constraints. By enforcing partial differential equations (PDEs), conservation laws, and constitutive relationships during optimization, PINNs enable the construction of models that are simultaneously data-efficient, physically consistent, and capable of operating in regimes where measurements are sparse or indirect. In contrast to conventional deep learning, where the loss is typically defined solely in terms of data misfit, the learning task in PINNs is reformulated as a constrained optimization problem in which admissible solutions are confined to the manifold defined by the underlying physics. This review provides a comprehensive assessment of recent developments in physics-informed machine learning with an emphasis on PINN-based formulations for forward modelling, inverse design, and equation discovery across nanophotonics, fluid mechanics, astronomy, and biomedical engineering. Particular attention is devoted to how physical knowledge is injected at different stages of the modelling pipeline, including synthetic data generation, non-dimensionalization and scaling, architecture selection, loss design, and post-training regularization. We highlight emerging strategies for multi-physics coupling, transfer learning across parameter and geometry spaces, and rigorous benchmarking against established numerical solvers. Finally, the review discusses interpretability, uncertainty quantification, and hardware acceleration, and articulates how physics-informed learning is reshaping engineering practice by enabling digital twins and design workflows that combine simulation and data in a unified differentiable framework.

# 1. Introduction

## 1.1 Background on PINNs and their theoretical foundation

The mathematical description of physical processes in science and engineering is predominantly expressed in terms of systems of PDEs derived from conservation principles, balance laws, and constitutive relations. Classical examples include the Navier–Stokes equations for viscous compressible and incompressible flows,[1] Maxwell's equations for electromagnetic phenomena in complex media,[2] and reaction–diffusion equations for electrophysiology and transport in biological tissues.[3] For decades, the dominant paradigm for solving such equations has been grid-based numerical discretization, particularly finite difference, finite element, and finite volume methods. These techniques approximate spatial and temporal derivatives on structured or unstructured meshes and reduce PDEs to large, often sparse algebraic systems for which well-developed solvers and error estimates are available.

Despite their maturity and high accuracy, traditional numerical methods encounter fundamental challenges in several practically relevant settings. Mesh generation in complex, multi-scale, or moving geometries remains labour-intensive and fragile, often becoming a bottleneck in industrial workflows and optimization loops. In high-dimensional problems such as kinetic formulations in plasma physics, quantum many-body systems, or stochastic PDEs in finance, the number of degrees of freedom required for mesh-based discretization grows exponentially with dimensionality, rendering brute-force simulation computationally prohibitive. Moreover, classical solvers are primarily forward tools: inverse problems, parameter identification, and data assimilation are typically handled by adjoint methods or repeated forward solves, which substantially increase computational cost and implementation complexity.[4-6]

In parallel, the resurgence of deep learning has demonstrated that deep neural networks can approximate complex, high-dimensional, nonlinear input–output mappings with remarkable success in areas such as computer vision, speech recognition, and language modelling.[7] The universal approximation theorem guarantees that sufficiently large networks can approximate any continuous function on a compact domain, but standard "black-box" neural networks are agnostic to the underlying physics. They rely heavily on large labelled datasets, which are often unavailable in scientific settings where each data point may require costly experiments or high-fidelity simulations. Purely data-driven models may violate basic physical principles such as conservation of mass, momentum, and energy, or causality and can extrapolate poorly outside the training distribution, limiting their utility in engineering design and safety-critical applications.[8]

PINNs were introduced to reconcile data-driven learning with first-principles modelling by embedding physical laws directly into the learning process.[9] In a typical PINNs, a neural network $u_\theta(\mathbf{x}, t)$ approximates the unknown solution field, and its derivatives with respect to space and time are evaluated using automatic differentiation. These derivatives are substituted into the governing

PDEs to form a residual, which is penalized in the loss function alongside boundary and initial conditions and, where available, measurement data. The learning task thereby becomes the minimization of a composite loss that enforces both data consistency and physics consistency. In essence, the PDE plays the role of a structured regularizer that restricts the hypothesis space to functions satisfying the governing equations to a prescribed tolerance. This construction transforms the solution of PDE-constrained problems into a differentiable optimization task suitable for modern deep learning toolchains while maintaining a clear connection to the underlying physics.[10]

## 1.2 Motivation: generalized models in data-scarce scientific problems

Many contemporary scientific and engineering challenges are characterized by an acute scarcity of high-quality data, as opposed to the data-rich environments that have driven advances in mainstream deep learning. In biomedical engineering, for example, acquiring detailed, patient-specific measurements of tissue properties, electrical conductivities, or mechanical stiffness is invasive, costly, and often infeasible in a clinical environment. Available data are typically limited to sparse signals such as electrocardiograms, body-surface potentials, or low-resolution imaging.[11] In astronomy and cosmology, observations are constrained to line-of-sight projections and "snapshots" in time; direct observation of the temporal evolution of galaxies, dark-matter structures, or stellar interiors is impossible, and inferred quantities must be reconstructed from incomplete and noisy measurements.[12] Similar constraints arise in high-temperature combustion, subsurface flows, and nuclear systems, where experimental access is limited and safety margins are tight.

In such data-scarce regimes, purely data-driven models risk overfitting noise and artifacts, providing spurious correlations that lack physical meaning and extrapolate unreliably beyond the training set. By contrast, PINNs exploit established governing equations such as bidomain models in cardiac electrophysiology, Burgers- or Navier–Stokes-type equations in fluid mechanics, radiative transfer and Vlasov–Poisson systems in astrophysics, or Helmholtz and Maxwell formulations in photonics as physics-based priors or inductive biases. The PDE residual acts as a continuous regularizer that suppresses unphysical behaviours, while the data term anchors the solution in observed states.[13, 14] This synergy allows PINNs to infer latent fields that are not directly observed (for example, intracellular potentials, velocity or pressure fields, or internal temperature distributions) from sparse measurements, thereby enabling digital twins and personalization in applications such as patient-specific cardiology, customized photonic devices, or adaptive flow control (Figure 1).

Generalizability and interpretability are equally crucial from an engineering perspective. Design and certification decisions must be grounded in models that not only fit existing data but also extrapolate safely to new operating conditions, off-design regimes, or rare events. Physics-informed architectures provide a structural safeguard: because their outputs are constrained to approximately satisfy known conservation laws and constitutive relationships, they are less prone to predicting physically impossible states such as negative densities, non-conservative fluxes, or unstable equilibria in nominally stable systems.[15] Moreover, the explicit presence of PDE operators and parameters within the learning formulation offers a natural basis for interpretability: parameters inferred by the model have direct physical meaning, and deviations between model predictions and measurements can be analysed in terms of missing physics, model inadequacy, or data quality rather than purely statistical metrics.

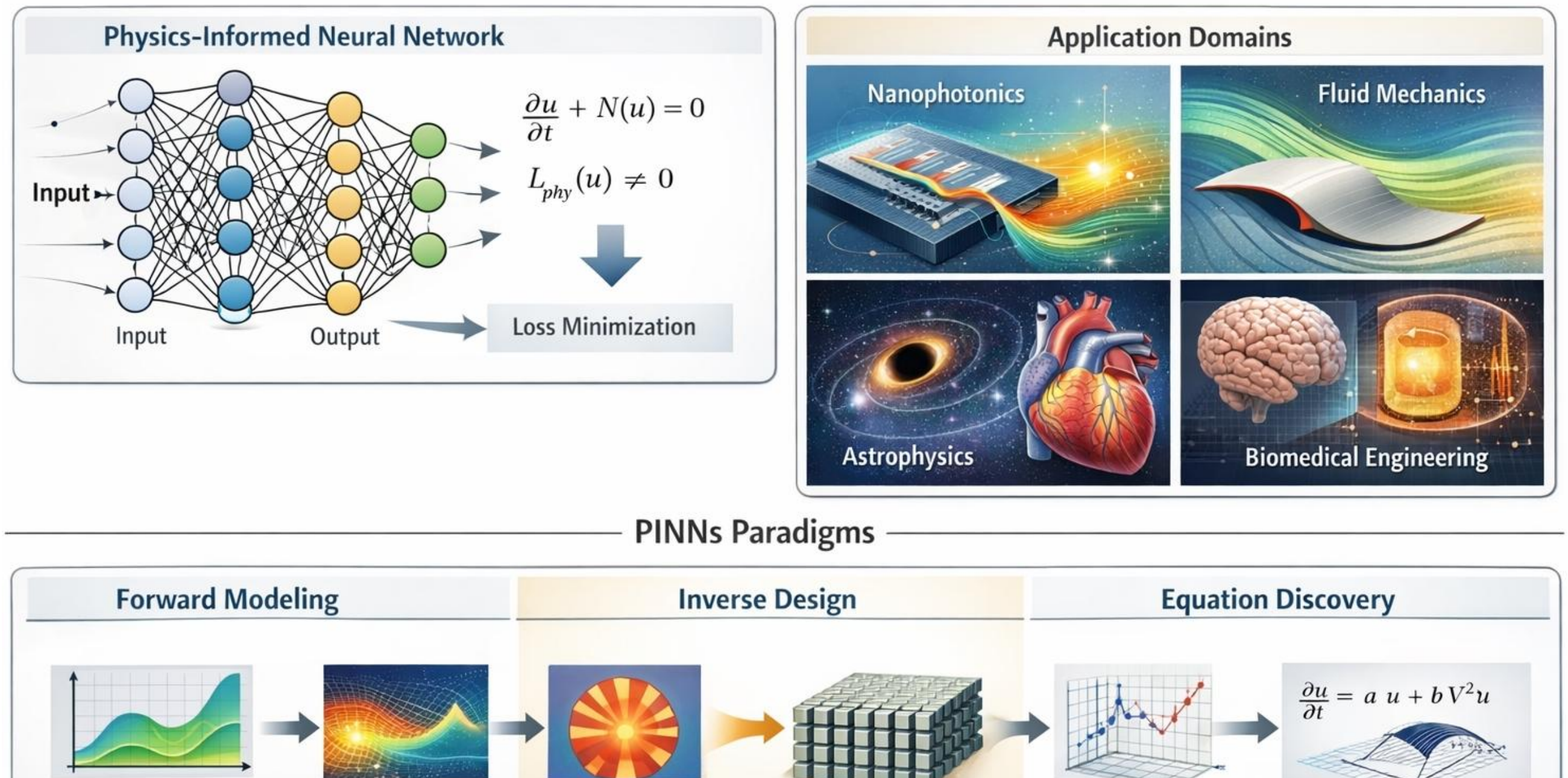

**Figure 1.** Overview of PINNs: applications and paradigms.

## 1.3 Challenges and opportunities in integrating different physical disciplines

While the formal PINNs framework is largely agnostic to the specific form of the governing equations, its practical deployment reveals domain-specific challenges arising from the analytical and numerical properties of different PDE systems. In fluid mechanics, the incompressible and compressible Navier–Stokes equations generate rich multiscale behaviour, including boundary layers, shear layers, and turbulent cascades.[16] Neural networks trained with standard activations exhibit spectral bias, favouring low-frequency components and struggling to capture sharp gradients, shocks, or fine-scale eddies. This complicates the application of naive PINNs to high-Reynolds-number flows and motivates specialized strategies such as adaptive activation functions, domain decomposition, conservative formulations, or hybrid couplings with turbulence models and lattice Boltzmann solvers.

In nanophotonics and integrated optics, the governing equations are Maxwell's equations or reduced Helmholtz formulations with complex-valued fields and strongly oscillatory solutions.[17-37] Engineered structures such as photonic crystals, metasurfaces, and waveguide couplers introduce abrupt refractive index contrasts and subwavelength geometric features that give rise to discontinuities or sharp kinks in the electromagnetic field, particularly in the presence of metals or high-index dielectrics.[38] Standard PINN formulations, which implicitly assume globally smooth solution manifolds, have difficulty resolving such features without excessive network capacity and training cost. This necessitates approaches based on weak formulations, interface-aware loss terms, or domain decomposition in which separate networks handle distinct material regions and continuity conditions across interfaces are enforced explicitly.

Biomedical applications often involve stiff, tightly coupled multi-physics models where characteristic time and length scales differ by several orders of magnitude. In cardiac electrophysiology, for example, ionic currents evolve on a sub-millisecond timescale, wavefront propagation on a millisecond scale, mechanical deformation on a scale of tens to hundreds of milliseconds, and vascular or hemodynamic responses on even longer timescales. Embedding all of these processes into a single monolithic PINNs can result in severe optimization stiffness: gradients associated with slow processes may dominate, leading the network to neglect fast transients, or vice versa.[39] Similar issues arise in coupled chemo-mechanical systems, electrochemical devices, and thermo-mechanical problems. Addressing these challenges requires careful non-dimensionalization, adaptive loss weighting, time-windowing, and possibly hybrid formulations in which different subsystems are treated with tailored numerical strategies within an overarching physics-informed framework.

Despite these difficulties, cross-disciplinary interactions have proven highly fruitful. Techniques developed to treat shocks and discontinuities in supersonic aerodynamics, such as artificial viscosity and shock-aware sampling, have been adapted to model sharp dielectric interfaces in nanophotonics and abrupt depolarization fronts in electrophysiology. Domain decomposition schemes originally devised for fluid dynamics (for example, extended PINNs and related space–time decompositions) are now applied to composite materials, thermo-mechanical systems, and structural health monitoring.[40] Conversely, advances in biomedical modelling, where uncertainty quantification and patient-specific inference are central concerns, have spurred the development of Bayesian and variational extensions of PINNs that are now being ported back to fluid, structural, and astro-physical applications.[41] These cross-pollinations suggest the emergence of a unified scientific machine learning framework that systematically combines physics, data, and numeric across domains.

## 1.4 Scope of the review

This review focuses on recent progress in physics-informed machine learning with an emphasis on PINNs as a unifying methodology for forward modelling, inverse design, and equation discovery in nanophotonics, fluid mechanics, astronomy, and biomedical engineering. The scope is deliberately restricted to approaches that incorporate explicit physical constraints usually in the form of PDEs, conservation laws, or variational principles into the learning objective. Purely data-driven architectures without physics embedding are outside the primary focus except where they serve as baselines or components of hybrid models.

We begin by formalizing the theoretical underpinnings of PINNs, with attention to their interpretation as function approximators constrained by PDE residuals, the role of automatic differentiation in evaluating high-order derivatives, and the formulation of forward, inverse, and hybrid learning paradigms. Subsequent sections examine algorithmic developments in multi-physics coupling, domain decomposition, and conservative formulations, with a particular emphasis on thermo-optical and bio-mechano-electrical systems that exemplify current multi-disciplinary challenges. We then discuss advances in transfer learning, few-shot adaptation, and hardware acceleration, highlighting how these strategies address the computational cost associated with training PINNs for realistic three-dimensional and time-dependent problems. Benchmarking results and critical comparisons with classical solvers are synthesized to delineate the regimes in which PINNs are currently competitive, where they remain inferior, and where hybrid solutions are most promising. Finally, we address interpretability, uncertainty quantification, and cross-domain opportunities, and articulate an outlook on how physics-informed learning may evolve into a standard component of engineering simulation and design ecosystems.

# 2. Theoretical Background of PINNs

## 2.1 Neural networks as function approximators

In physics-informed learning, deep neural networks are deployed as flexible parametric surrogates for solution fields of the form $u(\mathbf{x}, t)$, where $\mathbf{x}$ denotes spatial coordinates and $t$ time. For a feedforward architecture with $L$ hidden layers, the network defines a composite mapping.

$$\mathbf{z}_0 = (\mathbf{x}, t),\ \mathbf{z}_{k+1} = \sigma_k\,(\mathbf{W}_k \mathbf{z}_k + \mathbf{b}_k),\ k = 0, \dots, L-1 \tag{1}$$

with trainable parameters $\theta = \{\mathbf{W}_k, \mathbf{b}_k\}_{k=0}^{L-1}$ and a linear or nonlinear final layer producing the approximate state variables. In contrast to conventional computer vision applications, the choice of activation function $\sigma_k$ must be guided by the differential structure of the underlying governing equations, because higher-order spatial and temporal derivatives enter the physics loss.[42]

Piecewise-linear activations such as ReLU are widely used in classification and regression, but their non-smooth first derivative and vanishing *second* derivative render them ill-suited for PDEs involving Laplacians, biharmonic operators, or higher-order dispersion terms. Smooth activations with $C^\infty$ regularity, including hyperbolic tangent, Swish/SiLU, and sinusoidal functions, are therefore preferred in PINN architectures.[43] Tanh-based networks have been successfully employed in canonical fluid and solid mechanics benchmarks, but they may suffer from vanishing gradients for very deep architectures and sharp gradients in the solution. Swish-type activations, originally developed for computer vision, have demonstrated improved trainability for flows at moderate Reynolds numbers by mitigating gradient saturation while preserving differentiability of higher-order derivatives needed in Navier–Stokes residuals. For highly oscillatory wave phenomena, periodic activations such as the SIREN formulation $\sigma(z) = \sin\,(\omega z)$ enable faithful approximation of multi-frequency fields with large curvature, as encountered in nanophotonic resonators, plasmonic metasurfaces, and acoustic cavities.[44] Here, the network parameters $\theta$ effectively encode a parametric representation of the solution manifold across a design or parameter space, rather than a single deterministic realization, aligning with engineering objectives such as parametric sweeps and robust design under uncertainty.

## 2.2 Embedding physics via PDE constraints

The defining characteristic of a PINN is the reinterpretation of PDE-based modelling as a constrained optimization problem in function space. Consider a general nonlinear PDE system

$$\mathcal{N}(u(\mathbf{x}, t); \lambda) = 0 \text{ in } \Omega \times (0, T), \tag{2}$$

subject to boundary conditions $\mathcal{B}(u(\mathbf{x}, t)) = g(\mathbf{x}, t)$ on $\partial\Omega \times (0, T)$ and initial conditions $u(\mathbf{x}, 0) = h(\mathbf{x})$. The operator $\mathcal{N}$ may represent, for example, the incompressible Navier–Stokes system, Maxwell's equations in frequency or time domain, or reaction–diffusion systems used in cardiac electrophysiology. A neural network $u_\theta(\mathbf{x}, t)$ is introduced as a trial solution, and the physics residual is defined as

$$r_\theta(\mathbf{x}, t) = \mathcal{N}(u_\theta(\mathbf{x}, t); \lambda), \tag{3}$$

evaluated at scattered collocation points $\{(\mathbf{x}_f^i, t_f^i)\}_{i=1}^{N_f}$ in the spatio-temporal domain. The central ingredient enabling this formulation is automatic differentiation (AD), which computes derivatives such as $\partial u_\theta/\partial x_j$, $\partial^2 u_\theta/\partial x_j^2$, or $\partial u_\theta/\partial t$ exactly to machine precision via repeated application of the chain rule to the computational graph of the network, without resorting to finite-difference stencils or symbolic expansion. The training objective then combines data-consistency and physics-consistency terms into a single loss function,

$$\mathcal{L}(\theta) = \mathcal{L}_{\text{data}} + \mathcal{L}_{\text{PDE}} + \mathcal{L}_{\text{BC/IC}}, \tag{4}$$

with

$$\mathcal{L}_{\text{PDE}} = \frac{1}{N_f}\sum_{i=1}^{N_f} \left\| r_\theta\left(\mathbf{x}_f^i, t_f^i\right)\right\|^2, \tag{5}$$

$$\mathcal{L}_{\text{BC/IC}} = \frac{1}{N_b}\sum_{j=1}^{N_b} \left\| \mathcal{B}\left(u_\theta\left(\mathbf{x}_b^j, t_b^j\right)\right) - g\left(\mathbf{x}_b^j, t_b^j\right)\right\|^2, \tag{6}$$

In purely forward problems, $\mathcal{L}_{\text{data}}$ may be absent, whereas in data-assimilation or inverse tasks it penalizes deviations from experimental or simulation data points. From an engineering perspective, this embedding transforms the PDE solution into an unconstrained nonlinear least-squares problem in a high-dimensional parameter space. The resulting optimization landscape is non-convex and may exhibit multiple local minima, saddle points, and so-called "flat" directions, with direct implications for convergence speed and attainable accuracy. Nevertheless, the mesh-free nature of collocation and the ability to evaluate the residual at arbitrary locations enable flexible adaptation to complex geometries, moving boundaries, and multi-scale regions, provided that boundary conditions and material interfaces are encoded consistently in the loss term.[45]

## 2.3 Forward, inverse, and hybrid learning paradigms

Within this general formulation, three principal operational paradigms have crystallized: forward modelling, inverse design/parameter estimation, and hybrid equation discovery. In forward modelling, all physical parameters $\lambda$ are assumed known, and the objective is to approximate the state $u(\mathbf{x}, t)$ by minimizing a physics-dominated loss,

$$\min_\theta \mathcal{L}_{\text{PDE}}(\theta) + \mathcal{L}_{\text{BC/IC}}(\theta), \tag{7}$$

which yields a mesh-free surrogate for the PDE solution. For low-dimensional linear problems on simple domains, traditional finite element or finite volume solvers remain orders of magnitude faster and more accurate; however, PINNs become competitive when the dimensionality of the state space is high, the domain or parameter space is complex, or repeated solves are required within design or control loops. This scenario arises in high-dimensional quantum mechanics, financial mathematics, and parametric studies in fluid and structural mechanics where constructing and updating meshes is prohibitive.

In inverse design and parameter estimation, unknown coefficients $\lambda$ (e.g. transport coefficients, reaction rates, effective material parameters, boundary fluxes) are treated as additional trainable variables alongside $\theta$. The loss is augmented by sparse measurement data $\{u^{\mathrm{obs}}(\mathbf{x}_d^k, t_d^k)\}$,

$$\mathcal{L}(\theta,\lambda) = \frac{1}{N_d}\sum_{k=1}^{N_d} \left\| u_\theta\left(\mathbf{x}_d^k, t_d^k\right) - u^{\mathrm{obs}}\left(\mathbf{x}_d^k, t_d^k\right) \right\|^2 + \mathcal{L}_{\mathrm{PDE}}(\theta,\lambda) + \mathcal{L}_{\frac{\mathrm{BC}}{\mathrm{IC}}}(\theta), \tag{8}$$

This joint optimization implicitly replaces classical adjoint-based inverse methods and repeated forward–adjoint solves by a single training process, which is particularly attractive for parameter identification in fluid flows from scalar tracers, in thermo-optic devices from measured phase shifts, or in cardiac electrophysiology from body-surface potentials.

Hybrid learning and equation discovery address situations where the structure of $\mathcal{N}$ is partially unknown, as occurs in complex rheology, biological growth, or effective models for turbulent or astrophysical transport. Here, a library of candidate terms $\Theta(u)$ is constructed, containing polynomials, gradients, and nonlinear interactions, and a sparse coefficient vector $\xi$ is inferred from the relation

$$u_t \approx \Theta(u, \nabla u, \dots)\, \xi, \tag{9}$$

using techniques such as Sparse Identification of Nonlinear Dynamics (SINDy) or symbolic regression. The PINN provides a differentiable interpolant that denoises measurements and delivers accurate derivatives, while the sparse regression step extracts a compact, human-interpretable PDE model, as demonstrated in recent AI-assisted equation-discovery frameworks for systems biology and epidemiology.[46]

## 2.4 Handling multi-physics and coupled domains

Many engineering systems involve tight coupling between multiple physical fields and time scales, such as thermo-optical interactions in integrated photonics, aeroelasticity in aerospace structures, or bio-mechano-electrical feedback in the heart. In a multi-physics PINN, either a single multi-output network or a collection of subnetworks is used to represent the various state variables $\mathbf{u} = (u_1, \dots, u_M)$, with the total loss expressed as

$$\mathcal{L} = \sum_{m=1}^{M} w_m \mathcal{L}_{\mathrm{PDE},m} + \mathcal{L}_{\mathrm{BC/IC}} + \mathcal{L}_{\mathrm{data}}, \tag{10}$$

where each residual $\mathcal{L}_{\mathrm{PDE},m}$ corresponds to one of the coupled PDEs. For thermo-fluid systems, this may include a momentum equation, a continuity equation, and an energy equation, while in nanophotonics the coupling arises between thermal diffusion and Maxwell's equations through temperature-dependent refractive indices.

A key numerical challenge in such coupled formulations is loss stiffness and gradient imbalance: different equations may operate on widely separated scales and units, so that the gradients of some residuals dominate the optimization while others remain under-fitted. This can lead to solutions that satisfy, for example, boundary conditions and slow mechanical modes but neglect fast electrical wavefronts or localized thermal gradients. Engineering practice has adopted several strategies to alleviate this issue. Non-dimensionalization rescales all governing equations to order-unity magnitudes, improving conditioning and interpretability of the learned parameters. Adaptive loss weighting algorithms, such as GradNorm, SoftAdapt, or multi-objective balancing based on gradient norms, adjust the weights $w_m$ during training to equalize the contribution of each physics component, thereby mitigating the risk that one subsystem is effectively ignored.[47, 48]

Recent theoretical work using the Neural Tangent Kernel (NTK) has clarified that high-frequency components of the solution, which correspond to sharp layers, shocks, or narrow resonances, are associated with small eigenvalues of the NTK and thus evolve slowly under gradient descent. This insight has motivated frequency-dependent loss reweighting, curriculum sampling strategies that emphasize high-frequency regions, and hybrid formulations where stiff subsystems are treated by conventional time integrators inside the PINNs training loop.[49]

## 2.5 Recent algorithmic advances

To enhance robustness, accuracy, and scalability, a range of algorithmic extensions has been developed around the core PINN framework. Extended PINNs (XPINNs) apply the principle of spatial domain decomposition by partitioning $\Omega$ into subdomains $\{\Omega_i\}_{i=1}^{N}$ and assigning an individual neural network $u_{i,\theta_i}$ to each subdomain. The global loss includes local residuals and interface terms enforcing continuity of state and flux across subdomain boundaries,[50]

$$\mathcal{L} = \sum_{i=1}^{N} \mathcal{L}_{\text{PDE}}^{(i)} + \sum_{\Gamma_{ij}} \mathcal{L}_{\text{interface}}^{(ij)}, \tag{11}$$

where $\Gamma_{ij}$ denotes interfaces between subdomains. This approach enables parallel training on multi-GPU or multi-TPU platforms, local adaptation of network architectures to regions with shocks or boundary layers, and efficient treatment of non-convex geometries in fluid, structural, and photonic problems.

Conservative PINNs (cPINNs) address an important limitation of standard strong-form PINNs: the lack of strict enforcement of integral conservation laws.[51] For hyperbolic conservation laws or compressible flow with shocks, pointwise residual minimization does not guarantee global conservation of mass, momentum, or energy. cPINNs reformulate the physics loss in integral form over control volumes,

$$\mathcal{L}_{\text{PDE}}^{\text{cons}} = \sum_{K} \left\| \int_{K} \partial_t u_\theta \, \mathrm{d}V + \int_{\partial K} \mathbf{F}(u_\theta) \cdot \mathbf{n} \, \mathrm{d}S \right\|^2, \tag{12}$$

where $K$ denotes a cell and $\mathbf{F}$ the flux function. This strategy inherits desirable properties of finite volume methods, including robust shock capturing and exact balance of fluxes at interfaces, and has recently been extended to multi-physics and multiphase flows.

To reduce training cost, transfer PINNs and related meta-learning strategies exploit the fact that different parameter settings or geometries often share a common underlying operator structure. Networks pre-trained on a base configuration such as a baseline Reynolds number, reference geometry, or nominal material properties are fine-tuned for new conditions using significantly fewer optimization steps, enabling fast parametric sweeps in aerodynamic design, heat exchanger optimization, or nanophotonic inverse design. Complementary developments include accelerated training via second-order or quasi-Newton updates, NTK-guided preconditioning, and operator-learning approaches that approximate the solution map directly in function space.

# 3. PINN Forward Modelling

## 3.1 PINN Forward Modelling: Nanophotonics and Integrated Photonics

PINNs achieve mesh-free electromagnetic simulations by embedding Maxwell's curl equations or their spectral, scalar, or eigenvalue-problem reductions directly as residual penalty terms in the training loss, such that the network is constrained to produce field predictions that satisfy the governing physics without any labelled simulation data.[14, 52-54] Applied to forward electromagnetic problems, this strategy has been demonstrated across four thematically distinct application domains surveyed below: general EM fields, nanophotonics and metamaterials, integrated photonics and waveguide systems, and fiber-optic propagation. Across all four domains, the primary performance gains over conventional numerical methods are the elimination of mesh generation, removal of stability constraints such as the Courant-Friedrichs-Lewy (CFL) condition, and delivery of continuous, differentiable field representations that generalize across spatial coordinates and material configurations once training is complete.

### 3.1.1 PINNs for General Electromagnetic Fields

The time-domain forward solution of Maxwell's equations was first formulated as a PINN problem by Zhang et al.[55], who constructed a fully connected network that takes spatial and temporal coordinates as inputs and outputs the electric and magnetic field components, training it by minimizing a composite loss that simultaneously penalizes deviations from the curl equations

$$\frac{\partial B}{\partial t} = -\nabla \times \mathrm{E}, \quad \frac{\partial D}{\partial t} = \nabla \times \mathrm{H} \tag{13}$$

along with the initial and boundary condition residuals, thereby eliminating both spatial discretization and temporal stepping. The method was first validated on a one-dimensional cavity in homogeneous media and subsequently extended to media discontinuities by augmenting the loss with interface continuity conditions, and to conductive and nonlinear media by modifying the constitutive relations embedded in the residual, establishing a general architectural template for time-domain PINN-Maxwell solvers. The three-dimensional generalization of this approach was demonstrated by Qin et al. [56], who trained a fully connected neural network (FCNN) to reproduce the analytical field distribution of a metal resonant cavity directly from the 3D Maxwell system without any mesh discretization and without the time-step restriction of conventional finite-difference time-domain (FDTD), confirming that the same residual-constraint training strategy extends without modification from 1D cavities to full volumetric resonant structures. The convergence behavior under various architectural choices was systematically characterized by Shaviner et al. [57]across a suite of 1D free-space propagation and 2D Gaussian pulse problems using ablation studies that individually removed and reintroduced random Fourier feature embeddings, spatiotemporal periodicity enforcement, and temporal causality weighting. Their Neural Tangent Kernel (NTK) analysis exposed an intrinsic uneven-learning phenomenon in which loss gradients are largest where the error is already small, motivating subsequent adaptive sampling and weighting strategies to redirect the training effort toward high-error regions. For two-dimensional transient EM analysis under TMz polarization, Oh and Hong [58] addressed heterogeneous media problems by embedding the Helmholtz form of Maxwell's equations directly as the loss residual and using a collocation grid that conforms to material boundaries, producing mesh-free field predictions benchmarked against analytical references without any discretization of the spatial domain.

The treatment of material discontinuities, a persistent challenge for meshless methods that lack a natural mechanism for enforcing interface conditions, was addressed by Nohra and Dufour[59], who replaced discontinuous material property functions with smooth sigmoid-based approximations to restore the differentiability required by the automatic differentiation loss. They then deployed an overlapping domain decomposition strategy to partition the 3D domain into subregions, each handled by a dedicated subnetwork with shared residuals at the overlap boundaries, and validated the approach on both static and transient parametric EM problems. The NTK analysis of their formulation confirmed that the first-order Maxwell formulations outperform the second-order wave equation formulations at material interfaces because the lower-order derivatives reduce the gradient imbalance across subdomains. At high frequencies, Zhang et al. [60]proposed the Feature-Enhanced Physics-Informed Radial Basis Neural Network (FE-PIRBN), which combines a radial-basis function core network that provides localized spatial approximation with multi-resolution hash encoding borrowed from neural graphics to capture sub-wavelength field variations efficiently. This dual-enhancement strategy achieved relative errors of 1.40-5.82% for single- and double-cylinder scattering benchmarks while keeping the computational cost nearly independent of frequency, which is an important advantage over the finite element method (FEM) and FDTD, whose cost scales polynomially with frequency. In the low-frequency magnetostatic regime, Khan and Lowther [61] conducted a structured feasibility study covering electrostatic and magnetostatic benchmark problems of increasing geometric complexity, showing close agreement with the FEM ground truth and, critically, demonstrating that a network pre-trained on one problem instance can be fine-tuned for a geometrically related instance with substantially reduced training time through transfer learning, which is a practical consideration for device design workflows involving many similar configurations. Gong et al. [62]extended PINN formulations to 2D magnetostatic fields in electromagnetic devices, such as transformers and actuators, by encoding PDE residuals and boundary conditions as training objectives and introducing a mesh-assisted non-uniform collocation strategy that places more training points in regions of high field curvature identified from a coarse preliminary mesh, improving the accuracy in devices with concentrated flux regions compared to uniform sampling. Beltrán-Pulido et al. [63]addressed the parametric case by training a deep neural network to represent the magnetic field as a joint function of spatial coordinates and ten design parameters describing the geometry and operating point of an EI-core electromagnet, enabling field prediction across the entire design space after a single training run with an accuracy validated against finite element analysis (FEA) across a ten-dimensional parameter space. This demonstrates that PINN-based surrogate models can replace repeated FEA evaluations in parametric design exploration (Figure 2).

Structural extensions of the PINN paradigm beyond fully connected networks have been motivated by the difficulty in enforcing non-local physical constraints, such as interface continuity. Su et al. [64]addressed this by constructing an explicit graph over the computational mesh with nodes representing spatial degrees of freedom and edges encoding the adjacency topology, and training a physics-informed graph neural network (PI-GNN) using message-passing layers that allow information, and therefore interface conditions, to propagate explicitly along material boundaries; the resulting PI-GNN enforces EM field continuity at material interfaces without requiring additional penalty terms and additionally benefits from a variational Maxwell formulation embedded in the loss that accelerates convergence compared to strong-form PINN baselines. For nonlinear optical scattering problems, where the governing equations involve intensity-dependent refractive indices, Lim and Psaltis [65] demonstrated MaxwellNet, a convolutional U-Net architecture trained by minimizing the curl-of-curl Maxwell residual, computed via a Yee-grid finite-difference scheme applied to the network output, without requiring any pre-computed field data. By exploiting the translational equivariance of convolutional layers and the global field context captured by U-Net skip connections, MaxwellNet achieves millisecond-scale inference from refractive index distributions to scattered electric field components, approximately three orders of magnitude faster than CPU-based COMSOL, enabling the rapid evaluation of optical fields from arbitrary input permittivity maps. Building on this architecture, Gigli et al. [66]extended MaxwellNet to the nonlinear optical Kerr regime by adding a fully connected auxiliary module that takes the incident intensity as input and outputs dynamic weight adjustments to the convolutional kernels of the second encoding layer, allowing the network to capture the intensity-dependent refractive index $\mathrm{n}_0 + \mathrm{n}_2\mathrm{I}$ and achieved transverse electric (TE) and transverse magnetic (TM) relative errors of $1.71\text{-}3.93\times10^{-2}$ compared to COMSOL across a dataset of 5,000 glass diffuser geometries, with the differentiability of the physics-trained model exploited to perform gradient-based topology optimization of a microlens with respect to incident power. Liu et al. [67]further generalized the time-domain deep model to an unsupervised, mesh-free architecture that augments the forward field-prediction loss with a secondary inversion loss penalizing the difference between predicted and observed field values, thereby training a single network that simultaneously predicts EM field distributions in inhomogeneous media and recovers unknown material parameters from embedded observations, consolidating what are conventionally separate forward and inverse pipelines into a unified inference framework.

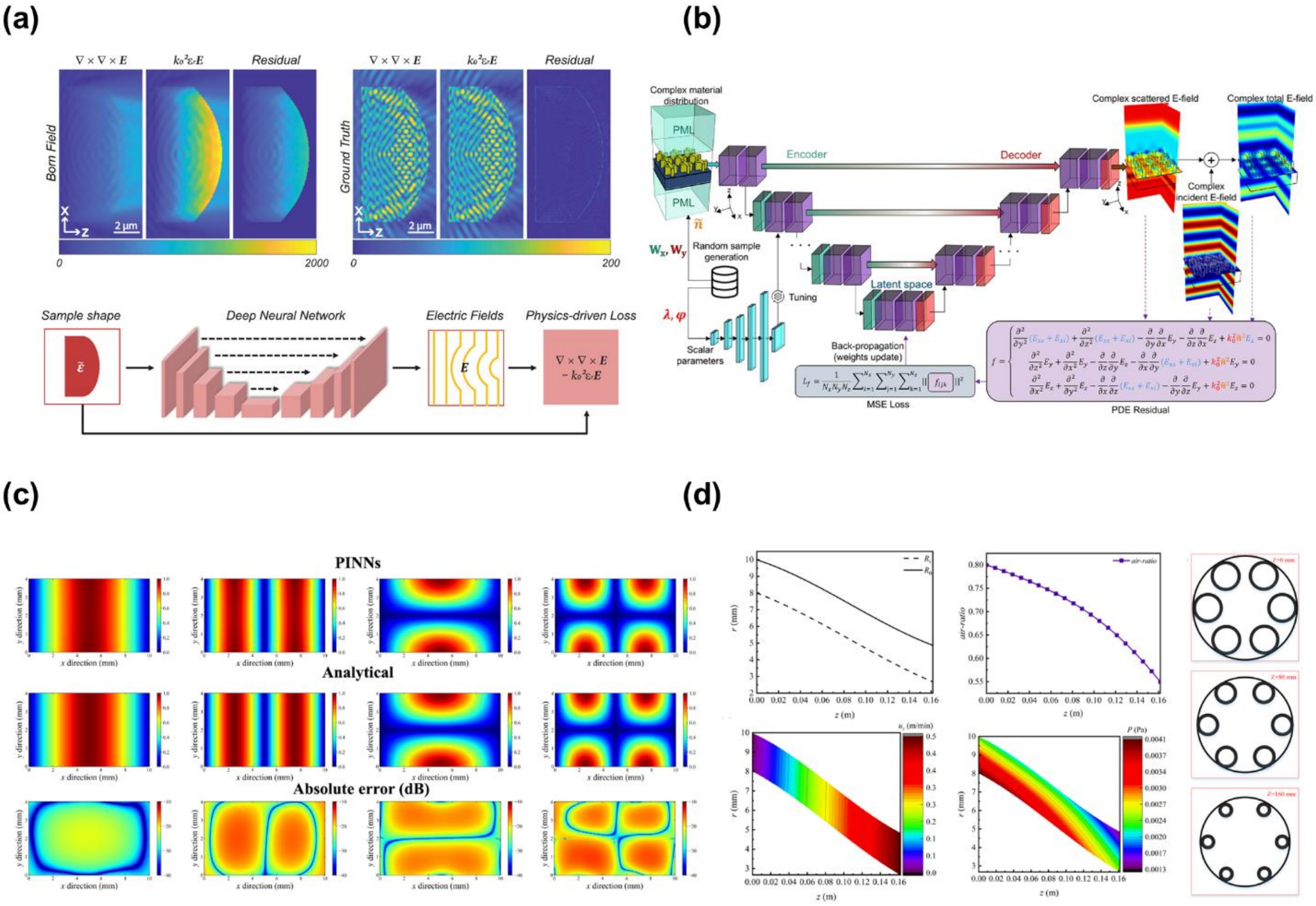

**Figure 2.** PINNs framework for forward electromagnetic modeling and representative results. (a) Illustration of the physics-driven loss formulation, where the residuals of Maxwell's equations are evaluated under two electric fields: the electric field obtained using the first-order Born approximation and the ground-truth solution computed using COMSOL.

The normalized electric permittivity distribution was used as the network input, and the residuals of Maxwell's equations calculated from the predicted electric field distribution served as the physics-driven loss for training the network.[65] **(b)** Schematic description of the PINN training process, where the 3D PINN model comprises a main U-Net architecture coupled with a tuning network to predict the electromagnetic field distribution.[68] **(c)** Electric field distributions of four eigenmodes ($TE_{10}$, $TE_{20}$, $TE_{11}$, and $TE_{21}$), showing the 4D-PINN solutions, corresponding analytical solutions, and absolute errors (dB) between the predicted and analytical results.[69] **(d)** Simulation results of the fiber drawing process, including the variation of inner and outer diameters along the fiber length, the evolution of $\xi$, the velocity distribution in the axial direction, the pressure distribution, and schematic fiber structures at different axial positions.[70]

### 3.1.2 PINNs in Nanophotonics and Metamaterials

Nanophotonic and metamaterial structures present a particularly demanding forward simulation challenge: sub-wavelength feature sizes require fine spatial resolution, complex material landscapes encompass dispersive, anisotropic, and nonlinear constitutive relations, and the absence of closed-form solutions for most device geometries makes every design iteration dependent on full numerical simulation.[71-78] PINNs address this challenge by replacing per-instance numerical solvers with networks whose physics-residual training encodes the governing EM equations once and whose inference operates in milliseconds, yielding the most favorable speed-accuracy trade-off in the regime where conventional solvers are most expensive. Ghosh et al. [79]established the physics-informed machine learning (ML) baseline for spatially periodic optical composites by formulating a PINN loss that penalizes violations of the Helmholtz eigenvalue equation derived from the periodic Maxwell system, alongside a labeled-data term guiding the network toward the physically correct mode, and demonstrating on datasets of 8,000 photonic and plasmonic composite configurations that physics-informed models reduce labeled data requirements by a factor of ~4 compared to meaning-informed models and ~20 compared to black-box networks, with the critical additional property that the network remains trainable for configurations with no known analytical solution, enabling mode prediction in metamaterial parameter regimes where ground truth is unavailable. For topological nanophotonic devices, Davoodi [80] introduced an active PINN-based surrogate framework that uniquely addresses the nonplanar wavefront excitation produced by localized sources such as electron beams and quantum dots, a regime not covered by the plane-wave assumptions of existing photonic PINN literature, by constructing a three-step training curriculum grounded in the Su-Schrieffer-Heeger (SSH) model topology, the universal plasmon ruler equation for inter-nanohole coupling, and the winding number criterion for topological phase identification, achieving a TSM architecture validation loss of $0.13\times10^{-3}$ and reducing the computational cost of designing unidirectional edge-mode ring-chain nanohole arrays by eliminating repeated FEM sweeps over the geometric parameter space. MaxwellNet [65] and its Kerr-effect extension [66] remain directly applicable to nanophotonic scattering, computing full-field Maxwell solutions for inhomogeneous media at the scale of the incident wavelength in under 21 ms, whereas Liu et al. [67]further consolidated forward scattering and material inversion within a single unsupervised deep model applicable to nanoscale inhomogeneous media. At the device-family level, NeurOLight [81] departed from the single-instance PINN paradigm by formulating a physics-agnostic neural operator that learns to solve an entire family of frequency-domain Maxwell PDEs simultaneously. It represents different device configurations through a joint PDE encoder that compresses the permittivity map using a wave-prior encoding derived from the dispersion relation and encodes incident sources via masked-source modeling inspired by masked language models, training on finite-difference frequency-domain (FDFD)-generated datasets for tunable multimode interferometer (MMI) and etched MMI photonic devices, and achieving a normalized mean absolute error (MAE) of 0.122 on unseen tunable MMI configurations, 54% lower prediction error than prior neural network baselines, and 140-200× speedup over FDFD, demonstrating that operator-level generalization over device families is achievable without per-instance retraining.

Building on these forward-scattering capabilities, PINN formulations have been specifically adapted for metamaterial media, requiring the simultaneous handling of multiple material complexity classes. Fang and Zhan [82] addressed the metamaterial parameter retrieval problem by embedding the frequency-domain TM Maxwell equation with perfect electric conductor (PEC) boundary conditions into the PINN loss and introducing a sinusoidal activation function $\sin(\pi s)$ in place of the standard tanh, a modification that was analytically motivated by the oscillatory nature of Helmholtz solutions and empirically validated to enable stable PINN convergence up to wavenumber k = 23, resolving the spectral bias limitation that causes standard PINNs to fail at high wavenumbers. Wang and Zhang [83] proposed a multi-receptive-field (MRF) PINN architecture employing six encoder-decoder pairs operating in parallel at different spatial scales with resolution-adaptive dilated convolutions, addressing the challenge that a single fixed receptive field cannot simultaneously capture fine-scale interface features and coarse-scale field envelopes. By embedding scalar TEz and TMz Helmholtz equations as training constraints, the MRF-PINN achieves sub-2.5% prediction error across five distinct nanoscale media classes, that is, dispersive, inhomogeneous, anisotropic, nonlinear, and chiral, with inference times under 70 ms, validating the multi-scale architecture as a general-purpose forward solver for complex metamaterial landscapes. Armbruster et al. [84]advanced beyond grid-based representations to a point-cloud formulation for all-dielectric metasurfaces with spatially varying inclined nanopillars, encoding each nanopillar as a set of unordered 3D points fed to a physics-informed PointNet (PIPN) that processes geometry without requiring any fixed spatial grid; the network embeds the 2D TE scalar Helmholtz equation and 3D vectorial Helmholtz equation with Floquet-Bloch periodic boundary conditions as loss constraints and is evaluated on metasurface geometries with varying pillar inclination angles and refractive indices, achieving near-field mean absolute percentage error (MAPE) of 1.69-3.04% while naturally accommodating the geometric variability introduced by fabrication imperfections in metasurface manufacturing. Medvedev et al. [68]extended the 2D metasurface PINN paradigm to fully three-dimensional vectorial electromagnetics by training a U-Net architecture with the complete vector Maxwell equations under Floquet-Bloch periodic boundary conditions and perfectly matched layer (PML) absorbers embedded as loss residuals, without any reference field data, on metasurface geometries spanning 700-800 nm; by including both TE and TM field components simultaneously, the model captures polarization-dependent near-field and far-field responses, including ellipticity and wavefront phase control, achieving a near-field MAPE of 0.12% for fixed illumination conditions and 1.45% for generalized illumination, making it the highest-accuracy fully vectorial data-free PINN for 3D metasurface simulation reported to date. Xiao et al. [85]demonstrated that a data-free PINN, trained by minimizing only the frequency-domain Maxwell residual over a domain containing two-dimensional etched cylindrical holes at nanometre scales without any labeled field data, achieves a scattered-field component prediction error of $\sim 10^{-6}$, outperforming supervised data-driven networks that require 40,000 training samples to reach comparable accuracy, confirming that the physics residual alone provides sufficient training signal for nanoscale EM forward simulation and establishing the data-free approach as a practical alternative to large-dataset deep learning for nanophotonic forward problems.

### 3.1.3 PINNs for Integrated Photonics and Waveguide-Based Systems

The design of integrated photonic circuits depends critically on accurate knowledge of the guided mode propagation constants, effective refractive indices, and transverse field distributions of the waveguide geometries used in interconnects, splitters, and modulators. Computing these quantities through conventional finite-difference or finite-element eigensolvers is straightforward for simple geometries but becomes computationally expensive for parametric sweeps over waveguide dimensions, material compositions, and wavelengths, the exact design exploration task for which PINN-based eigensolvers offer the most practical advantage through their continuous, data-free, and mesh-free field representations. Elsheikh et al. [86]established the foundational PINN slab-waveguide solver by embedding the 1D Helmholtz eigenvalue equation, augmented with a guided-mode existence constraint that prevents the optimizer from converging to radiation modes, into the loss, and training a four-hidden-layer FCNN on only 1,100 collocation points (1,000 interior plus 100 boundary) to predict the fundamental TE field profile and effective refractive index; the resulting model achieves prediction accuracy up to 99% and effective-index relative errors of $10^{-5}$~$10^{-6}$, validated against Lumerical FDTD and the variational method, while requiring no labelled simulation data. Mahmoud et al. [87]independently confirmed and quantified this data-free accuracy advantage, reporting that the same class of PINN slab-waveguide solver achieves a relative percentage error of 0.69% against the analytical solution, smaller than the 1.28% error of the FD solver operating on the same analysis grid of 349 points, and demonstrating simultaneous multi-mode prediction over user-specified effective-index intervals through a modified eigenvalue-constraining loss term that drives the PINN toward any targeted mode order. Khan et al. [88]extended PINN eigenanalysis to closed non-radiating waveguides with inhomogeneous and anisotropic filling media, handling material discontinuities through an eXtended PINN (XPINN) domain-decomposition strategy that assigns a dedicated sub-network to each homogeneous subdomain and enforces field continuity at their boundaries through interface residual terms; all identified eigenmodes show errors below −12 dB relative to High Frequency Structure Simulator (HFSS) analytical and full-wave references, and incorporating transfer learning from a pre-trained network reduced solution time by a factor of 23, demonstrating that PINN waveguide eigensolvers can be made practically fast through knowledge reuse across similar geometrical configurations.

Ünal and Durgun [89] addressed the more practical problem of strip waveguide effective-index prediction, where three-dimensional geometry prevents a simple 1D Helmholtz formulation, by designing a physics-aware neural network (EIMNet) that maps strip waveguide geometric parameters to an equivalent slab-waveguide effective medium through a learned physical transformation layer trained with a loss that penalizes violations of the effective-index method equations. By making the physics constraint explicit rather than implicit, EIMNet reduces effective-index prediction error by more than 50% compared to an equivalent black-box network of the same size and additionally predicts complete transverse field distributions and higher-order mode indices without additional training, capabilities beyond the reach of conventional single-value surrogate models. The full-vector eigenmode problem, simultaneously resolving all tangential electric and magnetic field components from the coupled Maxwell curl equations for photonic waveguides, was addressed by Xu et al. [69]through fourth-order derivative PINNs (4DPINNs), whose loss is derived by applying two successive curl operations to Maxwell's equations and simplifying, yielding

$$\nabla^2(\nabla^2 E) + k_o^2 \nabla^2(n^2 E) + k_o^4 n^2 E = 0 \tag{14}$$

as the residual constraint. The method additionally integrates fixed-point initialization strategies that seed the network with approximate mode shapes and adaptive learning rate schedules, achieving propagation constant errors below $10^{-4}$ and field distribution absolute errors below −12 dB relative to analytical benchmarks for strip waveguide geometries. The fundamental instability of eigenvalue-based PINN formulations was identified and resolved by Hares et al.[90], who formally proved that the combined loss function of conventional physics-guided neural networks (PGNNs), constructed as a weighted sum of physics residual, boundary condition, and eigenvalue-driving terms, contains local minima that do not correspond to true eigenpairs, causing gradient descent to diverge. Their three-component remedy consists of a multiplicative loss that provably preserves exact eigenpairs as global minima, a piecewise dynamic eigenvalue-driving term with a quadratic penalty below the substrate index and a shifted reciprocal above it, and trigonometric sinusoidal representation network (SIREN) activations in place of the standard tanh to exploit the periodic structure of mode profiles, reducing relative field-profile errors to 0.9-1.7% across six buried-channel and rib waveguide configurations, outperforming the reference FD solver that incurs up to 7.45% error on the same problems. Huang et al. [91]further reduced the training cost of waveguide eigenmode analysis by replacing iterative backpropagation with the analytical closed-form output-weight computation of extreme learning machines (ELMs), training a physics-informed ELM (PI-ELM) that satisfies the Helmholtz eigenvalue equation through a single matrix pseudoinverse rather than thousands of gradient steps, demonstrating applicability to waveguide cross-sections of varying geometry and to transmission line eigenmode analysis.

At the system and component levels, PINNs have been extended from isolated waveguide cross-sections to complete photonic circuit subsystems that operate under real measurement conditions. Gao et al. [92]addressed the calibration challenge of large-scale silicon optical phased arrays (OPAs), where the sidelobe suppression ratio degrades with channel count due to phase-shifter nonuniformity and thermal crosstalk; by embedding the far-field optical phased array (OPA) periodicity and translational invariance of the array response as a sinusoidal activation function and a translation-invariant loss term that penalizes phase distributions producing identical far-field patterns, resolving the inherent phase ambiguity in calibration, the PINN was pre-trained on 200,000 simulated far-field patterns and fine-tuned on 80,000 measured patterns through transfer learning, achieving a sidelobe suppression ratio of 10.8 dB on a 128-channel device while reducing real-measurement data requirements by approximately 70% compared to prior works. The physics-inspired causality-aware dynamic convolutional neural operator (PIC$^2$O-Sim) [93]approached time-domain photonic device simulation by analyzing the finite-difference time-domain (FDTD) update equations and encoding their inherent space-time causality, the principle that field values at a given time depend only on past and spatially local field values, into the receptive field of its dynamic convolutional backbone, which additionally adjusts its kernel weights based on the local permittivity map to capture device-specific propagation behavior. Validated on three challenging device types (tunable MMI, micro-ring resonator, and metaline), PIC$^2$O-Sim achieved 51.2% lower roll-out prediction error and 23.5× fewer parameters than state-of-the-art neural operators, with 133-310× speedup over single-process open-source FDTD. Teofilovic and Da Ros [94] demonstrated that the physics-informed approach extends naturally to thermo-optic effects by incorporating an exponential thermal decay physical law, derived from the diffusion equation governing heat transfer between waveguide phases, directly into the learning objective of a linear regression model trained on experimental resonance-shift measurements from a programmable photonic mesh circuit. They showed that even a simple physics-aware functional form significantly improves the model's ability to predict thermal crosstalk between spatially separated mesh elements compared to purely data-driven regression. Qi and Sarris [95] further expanded the PINN scope to multiphysics systems by training a network to satisfy the heat equation residual as an unsupervised physics constraint, demonstrating that this PINN thermal surrogate can directly replace the thermal solver in a coupled EM-thermal co-simulation, with temperature field predictions consistent with FEM references and inference sufficiently fast to reduce the overall multiphysics simulation time.

### 3.1.4 PINNs in Fiber Optics and Propagation Modelling

Optical fiber propagation spans a hierarchy of governing equations, from the scalar slowly varying field Helmholtz equation for weakly guided modes, through the multimode NLSE for nonlinear multichannel propagation, to the Navier-Stokes equations governing the molten-glass dynamics of fiber drawing. PINNs have been applied across this entire hierarchy, providing mesh-free solutions whose continuous spatial and propagation-distance representations make them particularly well-suited to problems with complex free boundaries or long-distance generalization requirements. Zhang et al. [96]demonstrated a physics-informed mode decomposition network for multimode fibers by embedding orthogonality and propagation-equation constraints into the training loss, training the network without any pre-computed mode solutions and enabling decomposition of 55 modes from a structured light field in a single forward pass, a capability directly applicable to the real-time mode characterization needed in space-division multiplexing links for increasing fiber transmission capacity. Luo et al. [97] introduced PINN-beam propagation method (BPM), a physics-informed beam propagation method that solves the slowly varying-field Helmholtz equation (SVHE) governing light-field envelope propagation,

$$\frac{\partial u}{\partial z} = \frac{i}{2k_o n_o}\nabla_\perp^2 u + ik_o[n(x,y,z) - n_o]u \tag{15}$$

where $u$ is the slowly varying field envelope, $n_o$ is the reference refractive index, and $\nabla_\perp^2$ is the transverse Laplacian, by constructing a composite training loss that combines the SVHE residual with a residual-based adaptive sampling strategy, which dynamically places more collocation points in regions of high local PDE residual, a self-adaptive weight scheduler that balances the relative magnitude of boundary and interior loss terms, and causal weights that enforce temporal propagation order along the z-axis; the resulting PINN-BPM yields enhanced accuracy and improved generalization compared to standard PINN baselines for both single-mode and multimode-fiber propagation scenarios. Zang et al. [98]investigated the generalization properties of a PINN trained on the nonlinear Schrödinger equation governing communication fiber propagation, demonstrating through systematic distance extrapolation experiments that the network produces accurate signal and pulse evolution predictions beyond the propagation distances present in the training data, a distance-generalization capability that conventional data-driven models lack and that reduces the need for per-link retraining in optical fiber communication system simulation. Ding et al. [70]addressed the virtual fabrication of microstructured optical fibers (MOFs) by incorporating the full free-boundary Navier-Stokes system governing molten glass dynamics, including continuity, momentum, and energy equations with surface-tension boundary conditions at the inner and outer fiber surfaces, into the PINN loss alongside a dedicated secondary network that tracks the time-evolving free-boundary positions; the resulting model captures capillary collapse dynamics, velocity and pressure distributions within the preform, and the coupled effects of temperature, feed speed, and draw speed on the inner and outer diameter profiles, with predictions validated against experimental fabrication measurements and enabling quantitative virtual process design without physical trials. Table 1 summarizes forward PINNs design applications in electromagnetism.

**Table 1.** PINNs for Forward Modeling

| Ref | Year | Application | Input | Output | Architecture | Parameter Size | Dataset | Governing Loss | Accuracy |
|---|---|---|---|---|---|---|---|---|---|
| [65] | 2022 | 2D linear/nonlinear optical scattering (microlenses, diffusers) | Refractive index distribution $n_s(z, x)$ | Scattered electric field components | MaxwellNet: U-Net | ~$8\times10^6$ parameters | 5,000 synthetic refractive index distributions | Maxwell's equations (curl-of-curl) residual | TE relative error 1.71-$3.02\times10^{-2}$ |
| [68] | 2025 | 3D metasurface diffraction (polarization, wavefront) | 3D refractive index map, illumination parameters | Scattered E-field components (Ex, Ey, Ez) | U-Net with tuning network | - | No labeled data | Vectorial Helmholtz equation residual with Floquet-Bloch BCs and PML | Near-field MAPE 0.12% (fixed illum.), 1.45% (generalized) |
| [69] | 2025 | Full-vector waveguide eigenmodes (strip waveguide) | (x, y) coordinates | Tangential E-field components, propagation constant $\gamma$ | 4DPINN: 5-layer FCNN, 16 neurons/layer | <1k parameters | 4000 collocation points | Fourth-order eigen-equation from Maxwell's equations | $\gamma$ errors $<10^{-4}$; field max AE $<-12$ dB |
| [70] | 2024 | Microstructured optical fiber (MOF) drawing | (r, z) coordinates | Velocities, pressure, inner/outer radii | Two networks: PINN (30×250) + free-boundary network (5×200) | - | No external data; uses drawing parameters | Navier-Stokes residuals + kinematic and normal stress BCs | Qualitative agreement with fabrication trends |
| [55] | 2021 | Time-domain EM fields in 1D cavity | (x, t) coordinates | Electric and magnetic fields | FCNN | - | Collocation points from 1D cavity | Time-domain Maxwell's curl equations | - |
| [56] | 2025 | 3D metal resonant cavity fields | (x, y, z) coordinates | Electric and magnetic field distributions | FCNN | - | Collocation points for 3D cavity | 3D Maxwell's equations residual | Good agreement with analytical solution |
| [57] | 2025 | 1D/2D unsteady EM wave propagation | (x, t) or (x, y, t) coordinates | EM field values | FCNN with Fourier features, periodicity, causality | - | Collocation points from 1D/2D test cases | Time-domain Maxwell's equations residual | - |
| [58] | 2025 | 2D transient EM fields in inhomogeneous media | (x, y, t) coordinates, $\varepsilon_r(x, y)$, $\mu_r(x, y)$ | Electric field $E_z$ | FCNN with 5-20 hidden layers, 64 neurons/layer | - | 10,000 IC/BC points; $N_{PDE}$ collocation points | Time-domain Helmholtz equation residual | Relative L2 errors $8.42\times10^{-2}$ to $1.53\times10^{-1}$ |
| [59] | 2024 | 3D EM problems with material discontinuities | (x, y, z, t) and material parameters | Magnetic field H, electric field E | FCNN with smoothed Heaviside, overlapping domains | - | Manufactured solutions and FEM comparisons | First-order Maxwell's equations residuals | Max relative error 0.4% (magnetostatics); $2\times10^{-3}$ (transient) |
| [60] | 2025 | High-frequency EM scattering (cylinders) | (x, y) coordinates | Electric field E | FE-PIRBN (RBF core + hash encoding) | - | 900 collocation points; data-free | Helmholtz equation with PML | Relative error 1.40-5.82% vs. numerical solvers |
| [61] | 2022 | Electrostatic/magnetostatic benchmarks | (x, y) coordinates | Electric or magnetic potential | FCNN | - | FEM ground truth for comparison | PDE residual + boundary conditions | Close agreement with FEM; transfer learning reduces training time |
| [62] | 2023 | 2D magnetostatic fields in devices (transformers, actuators) | (x, y) coordinates | Magnetic vector potential A, field intensity H | FCNN with mesh-assisted non-uniform sampling | - | Collocation points from coarse FEM mesh | 2D magnetostatic PDE residuals | Comparable to FEM |
| [63] | 2022 | Parametric 2D magnetostatic (EI-core electromagnet) | (x, y) and 10 design parameters | Magnetic field as function of space and parameters | DNN | - | FEM-generated data | Physics-informed energy functional | Validated against FEA across 10D parameter space |
| [64] | 2023 | 2D EM simulations with material interfaces | Graph nodes | EM field values | Physics-Informed Graph Neural Network (PI-GNN) | - | Mesh-based graph | Variational Maxwell formulation residual | - |
| [66] | 2023 | 2D nonlinear optical scattering (Kerr effect) | $n_s(z, x)$ and incident intensity $I_0$ | Scattered electric field | MaxwellNet with tunable convolutional kernels | ~$8\times10^6$ parameters | 5,000 diffuser geometries | Maxwell's equations residual with Kerr nonlinearity | TE error 1.71-$3.02\times10^{-2}$; TM error 2.52-$3.93\times10^{-2}$ |

## 3.2 Fluid Mechanics

PINNs provide a revolutionary approach in fluid mechanics, integrating the data-driven adaptability of deep learning with the stringent requirements of physical laws, such as the Navier-Stokes equations. In contrast to conventional "black-box" models, PINNs optimize a loss function that incorporates the residuals of partial differential equations (PDEs), enabling their operation despite sparse or noisy experimental data.

### 3.2.1 Navier–Stokes equations in laminar/turbulent regimes

PINNs were developed as an effective technique for solving the Navier-Stokes equations (NSE) that describe the fluid flows. These equations are crucial in laminar, transition, and turbulent flow regimes, and PINNs have shown tremendous potential in solving these complicated fluid dynamics flows. The numerical solution of the compressible/ incompressible Navier-Stokes equations (NSE) utilizing finite elements, spectral, and even meshless approaches has advanced significantly in computational fluid dynamics (CFD) over the past 50 years[99-102]. However, mesh generation is still an art and takes a lot of time for industrial complexity problems, and we are yet unable to seamlessly integrate data into current algorithms for real-world applications.

There are several models, developed for solving Navier–Stokes equations[103]: (1) Direct Numerical Simulations (DNS), solving NSE completely without any turbulence modelling, but consume more expensive computational cost (run time, and hardware). (2) Large Eddy Simulations (LES), provide a good balance in computational cost with an accurate result. Finally, (3) Reynolds-Averaged Navier-Stokes (RANS), is the good choice for industrial applications. Pure regression is used for solving spatial-temporal fluctuations in flow topologies[104]. PINNs has gotten a lot of interest for pure regression. It uses well-posed PDE information to produce data-driven predictions, even in data-scarce conditions[105]. PINNs and their derivatives have been studied extensively [106-116] for spatiotemporal and inverse modeling purposes. The most common use of hybrid regression is turbulence modeling[117]. Because of the broad spectrum of spatiotemporal scales in turbulent flows, resolving the smallest scales in a practical model remains a difficult job. As a result, it is usual to only resolve larger scales, such as in Reynolds averaged Navier-Stokes (RANS), or large-eddy simulations (LES), and then use an analytical model to describe the effects of small scales on larger scales. Machine Learning (ML) has been widely applied to closure modeling, including the evaluation of Reynolds stresses in RANS [118-124] and the generation of subgrid-scale statistics in LES[125-128].

As shown in Fig. 3a, illustrates that average velocity fields built on the Reynolds stress vector exceeded those based on the Reynolds stress tensor in a turbulent square duct flow (Re = 3200). Cruz et al. [129]pushed for the deviation of the Reynolds stress tensor. This is found on the discovery that mistakes in mean velocity emerge when the Reynolds stress tensor (obtained from DNS databases) is incorporated in the average momentum balancing computations. Matai & Durbin developed zonal $k-\omega$ model for incompressible adverse pressure gradient (APG) flow test cases. This model provides better performance than the standard model and once the result convergence occurs, the zones are unified. Proper orthogonal decomposition (POD)[130], a famous method for evaluating turbulence[131], has been used to reduce dimensions and extract features for ML models. Lui and Wolf [132] used POD to lower the dimensionality of turbulent airfoil data before training a neural network for reduced-order modeling applications. Other group [133, 134] proposes a training model enhancing the existing RANS turbulence model to forecast separation-induced transitional flows over low Reynolds number airfoils, as well as dynamic stall onset for pitching airfoils. This is training simulates steady state airfoil problems at various angles of attack. The learned model is applied to previously unseen steady-state cases with varied airfoil portions. As shown in Fig. 3b (Surface pressure coefficient for NACA 0012 airfoil), the training model can success to capture transition flow and separation for all airfoil cases. Sharma et al [104] perform PDE-preserving neural network (PPNN) for a lid-driven cavity with unseen parameters. Fig. 3c displays the velocity contours from PPNN, PPNN-partial, and baseline models, along with actual data, highlighting the benefits of PPNN at Re = 400. The base model predicts accurately for the first few steps, but errors quickly mount, resulting in a random, noisy, and unphysical solution field. PPNN, a PIML technique, outperforms the baseline ML model in terms of adaptability, consistency, and forecasting over the long run. Recently, Yazdani & Tahani invistigate a novel turbulence model $k$-$\omega$-PINNs turbulence model to Improves the precision and effectiveness of chaotic flow computations. As shown in Fig. 3d reduces the overestimation of turbulent kinetic energy and enhances the prediction of separation in the near-wall region as compared to the base RANS simulation. This is hybrid model that demonstrates the utility of PINNs for recognizing parameters and improvement for fluid dynamics.

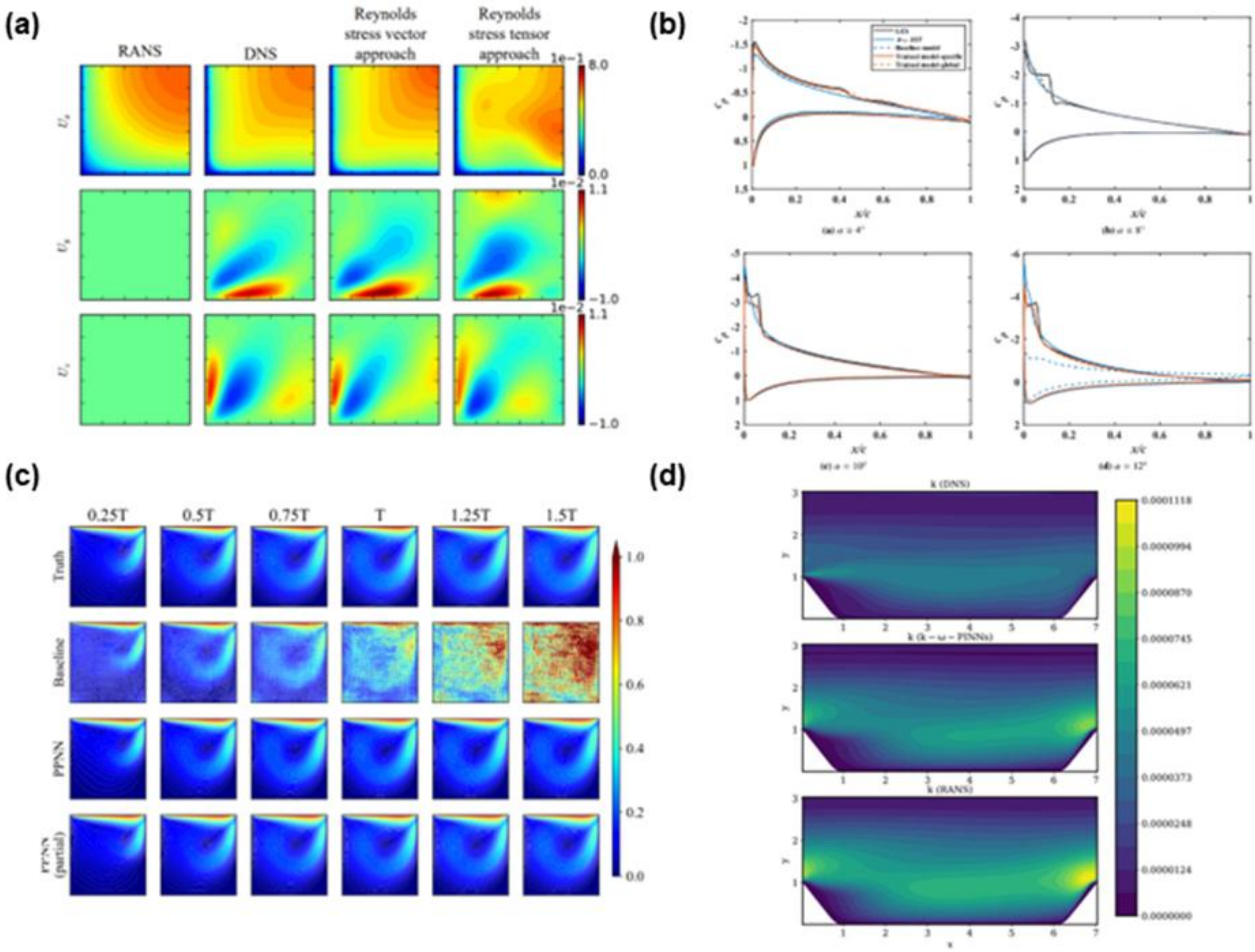

**Figure 3.** Application of ML for solving NSE: a) average velocity contour for a turbulent square duct flow at Re = 3200. b) Data comparison for Surface pressure coefficient for NACA 0012 airfoil. c) Comparison of velocity contour for a lid-driven cavity at Re = 400. d) Turbulent kinetic energy contour for $k$-$\omega$-PINNs compared with DNS and $k$-$\omega$ turbulence model.

### 3.2.2 Vortex dynamics and boundary-layer flows

Vortex dynamics refers to concentrated areas of vorticity that evolve throughout time. PINNs succeed here because they preserve the underlying physics more effectively than solely data-driven models. PINNs can accurately model Von Kármán vortex streets behind cylinders. Mao et al. (2020) [135]found that PINNs are especially successful for high Reynolds number flows, where standard solvers may require exceedingly fine meshes.

Early studies showed that PINNs can directly approximate solutions to incompressible Navier-Stokes equations in both laminar and turbulent regimes. The NSFnet [105, 112, 115] architecture demonstrated precise reconstruction of instantaneous flow fields without requiring pressure data as input by introducing velocity–pressure and vorticity–velocity formulations within PINNs. This study demonstrated that convergence and forecast accuracy depend on the proper balancing of physics and data losses[114].

Navier–Stokes Momentum Equations (2D Unsteady)

$$\frac{\partial u}{\partial t} + u\frac{\partial u}{\partial x} + v\frac{\partial u}{\partial y} = -\frac{\partial p}{\partial x} + \nu\left(\frac{\partial^2 u}{\partial x^2} + \frac{\partial^2 u}{\partial y^2}\right) \quad (16)$$

$$\frac{\partial v}{\partial t} + u\frac{\partial v}{\partial x} + v\frac{\partial v}{\partial y} = -\frac{\partial p}{\partial y} + \nu\left(\frac{\partial^2 v}{\partial x^2} + \frac{\partial^2 v}{\partial y^2}\right) \quad (17)$$

Continuity Equation

$$\frac{\partial u}{\partial x} + \frac{\partial v}{\partial y} = 0 \quad (18)$$

PINN Physics Residuals

$$f_u = u_t + uu_x + vu_y + p_x - \nu\,(u_{xx} + u_{yy}) \quad (19)$$

$$f_v = v_t + uv_x + vv_y + p_y - \nu\,(v_{xx} + v_{yy}) \quad (20)$$

$$f_c = u_x + v_y \quad (21)$$

Total Loss Function

$$L = \lambda_d\, L_{data} + \lambda_p\, L_{physics} + \lambda_b\, L_{BC} + \lambda_i\, L_{IC} \quad (22)$$

Streamfunction Formulation

$$u = \frac{\partial \psi}{\partial y} \quad (23)$$

$$v = -\frac{\partial \psi}{\partial x} \quad (24)$$

Vorticity Transport Equation

$$\omega = \frac{\partial v}{\partial x} - \frac{\partial u}{\partial y} \quad (25)$$

$$\frac{\partial \omega}{\partial t} + u\frac{\partial \omega}{\partial x} + v\frac{\partial \omega}{\partial y} = \nu\,\nabla^2 \omega \quad (26)$$

Later research investigated vortex-dominated flows, including vortex-induced vibration (VIV). With minimal training datasets, PINN-based models were able to rebuild unstable wake structures surrounding bluff bodies and forecast flow-induced forces with force-coefficient prediction errors of only a few percent. These findings demonstrated how physics-constrained learning may capture periodic vortex shedding mechanisms at a lower computing cost than high-fidelity CFD[136].

Reynolds-averaged Navier-Stokes (RANS) equations for laminar and turbulent boundary layers can also be solved using PINNs. The durability of physics-based learning even when turbulence closures are not explicitly mandated is demonstrated by the results, which revealed prediction errors below 1% in flows with severe pressure gradients[112].

Forward simulation of free-convective boundary layers has been significantly enhanced by recent advances utilizing derivative networks guided by physics. Compared to conventional numerical solvers, these compact neural models showed promise for lower computational costs by capturing steep gradients and boundary behavior over a range of parameter regimes[137].

Therefore, the research shows that PINNs offer mesh-free simulation, lower data requirements, and excellent physical consistency, making them a suitable framework for forward modeling of vortex dynamics and boundary-layer flows. Although laminar boundary layers, vortex-induced vibration, and interface-driven flows have all been accurately predicted, more methodological developments are needed to accurately simulate totally turbulent and extremely unstable vortex regimes.

## 3.3 Astronomy and Space Physics

### 3.3.1 PINNs in Astronomy

PINNs were applied to study pulsar magnetospheric models, focusing on axisymmetric cases. They successfully reproduced both axisymmetric and non-dipole configurations, while showing that energy losses differed by no more than a factor of three from the classical dipole case. This work establishes a reliable elliptic PDE solver for astrophysical problems and highlights PINNs as an efficient and generalizable approach for future three-dimensional magnetospheric modeling.[138]

This study expands the application of PINNs to eigenvalue problems in black hole perturbation theory. A supervised learning approach is applied to the Regge-Wheeler and Teukolsky equations, which describe gravitational perturbations of Schwarzschild and Kerr black holes. While earlier unsupervised PINN methods successfully calculated quasinormal mode frequencies, this work investigates their ability to compute higher overtones beyond the fundamental modes. The results show that PINNs can recover complex frequencies for different spin values and mode parameters, although approximation errors increase with higher rotation and overtone numbers due to residuals in the training data.[139]

We modelled radiative transfer in exoplanetary atmospheres using PINNs, focusing on scattering, by integrating differential equations into the loss function for improved accuracy and adaptability. A 1D isothermal model with pressure-dependent absorption and Rayleigh scattering was tested: PINNs accurately reproduced transmission spectra under pure absorption and computed both direct and diffuse stellar light under Rayleigh scattering. Preliminary results are promising, though errors from simplifications and challenges with complex conditions (e.g., clouds and hazes) remain for future development.[140]

The study examines linear matter perturbations in the Standard Cosmological Model and a modified gravity model using PINNs to integrate differential systems. The method is applied to the matter perturbation equation and compared with structure growth data ($f\sigma 8$), while also allowing the calculation of an exact error bound directly from the network outputs and residuals. Using updated datasets, the results provide tighter constraints on the ($\Omega m - \sigma 8$) plane than in previous works.[141]

### 3.3.2 PINNs in Cosmology

PINNs are powerful tools for solving differential equations by embedding physical laws into learning. In this study, a Schrödinger–Poisson informed neural network is developed to simulate the gravitational collapse of fuzzy dark matter in 1D and 3D by solving the nonlinear SP equations. The model accurately reproduces mass conservation, density profiles, and structure suppression, validated against analytical and numerical benchmarks. The results highlight the potential of PINNs for efficient and scalable modeling of fuzzy dark matter and other astrophysical systems, overcoming the limitations of traditional numerical solvers in handling nonlinear, multiscale phenomena, particularly the fine wave features of FDM on cosmological scales.[142]

PINNs provide a predictive framework that merges statistical patterns with physical constraints by enriching the loss function. Hydrodynamic simulations are essential in cosmology but computationally expensive, while dark matter simulations are much faster. This gap motivated machine learning approaches for baryon inpainting, though reproducing hydrodynamic scatter remains challenging. Here, PINNs are applied for the first time to baryon inpainting, combining advanced neural architectures with baryon conversion efficiency theory in the loss function. A Kullback-Leibler–based comparison further enforces scatter reproduction. Using the simba suite of cosmological simulations, the method extracts baryonic properties and achieves accurate predictions from dark matter halo features. The results show recovery of the fundamental metallicity relation and reproduction of scatter consistent with target simulations.[143]

PINNs are widely applied to solve differential equations but lack mechanisms for uncertainty quantification. To address this, a two-step training procedure with Bayesian Neural Networks is used, incorporating error bounds from PINNs to define a heteroscedastic variance and improve uncertainty estimates. The method is applied to forward problems and parameter estimation in cosmological inverse problems.[144]

The Square Kilometre Array Observatory (SKAO) will enable direct observation of the Epoch of Reionization by mapping neutral hydrogen across redshifts. While radiative transfer simulations can reproduce these processes, they are computationally expensive and limited in scale and resolution. To address this, the study introduces PINION, a physics-informed neural network trained on C2-Ray outputs with a physics constraint on the reionization chemistry equation. PINION can accurately predict the full 4D hydrogen fraction history between $z = 6$ and 12 using only five snapshots. Its predictions closely match C2-Ray results for $z > 7$, but accuracy decreases for $z < 7$. The work highlights how PINION could be further improved with additional inputs and generalized to large-scale simulations.[145]

Machine learning has become important in many fields due to its success in solving diverse problems. One promising application is training artificial neural networks (ANNs) to solve differential equations without the need for numerical solvers. This approach offers an alternative to traditional numerical methods, with reduced memory requirements, easier parallelization, and lower computational cost. In this work, ANNs are trained to represent a set of solutions to the differential equations governing the cosmic background dynamics for four different models: the $\Lambda$CDM model, the Chevallier-Polarski-Linder parametric dark energy model, a quintessence model with exponential potential, and the Hu-Sawicki f(R) model.[146]

### 3.3.3 PINNs in Space Weather

Studying the magnetic field on the solar surface is crucial for understanding solar and heliospheric activities that drive space weather. Surface flux transport (SFT) modeling enables the simulation and analysis of magnetic flux evolution, providing insights into solar activity mechanisms. In this work, we develop a novel PINN-based model to study the evolution of bipolar magnetic regions using SFT in one- and two-dimensional settings. The efficiency and computational feasibility of the model are validated against a numerical Runge–Kutta scheme. The mesh-independent PINN accurately reproduces the observed polar magnetic field with improved flux conservation. This advancement allows for more efficient and accurate simulations of solar magnetic flux transport and demonstrates the applicability of PINNs in solving advection–diffusion equations with a focus on heliophysics.[147]

Space weather refers to the dynamic conditions in the solar system, specifically the interactions between the solar wind—a stream of charged particles emitted by the Sun—and Earth's magnetic field and atmosphere. Accurate forecasting is crucial to mitigating potential impacts on satellite operations, communication systems, power grids, and astronaut safety. However, current coronal solar wind models, such as MULTI-VP, demand substantial computational resources. This paper introduces PINNs as a faster yet accurate alternative that respects physical laws. By combining physics and data-driven techniques, PINNs deliver rapid and reliable forecasts. The study shows that PINNs can significantly reduce computation times while maintaining results comparable to MULTI-VP, offering an efficient and dependable approach for solar wind forecasting (Figure 4).[148]

This paper presents a new numerical method for calculating the equilibria and dynamic structures of magnetized plasmas in coronal environments. The technique employs PINNs, which integrate the partial differential equations of the model. PINNs are applied to compute various magnetohydrodynamic (MHD) equilibrium configurations and to obtain exact two-dimensional steady-state magnetic reconnection solutions. The advantages and limitations of PINNs compared to conventional numerical codes are discussed, with suggestions for improvements. As a mesh-free method, PINNs can generate solutions and their derivatives almost instantaneously at any point in the spatial domain. The results pave the way for future developments of time-dependent MHD codes based on PINNs.[149]

### 3.3.4 PINNs in Gravitational Waves

To address the agile capture problem for a test mass (TM) inside a drag-free satellite,an essential challenge for space-based gravitational wave detection missions, a new control framework called PINN-DDPG was introduced. After orbital deployment, the TM becomes difficult to control due to its risk of colliding with the satellite cavity, the extremely small electrostatic actuation forces (on the order of micronewtons), and possible gaps in sensor measurements.

The proposed approach combines a PINNs with a Long Short-Term Memory (LSTM) network to form a PINN-LSTM module that accurately forecasts the TM’s future state by integrating physical laws with time-series observations. These predictions are then supplied to a Deep Deterministic Policy Gradient (DDPG) controller, which selects optimal control actions using both current and anticipated system states.

Simulation results showed clear advantages. The PINN-LSTM converged 60% faster than a standard PINN, and the full PINN-DDPG controller reduced the time required to stabilize the TM’s position and velocity errors by 70% compared with a conventional DDPG controller. The algorithm also generalized well across diverse initial conditions and outperformed alternatives such as Model Predictive Control (MPC). The authors further used visualization techniques to make the neural network’s training dynamics more interpretable. These results demonstrate that the method provides precise and agile control in a highly complex environment, making it promising for future space missions.[150]

Another domain where PINNs have proven useful is gravity-field modeling in astrodynamics. Traditional models rely on spherical harmonics, which require millions of coefficients, are not compact, and struggle to represent discontinuities such as mountain ranges. They also diverge near a body's surface, limiting their use for landing missions or small-body navigation. As an alternative, a PINN was trained to learn a single scalar gravitational potential $\hat{U}$, embedding the physical constraint $a = -\nabla U$directly into the loss function to guarantee a conservative field and improve data efficiency.

For Earth’s gravity field, characterized by sparse, discontinuous structures, the PINN approach offered “strong advantages,” achieving an order-of-magnitude reduction in parameters for comparable accuracy and providing significantly faster evaluations suitable for onboard computation. For the Moon, however, the benefit was “less prominent”: its gravity field, shaped by a highly irregular, crater-dominated surface, has higher entropy, making spherical harmonics relatively more effective. The authors conclude that PINNs are especially powerful for bodies with structured, discontinuous features, while high-entropy fields may require additional data or physics-based constraints.[151]

PINNs have also been applied to wave propagation and, in particular, to Full Waveform Inversion (FWI) in seismology. FWI is a powerful but computationally expensive imaging technique, and deep learning approaches are often hampered by limited observational data, since seismic measurements are typically sparse. This limitation was addressed by embedding the 2D acoustic wave equation directly into the network's loss function, reducing dependence on labeled data. One network predicts the wavefield over space–time, while a second, smaller network estimates the unknown wave speed.

Tests on synthetic case studies showed that although PINNs deliver "good results" for the forward problem, traditional numerical solvers remain more "efficient and accurate." However, for the inverse FWI problem, the PINN framework "yields excellent results" with "limited computational complexity." Its mesh-free nature provides substantial benefits, including automatically satisfying absorbing boundary conditions—a major challenge for classical solvers—naturally enforcing free-surface constraints, and efficiently modeling multiple sources through superposition.[152]

### 3.3.5 PINNs in Stellar

Understanding stellar properties requires detailed modeling of observed spectra using stellar atmosphere models. Most observable absorption features originate in the photosphere, the layer where photons begin to escape, making it the primary region that spectral models must characterize. Stellar spectral modeling typically involves two main steps: constructing a stellar atmosphere model and generating synthetic spectra.

The first step determines the atmospheric structure by computing pressure, temperature, and electron density as functions of optical depth. This is traditionally done using pre-computed model grids such as ATLAS (Castelli & Kurucz, 2003), MARCS (Gustafsson et al., 2008), and PHOENIX (Allard, 2016). The second step, spectral synthesis, uses the fixed atmospheric structure to perform radiative transfer calculations and produce synthetic spectra (e.g., Kurucz & Avrett, 1981; Sneden et al., 2012; Gerber et al., 2023; Piskunov & Valenti, 2017).

A central tool in this process is the ATLAS suite of codes (Kurucz, 1970; Kurucz & Avrett, 1981; Kurucz, 1996; 2014), developed by Robert Kurucz. ATLAS-12 computes six atmospheric quantities, temperature, pressure, density, electron density, opacity, and radiative acceleration, across 80 depth layers, resulting in a 480-dimensional output. Mapping stellar parameters to such high-dimensional atmospheric structures is challenging for standard neural networks, which lack the inductive biases needed to capture the underlying physics.

However, stellar atmospheres are governed by well-established physical laws, particularly hydrostatic equilibrium and radiative transfer. These constraints provide the necessary structure that conventional architecture fails to exploit. PINNs address this limitation by embedding the governing equations directly into the loss function, ensuring both data fidelity and physical consistency.

One such application is Kurucz-a1, a PINN designed to emulate 1D LTE stellar atmosphere models. Kurucz-a1 uses a dual-encoder architecture (Figure 4) that separates global stellar properties from local depth-dependent information, reflecting the true physical structure of stellar atmospheres. Four fundamental stellar parameters—effective temperature (Teff), surface gravity (log g), metallicity ([Fe/H]), and α-enhancement ([α/Fe])—are processed through a multilayer perceptron to generate 512-dimensional embeddings.

A second encoder transforms the 80 Rosseland optical depth points (τRoss) into 512-dimensional depth embeddings. These are combined with the stellar parameter embedding to produce 1024-dimensional representations at each depth layer. A three-layer MLP emulator (with hidden sizes [1024, 512, 256]) then predicts six atmospheric parameters for each layer: column mass density (ρx), temperature (T), gas pressure (P), electron density (XNE), Rosseland mean opacity (κRoss), and radiative acceleration (ACCRAD).

Kurucz-a1 achieves hydrostatic equilibrium at a level comparable to the ATLAS-12 solver and significantly outperforms an unconstrained MLP baseline, demonstrating the critical importance of physics-informed loss terms. Remarkably, Kurucz-a1 even attains slightly better hydrostatic balance on average than ATLAS-12, likely because ATLAS-12 relies on an older finite-difference scheme whose ~0.1% deviations from perfect equilibrium arise from discretization and legacy optimization techniques.[153]

### 3.3.6 PINNs in Astrophysics

The termination shock marks the boundary between the supersonic and the subsonic solar wind outflow, as it is the region at which the solar wind speed decreases sharply as a result of its interaction with the interstellar medium. It is a standing shock wave at the distance of about ∼100 AU as measured by Voyager 1 and Voyager 2 space probes. A complete 3D coverage of the entire heliosphere is impossible and especially for the outer parts of the Solar System the observational coverage is very limited. Thus, numerical modeling and simulations are often necessary to study the solar wind evolution throughout the entire heliosphere. In a number of recent works PINNs have been used for modeling simpler and classic shock wave problems.[135, 154-156].

While there have been machine learning studies on interstellar shock prediction or Coronal Mass Ejection transit times[157, 158], there have not been any PINNs studies investigating astrophysical shocks in the presence of a gravitational field.

The first paper using PINNs for modeling shocks while accounting for the gravitational force was (S P Moschou et al 2023 Mach). Their paper had three parts, inverse problem, data assimilation, and forward problem. For the forward problem, since they were modeling a shock wave, which appears as a discontinuity and because neural networks has problem prediciting discontinuous functions, a PINN term $\lambda_{WE} = \frac{1}{\epsilon(|\nabla\cdot\hat{u}| - \nabla\cdot\hat{u}) + 1}$ was introduced where $\hat{u}$ is the predicted velocity and ε is the parameter that controls the strength of the down-weighting. The weighting term reduces the importance of the PDE residuals near the shock compression area ($\nabla \cdot \hat{u} < 0$), thus allowing the neural network to gradually learn to predict the smooth regions before and after the shock with high accuracy and shrinking their distance until a thin discontinuity region is formed. During their study they perform some forward experiments in the time-dependent case. In the figure in the left column, we see a collection of experiments each of which uses as initial condition one of the intermediate steps computed by PLUTO and solves the forward problem until t = t_[150] = 200.

This work has shown that PINNs are still in the early stages of development and cannot yet match the robustness or accuracy of traditional physics-based PDE solvers for forward modeling. However, they can still deliver reasonable results when applied to short-time integrations. A major challenge is that PINNs often struggle with competing loss terms during gradient descent, which can lead to degraded performance, especially in systems with strong multi-scale variability, such as the density and pressure stratification caused by gravity in the solar atmosphere studied here. To address this issue, the conservative PDE system was reformulated to account for the $r^{-2}$ falloff in pressure and density, which significantly improved convergence and the recovery of the ground-truth solution, in some cases by factors of 2 to 16. In conclusion, PINNs, and machine learning methods more broadly, are beginning to see increasing adoption across various heliophysics applications.[159]

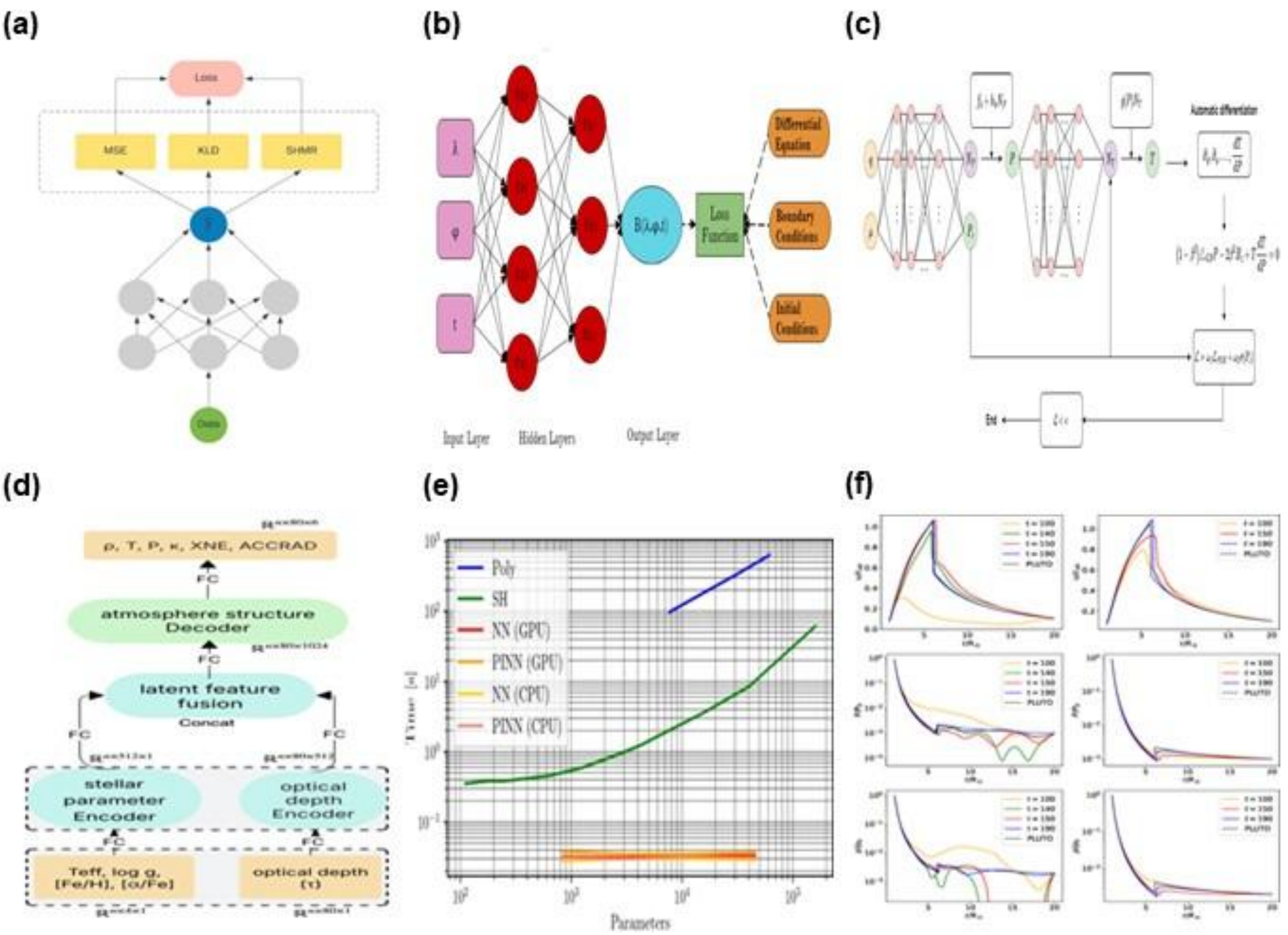

**Figure 4.** Some of the Forward Design PINNs architectures in Various specializations, and comparing the PINN model with other models. (a) PINN architectures In Cosmology.[143] (b) PINN architectures in Space Weather.[147] (c) PINN architectures in Astronomy.[138] (d) Kurucz-a1 neural network architecture with dual-encoder design.[153] (e) Comparing the total evaluation time of PINN with other gravity models.[151] (f) The predictive velocity for the Pluto simulation.[159]

## 3.4 Biomedical Applications

The application of PINNs in biomedicine has offered many advances in traditional forward solvers that make them more accurate and efficient for biophysical computations (such as electrophysiology and hemodynamics predictions.) Such fields traditionally use solvers like finite element analysis (FEA) which can be computationally heavy, especially if the computations are patient-specific, or time-dependent.[160]

### 3.4.1 Modelling of Hemodynamics in Arteries

PINNs are useful in cardiovascular flow modeling because they are able to work with data scarcity, and unknown boundary conditions, without the need for patient-specific solutions. They are able to work by embedding the governing Navier-Stokes equations into the neural network's loss function to make sure that the solution is within the accurate boundaries, enabling accurate predictions from sparse clinical data. Such models are used in one-dimensional pulse wave analysis, full three-dimensional hemodynamic reconstruction, and hybrid models as well.

One-dimensional (1D) models, which simplify vessels to their centerlines, are highly effective for simulating pressure and flow wave propagation across large arterial networks with minimal computational cost.[161] PINNs excel in this area by estimating complete hemodynamic states from limited clinical data. In a paper by,[162] PINNs were demonstrated to solve nonlinear PDEs (Burgers and Korteweg-de Vries equations) that can model phenomena such as pulse waves in viscoelastic arteries. Their key innovation was the use of Bayesian hyperparameter optimization to automatically tune the network, achieving accurate solutions for periodic and solitary waves. Their work provides a foundational example of using deep learning to solve complex PDEs with forms relevant to physiology. Furthermore, researchers were able to develop a multi-domain PINN framework to model pulsatile flow in a human aortic-carotid bifurcation.[163] This approach used sparse 4D flow MRI data measured at only a few cross-sections to reconstruct the absolute pressure field while avoiding the need for prescribed outlet boundary conditions, which are typically major sources of error in conventional models. This study was among the first to apply PINNs to real clinical hemodynamic data (Figure 5).

Another area of application is the diagnosis of localized vascular diseases like aneurysms or stenosis, which require the capture of 3D flow fields. Here, PINNs serve as mesh-free solvers for the 3D Navier-Stokes equations, but their performance becomes really sensitive to network architecture. In a paper by,[164] researchers were able to conduct a systematic comparison of seven neural network architectures (including CNNs, Deep Galerkin Method (DGM), and Fourier Networks) for solving steady 3D blood flow in idealized and aneurysm (patient-specific) geometries. They concluded that although standard fully connected networks offered a good balance of speed and accuracy, more sophisticated architectures like the DGM yielded superior accuracy in simpler geometries. However, all architectures struggled with accuracy in complex aneurysm sacs, which is a potential that can still be solved using PINNs. Moreover, researchers were able to offer a prediction of 4D hemodynamics (3D space + time) over a single cardiac cycle.[165] Using the specific models of some patients' aorta shape, researchers used PointNet-based architectures and integrated the Navier-Stokes equations in the loss function of the PINN. This research was useful because it emphasized that the choice of architecture of the neural network depends on vascular morphology, and they provided a valuable analysis linking prediction errors to regions of complex vortex formation (areas where the smooth flow of blood breaks down and forms swirls).

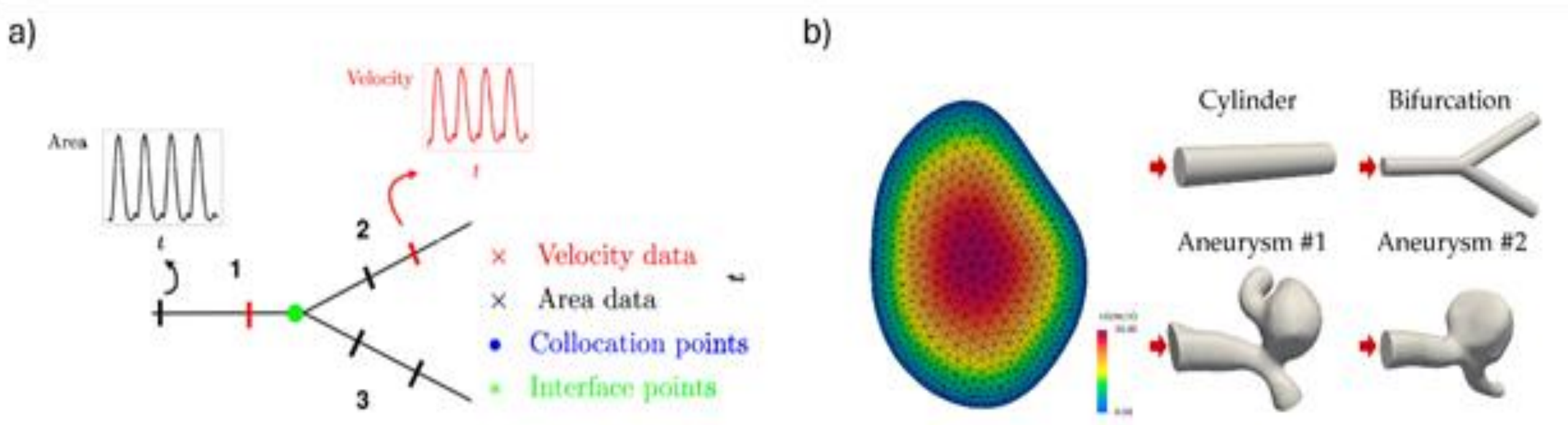

**Figure 5.** PINN approaches for arterial hemodynamics modelling. a) Topology of arterial network with 3 vessels shown and 1 bifurcation point.[163] b) The implementation of the parabolic inlet velocity profile and evaluated PINN architectures on four distinct geometries. These included two idealized models (a cylinder and a bifurcation) and two patient-specific intracranial aneurysm models.[164]

Beyond the modelling of pulse waves and the localized patient-specific anatomy, there has been some hybrid methodologies that combine the reduced-order physical models with the pattern-recognition power of neural networks. For instance, researchers trained neural networks not to replace, but to augment a reduced-order physics model (ROM) for predicting pressure loss across stenotic segments[166]. Their NN-augmented ROM used the fast ROM prediction as an input feature, allowing a subsequent NN to learn and correct the ROM's residual error against high-fidelity 3D CFD data. This hybrid model was useful because it required significantly less training data than a pure data-driven NN.[166] At the same time, this model achieved accuracy in predicting fractional flow reserve (FFR) comparable to the measurement variability of invasive FFR.

### 3.4.2 Thermal Characterization of Tissues for Cancer Detection and Treatment

The application of PINNs in the modelling of heat transfer in living tissues is useful because of its ability to adapt to the biological constraints of the body, which include tissue heterogeneity, and dynamic blood perfusion (which can be modelled by the Pennes' bioheat transfer equation) without needing to create a mesh around the area being studied.[167] Many studies have applied a PINN model using the Pennes bioheat transfer equation (BHTE) as the loss function of the network and were able to produce promising thermal characterization models for tissues. For instance, Pratama et al. applied a PINN with a restarting strategy to solve the steady-state Pennes equation in 2D cylindrical coordinates for breast models containing tumors.[168] Their results, which used a restarting strategy to solve the BHTE, were able to produce results comparable to other methods. This is because their restarting strategy constantly updated the weights of the neural network and minimized the loss function, leading to accurate results with less computational time. In another work by Bowman et al., researchers were able to demonstrate the effectiveness of PINNs in surrogating a numerical laser-tissue interaction solver (called Python Ablation Code, or PAC1D) for the 1D heat equation with a source term.[169] Their work critically established that PINNs could handle both insulated and convective boundary conditions, and that the choice of activation function in the neural networks significantly impacted performance. This work shows the potential of PINNs as valid alternatives for traditional solvers dealing with 1D tissue interactions with electromagnetic radiations (Figure 6).

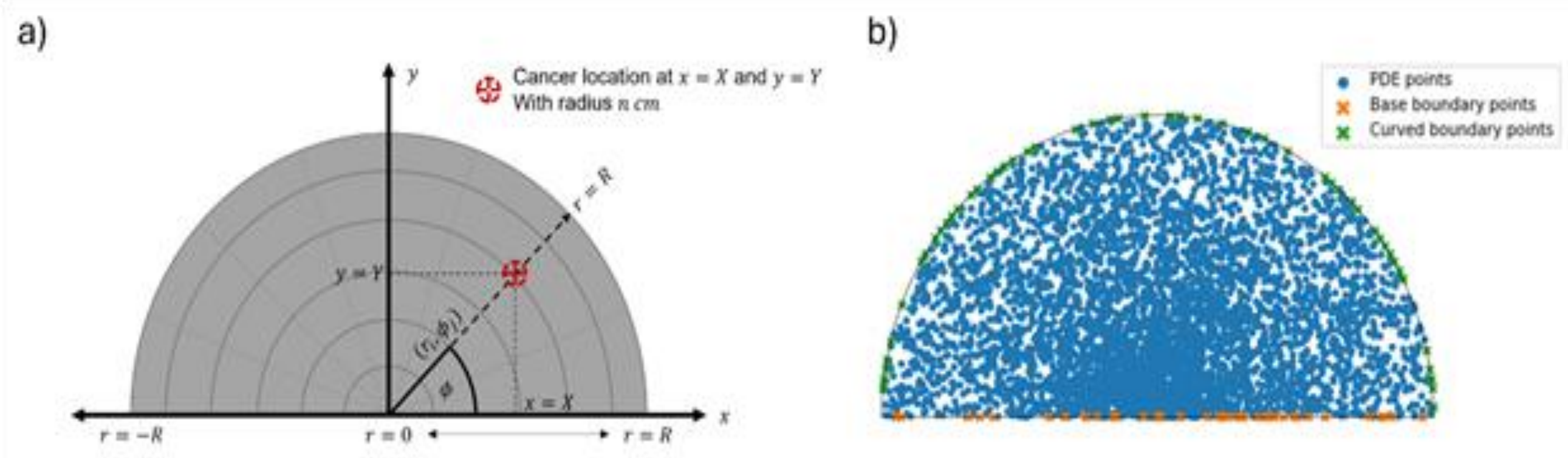

**Figure 6.** Problem setup of Ref [168] for cancer detection using thermal analysis. a) 2D cylindrical computational domain used for thermal analysis. b) Example of the random sampling of interior collocation points (blue) and boundary points (orange) used to train the PINN solver for the bioheat equation.

In addition to cancer treatments and predictions, PINNs were also used in thermal forward modelling, which is used to plan and study hyperthermia treatments. A 2025 paper developed a parameterized, multi-domain PINN for 3D triple-layered skin tissue containing an embedded tumor, gold nanorods, and a complex vascular network.[170] The model simulated nanoparticle-assisted photothermal therapy (NPTT) for over 600 seconds and showed 3D temperature fields and subsequent thermal damage metrics (namely the Arrhenius integral and CEM43) for both tumor and healthy tissue[170]. Their model was able to identify treatment optimization options to achieve tumor cell death while preserving surrounding healthy tissue over the treatment duration, which can be very useful for personalized treatment planning. Despite the increasing accuracy of PINNs in the characterization of thermal diffusion of tissues, one key advantage that PINNs offer are their computational efficiency after they have been trained. A 2023 paper demonstrated this by comparing their PINN for 3D breast cancer modeling against FEA, finding that while total training time was comparable, the simulation (prediction) time was approximately 12 times faster than the FEA solve time.[167] This rapid inference of PINN models makes them potentially deployable as core engines for real-time or iterative clinical treatment planning systems.

# 4. PINN Inverse Design

## 4.1 PINN Inverse Design: Nanophotonics and Integrated Photonics

Inverse problems in electromagnetics and photonics require the estimation of unknown quantities, distributed material parameters, device geometry variables, or source configurations from partial observations of the resulting field, a task that is inherently ill-posed and highly sensitive to noise because many different material distributions can produce similar field signatures. PINNs are structurally well-suited for inverse EM problems: unknown parameters are promoted to additional trainable variables alongside the network weights and are optimized jointly with the physics residual and data-fidelity loss, allowing the physical governing equations to regularize the inversion without requiring large collections of labelled (input, output) pairs. This section organizes PINN-based inverse contributions into two thematically distinct categories that reflect the two primary types of unknowns encountered in EM and photonic inverse problems: inverse EM scattering and permittivity profiling, which addresses the recovery of distributed material properties from observed scattered or near fields; and photonic device inverse design, which addresses the synthesis of device geometries and configurations from target spectral responses or field performance specifications.

### 4.1.1 Inverse Permittivity Profiling

The foundational PINN framework for inverse nano-optical scattering was established by Chen et al.[171], who constructed a PINN that embeds the frequency-domain TM Helmholtz equation with PEC boundary conditions as the physics residual and adds a data-fidelity term penalizing the difference between the predicted near-field and synthetic near-field observations generated by FEM, promoting the unknown permittivity $\varepsilon_r$ to a trainable scalar variable optimized simultaneously with the network weights; the approach was first validated on finite-size nanostructure arrays and multi-component nanoparticles for cloaking layer design, establishing the inverse-scattering loss architecture that Chen and Dal Negro [172] subsequently extended to the full-vector Maxwell system for simultaneous 2D and 3D retrieval of both complex electric permittivity $\varepsilon_r$ and magnetic permeability $\mu_r$ of scattering objects in the resonance regime. In the extended framework, the training loss combines three terms with adaptive trainable weights,

$$L(\theta, W_f, W_i, W_b, \varepsilon_r) = W_f\, L_f + W_i\, L_i + W_b\, L_b \tag{27}$$

where $L_f$ is the full-vector Maxwell PDE residual $L^2$ norm, $L_i$ is the complex near-field observation $L^2$ norm at measurement points, and $L_b$ is the boundary condition $L^2$ norm. the weights $W_f$, $W_i$, $W_b$ are updated via gradient ascent at each training step in the adaptive PINN variant, allowing the optimizer to dynamically reallocate training effort between physics satisfaction and data fidelity. The resulting framework achieves $L^2$ errors of ~$10^{-4}$ for the complex electric field, recovers both real and imaginary parts of $\varepsilon_r$ (including lossy materials) and $\mu_r$ simultaneously, and maintains relative noise errors below 10% for Gaussian noise amplitudes up to 40% of the maximum field amplitude, robustness that makes it applicable to experimental SNOM data with realistic noise levels; importantly, the adaptive PINN variant resolves high-index scatterers ($\varepsilon_r = 5 + j$) where standard PINNs with fixed loss weights diverge during training, and the 3D extension retrieves $\varepsilon_r = 2.57 \pm 0.45$ for a spherical object of known true permittivity $\varepsilon_r = 3$. Applications demonstrated include scanning near-field optical microscopy (SNOM), biomedical imaging, and characterization of metamaterial devices (Figure 7).

Hu et al. [173]generalized unsupervised PINN-based inverse scattering to the multi-frequency regime by formulating the inverse problem as a PINN optimization over scattered field observations at multiple frequencies simultaneously, introducing a frequency scale factor that normalizes the gradient magnitudes across frequency channels to correct the imbalance caused by wavelength-dependent field amplitudes, and deploying a dynamic sampling strategy that concentrates collocation points in high-residual regions during training; the method was validated on four numerical and one experimental benchmark encompassing electrically large and high-contrast cylindrical scatterers, showing robustness to measurement noise and superior accuracy under distribution shift compared to traditional data-driven deep learning methods that require large training datasets. A two-step cascade extension[174] further improved accuracy by first computing a prior permittivity estimate using the classical distorted finite-difference frequency-domain iterative method (DFIM), which provides a physically plausible initialization, and then embedding this prior as an additional data term in the PINN loss to constrain the solution toward physically realizable permittivity distributions. This combination inherits the noise robustness of physics-informed optimization while exploiting the structured prior knowledge from classical algorithms, reducing convergence time, and improving accuracy for high-contrast scatterers. Piao et al. [175]addressed the heterogeneous inverse problem with unknown interface locations, where neither the material values nor the boundaries between them are known a priori, by proposing a domain-adaptive PINN (Da-PINN) that introduces a learnable interface-location parameter controlling a soft domain decomposition; during training, the subdomain boundaries are updated simultaneously with the material parameters and network weights through adaptive gradient descent, with electromagnetic interface conditions (continuity of tangential field components) enforced in the loss to ensure physical consistency across the discovered boundaries. Zeng et al. [176]formulated permittivity inversion for electric field distribution problems under the Poisson equation framework in 2D and 3D, using a sigmoid-function-based dielectric distribution representation that constructs a smooth spatial permittivity model from a small set of learnable region parameters, recovering the relative permittivity from sparse electric field measurements at a set of collocation points with results consistent with FEM benchmarks and demonstrating that PINN-based inversion under the Poisson equation is viable for engineering dielectric structures from low-density field observations. Baldan et al. [177]addressed the parametric scalability limitation inherent in per-instance PINN training, where each new geometry or material combination requires a full retraining from scratch, by proposing a hypernetwork architecture that takes the system parameters (geometric dimensions and material properties) as input and outputs the complete set of PINN weights, enabling the hypernetwork to be trained once across a distribution of inverse EM problem instances and then instantly generate task-specific PINN weights for any new configuration through a single forward pass of the hypernetwork, eliminating the need for repeated retraining.

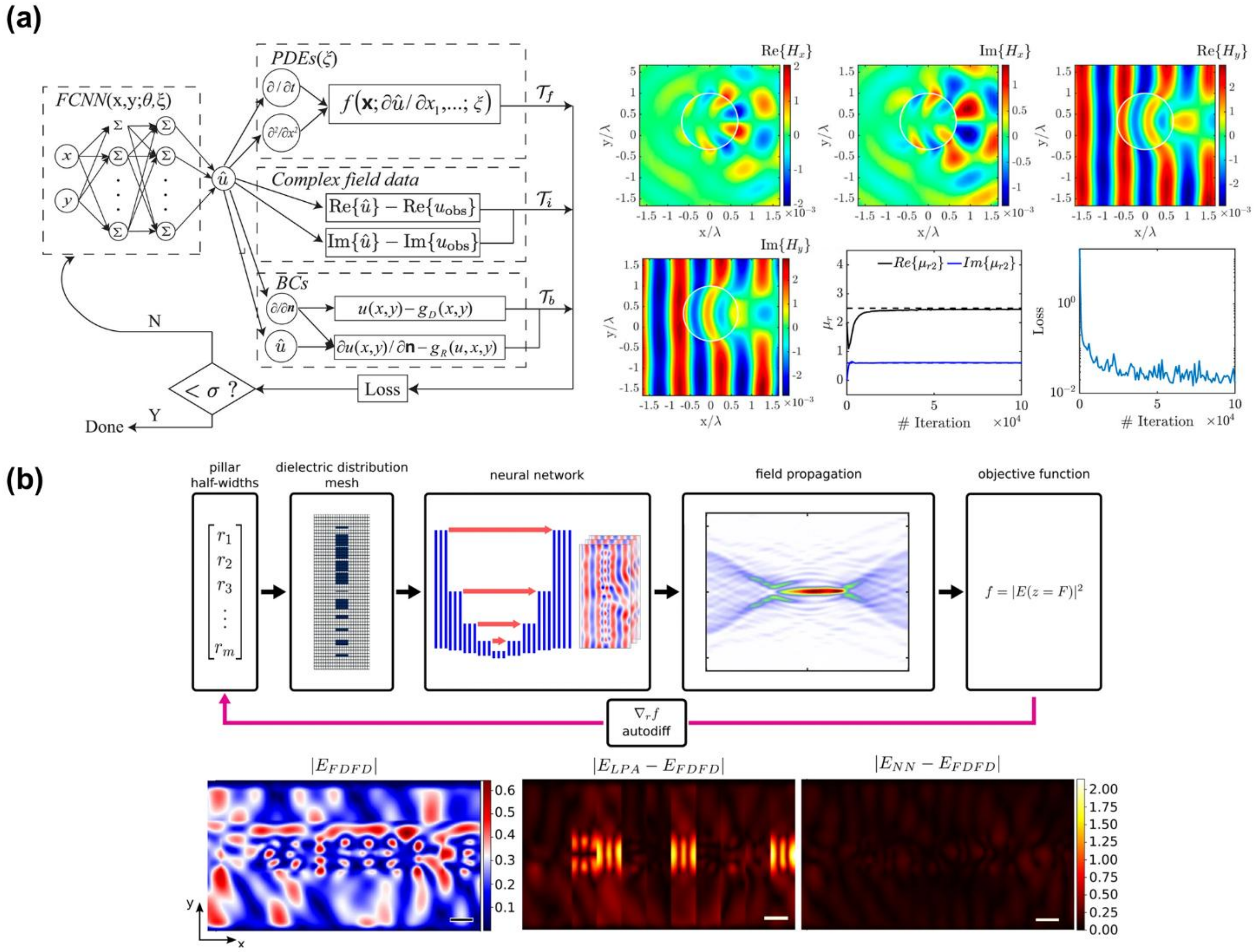

**Figure 7.** PINNs frameworks for inverse electromagnetic modeling and parameter retrieval. (a) Schematic illustration of a PINN used to solve the parameter retrieval problem in near-field microscopy. A fully connected neural network (FCNN) takes spatial coordinates and trainable parameters as inputs and outputs the surrogate solution of the governing partial differential equation (PDE). The loss function enforces consistency with Maxwell's equations, complex field measurements, and boundary conditions while simultaneously optimizing the network weights and unknown material parameters until the loss falls below a predefined threshold. Real and imaginary components of the complex magnetic fields $H_x$and $H_y$are used as training data, and the convergence of the retrieved complex permeability $\mu_{r2}$together with the total loss evolution illustrates the optimization process. [172](b) Example of an inverse design workflow using a neural-network-assisted electromagnetic solver. The structural parameters describing the meta-optic were grouped into overlapping batches and converted into dielectric distributions that served as inputs to the neural network. The network predicts local field patches that are stitched together and propagated using the angular spectrum method. The resulting field was compared with the target response to form the objective function, which was backpropagated using automatic differentiation to update the structural parameter distribution. [178]

### 4.1.2 Inverse Design of Photonic Devices

At the device level, PINN-based inverse design embeds physics constraints specific to the target device class, waveguide transfer functions, resonance conditions, diffraction propagation models, or fiber mode equations directly into the learning objective, constraining the optimization to explore only physically realizable device geometries and thereby regularizing the otherwise highly underdetermined mapping from target performance to design parameters. Pan et al. [179]proposed EAM-PINN (PINN with Embedded Analytical Models) for the inverse design of multilayer dielectric-loaded rectangular waveguide devices, deriving a closed-form $S_{21}$ transfer function model from waveguide theory and embedding it as a hard analytical constraint that replaces part of the PINN loss, reducing the loss scale and improving optimization conditioning, while additionally enforcing physical bounds on the permittivity through a softmax-based hard-constraint activation function; the resulting EAM-PINN was demonstrated on three increasingly complex inverse tasks, analytical solution recovery, bandpass filter design, and dispersive delay line synthesis, achieving higher solving efficiency and stronger convergence stability than a standard PINN without the analytical model embedding. Torabi et al. [180]introduced a physics-informed graph neural network (PI-GNN) for the inverse design of microring resonator biosensors at 1550 nm, representing the device as a heterogeneous graph in which nodes correspond to physical components (ring, bus waveguide, coupling region, surrounding medium, substrate) with edges encoding their physical interactions, and embedding both the resonance condition $m\lambda_{res} = n_{eff}\ 2\pi R$ and the spectral sensitivity requirement $\frac{\Delta\lambda}{\Delta n_{clad}}$ as physics-informed loss terms that penalize designs violating sensing constraints; trained on 2,500 analytically generated and FDTD-simulated device geometries, the PI-GNN achieves an RMS relative geometry error of 3.46% on the FDTD test dataset, exhibiting stable monotonic convergence superior to diffusion-model baselines while satisfying both resonance alignment and sensitivity requirements throughout optimization. Luo et al. [181]demonstrated that a hybrid physics-informed mode solver trained to satisfy the vectorial Helmholtz equation for fiber eigenmodes can be coupled with particle swarm optimization (PSO) to invert the fiber refractive-index profile from target mode performance specifications: for three-ring and four-ring multi-step-index fibers, PSO navigates the large discrete-valued refractive index (RI) parameter space guided by the differentiable PINN mode solver to identify optimized profiles ensuring low inter-mode crosstalk; for graded-index fibers, the same framework fine-tunes continuous RI profiles to minimize root-mean-square mode group delay, targeting long-haul transmission requirements without any training data from full wave simulations.

Liang et al. [182]applied a two-stage physics-guided neural network to the inverse design of photonic-plasmonic nanodevices for superfocusing applications, addressing the severe input-output dimensionality mismatch between the target far-field specification (a few focal spot parameters) and the design space (dozens of geometric variables) by using improved coupled-mode theory (CMT) for optical waveguides to compute physical intermediate quantities, coupling coefficients and propagation constants, that bridge the two stages and inject physically meaningful constraints at each training step, enabling accurate prediction of nanodevice geometries that achieve prescribed superfocusing performance. Pan et al. [183]extended physics-informed inverse design to polarization-multiplexed metasurface beam splitters, devices that simultaneously control right-handed and left-handed circular polarization channels, by training a modified U-Net with two decoupled decoder branches, one per polarization channel, embedding the Jones matrix formalism and angular spectrum propagation derived from scalar diffraction theory as physics loss terms that ensure the predicted phase distributions are consistent with the target far-field patterns; the resulting network achieves a phase retrieval mean squared error (MSE) of $4.3\times10^{-3}$ and a mean polarization extinction ratio of 34.11 dB at 1550 nm, outperforming purely data-driven U-Nets and demonstrating generalization to arbitrary target patterns not seen during training. Zhelyeznyakov et al. [178]addressed the fundamental limitation of conventional metalens inverse design, where the local phase approximation (LPA) independently optimizes each meta-atom and ignores inter-element optical coupling, by training a data-free PINN surrogate purely on 2D frequency-domain scalar Helmholtz residuals without any FDFD simulation data, and using its differentiability to perform gradient-based large-area metalens optimization that accounts for coupling effects across the entire aperture. The surrogate achieved a relative field prediction error of ~0.21 compared to FDFD, and the metalens designs produced through PINN-guided optimization delivered experimentally validated intensity improvements of up to 53% over LPA-based designs at NA = 0.9. The data-free PINN framework of Xiao et al.[85], which in Section 3.1.2 solved the forward EM scattering of etched cylindrical holes with an error of $\sim 10^{-6}$, was shown to extend naturally to inverse design by repurposing the same physics-trained network to optimize the dielectric constant distribution that produces a target scattered field, achieving retrieval errors of $\sim 10^{-4}$ and confirming that the data-free approach functions as a unified forward-inverse platform that obviates the need for separate forward and inverse solvers for nanoscale EM structures. Table 2 listed recent PINNs inverse design application in nanophotonics

**Table 2.** PINNs for Inverse Design

| Ref | Year | Application | Input | Output | Architecture | Parameter Size | Dataset | Governing Loss | Accuracy |
|---|---|---|---|---|---|---|---|---|---|
| [172] | 2022 | Imaging and parameter retrieval of photonic nanostructures from near-field data | (x, y) or (x, y, z) coordinates, near-field observations | Complex E/H fields, retrieved $\varepsilon_r$ and $\mu_r$ | FCNN (2D: 4×64; 3D: 3×20) with adaptive loss weights | - | Synthetic FEM data (2D 200×200 grid, 3D 50×50×30) | $w_f L_f$ (PDE) + $w_i L_i$ (data) + $w_b L_b$ (BC) | L2 errors for fields ~$10^{-4}$; retrieved $\varepsilon_r$ within few % of true |
| [178] | 2023 | Large-area metalens inverse design (1 mm aperture) | Target focal spot intensity | Optimized pillar half-widths | U-Net (differentiable surrogate) | - | No training data; PINN trained by minimizing PDE residual | L1 norm of PDE residual (Maxwell's equations) | Relative field error μ=0.21 (vs FDFD) |
| [171] | 2020 | Inverse scattering in nano-optics (periodic arrays, coated cylinders) | (x, y) coordinates, scattered field data | Electric field $E_z$, retrieved $\varepsilon_r$ | FCNN (4×64) or ResNet | - | Synthetic FEM data (e.g., 201×201 grid) | $w_f L_f$ (PDE) + $w_i L_i$ (data) + $w_b L_b$ (BC) | L2 errors 2.82% (periodic ε=3), 5% (ε=12), 0.51% (single cylinder) |
| [173] | 2023 | Multi-frequency electromagnetic inverse scattering | (x, y) coordinates, scattered field at multiple frequencies | Total fields and $\varepsilon_r$ distribution | PINN with frequency scale factor and dynamic sampling | - | Synthetic and experimental scattered field data | PDE residual + data misfit | Good accuracy and robustness |
| [174] | 2024 | EM inverse scattering with a priori knowledge | (x, y) coordinates, scattered field + prior $\varepsilon_r$ estimate from DFIM | Total fields and refined $\varepsilon_r$ | Cascaded PINNs | - | Synthetic and experimental data | Data loss (including prior) + PDE residual | Improved accuracy over DFIM and standard PINN |
| [175] | 2024 | Inverse problems in heterogeneous media with unknown interface | (x, y) coordinates, scattered field | Fields, $\varepsilon_r$, and interface location | Domain-adaptive PINN (Da-PINN) with learnable interface parameter | - | Synthetic data for 2D heterogeneous media | PDE residual + interface conditions + data misfit | - |
| [176] | 2023 | Electric field distribution and permittivity inversion from sparse measurements | (x, y) or (x, y, z) coordinates, sparse E-field measurements | Electric field and $\varepsilon_r$ distribution | PINN with sigmoid-based material model | - | Sparse E-field measurements from FEM benchmarks | Poisson equation residual + data misfit | Consistent with FEM benchmarks |
| [177] | 2023 | Inverse electromagnetic problems (parametric) | System parameters (geometry, material) | PINN weights for a specific instance | Hypernetwork that outputs weights of target PINN | - | Distribution of inverse problem instances | PDE residual for the target problem | - |
| [179] | 2024 | Inverse design of multilayer dielectric-loaded rectangular waveguide devices | Target $S_{21}$ response | Layer thicknesses $s_i$ | EAM-PINN: FCNN + hard constraint + embedded analytical $S_{21}$ model | - | 1000 frequency sampling points | Target loss (e.g., $S_{21}$ deviation) + $L_{sf}$ | Retrieving case: relative L2 error <0.33% |
| [180] | 2026 | Inverse design of microring resonator biosensors | Target spectral characteristics ($\lambda_o$, Δλ) | Geometry (R, w, h, g) | Physics-Informed Graph Neural Network (PI-GNN) | - | 2,500 geometries from analytical and FDTD models | $L_{data}$ (MSE) + $L_{phys}$ (resonance condition) | RMS-RGE 5.14% (analytical), 3.46% (FDTD) |
| [181] | 2025 | Inverse design of optical fibers (MSIFs, GRIFs) | Target mode properties | Optimized refractive index profile | Hybrid physics-informed + data-driven neural operator (DeepONet-like) | - | 500 labeled + 4000 unlabeled RIPs | $w_{data} L_{data}$ + $w_{res}$-hyb $L_{phy}$-$R_{hyb}$ + $w_{res}$-phy $L_{phy}$-$R_{phy}$ | $n_{eff}$ correlation >0.99987; field correlation ~0.997-0.999 |
| [182] | 2022 | Inverse design of photonic-plasmonic nanodevice for superfocusing | Taper angle θ and AgNW radius r (or Bezier curve parameters) | TM0 intensity spectrum (161-dim) | Physics-guided two-stage NN (first stage: coupling coefficients) | - | 255 (cone) + 4500 (custom shape) COMSOL simulations | MAE between predicted and target spectrum | Predictive accuracy 98.5%; coupling efficiency improved from 63.9% to 83.5% |
| [183] | 2025 | Inverse design of polarization-multiplexed metasurface beam splitters | Target far-field intensity pattern | Phase distributions $\varphi_{RCP}$, $\varphi_{LCP}$ | Modified U-Net with dual decoders | 34.5M parameters | Synthetic target patterns | $L_{MSE} + \lambda_1 L_{energy} + \lambda_2 L_{smooth}$ | Phase retrieval MSE $4.3\times10^{-3}$ (three-spot), 0.046 (chess knight) |

## 4.2 Fluid Mechanics Inverse Design Using PINNs

Inverse design utilizing PINNs has emerged as a preeminent approach for addressing topology optimization and parameter identification challenges across several application fields. The concept redefines conventional optimization issues as variational problems in which neural networks acquire parameterized design representations that minimize objective functions while adhering to restrictions imposed by physical laws[184]. This method obviates the necessity for pre-established parameterization schemes and facilitates a more efficient exploration of high-dimensional design spaces compared to traditional techniques.

PINNs include governing PDEs into neural network loss functions as shown in Eq.1, allowing for meshless solution and inference of fields and parameters from sparse or indirect data. PINNs have been frequently used to solve fluid inverse issues where boundary conditions, geometry, or constitutive characteristics (such as viscosity and permeability) are unknown or difficult to measure directly[154]. The following sections discuss their applications in nanofluidic and microfluidics inverse design, with a focus on reconstructing channel configuration from flow data and predicting material or transport attributes.

Flow behavior is regulated by high surface-to-volume ratios and delicate pressure-velocity correlations at the micro and nanoscales. Traditional inverse design, which involves determining a shape or fluid property that produces a desired flow, frequently necessitates thousands of costly CFD simulations. PINNs get around this by treating the geometry or fluid properties as learnable parameters in the neural network optimization loop.

### 4.2.1 Channel geometry reconstruction from flow fields

A critical inverse problem is the reconstruction of the configuration of a microchannel or the detection of concealed solid boundaries from sparse velocity measurements, particularly when direct visualization is obstructed. A recent study [185] has shown that PINNs can infer the presence and geometry of static and moving solid boundaries by treating the "body fraction" or the boundary location as an unknown variable as shown in Fig. 8a. Raissi et al. (2020) [115]were the first to employ PINNs to reconstruct 3D flow fields and pressure distributions around complex, unknown geometries using sparse spatial-temporal velocity data as illustrated in Fig. 8b.

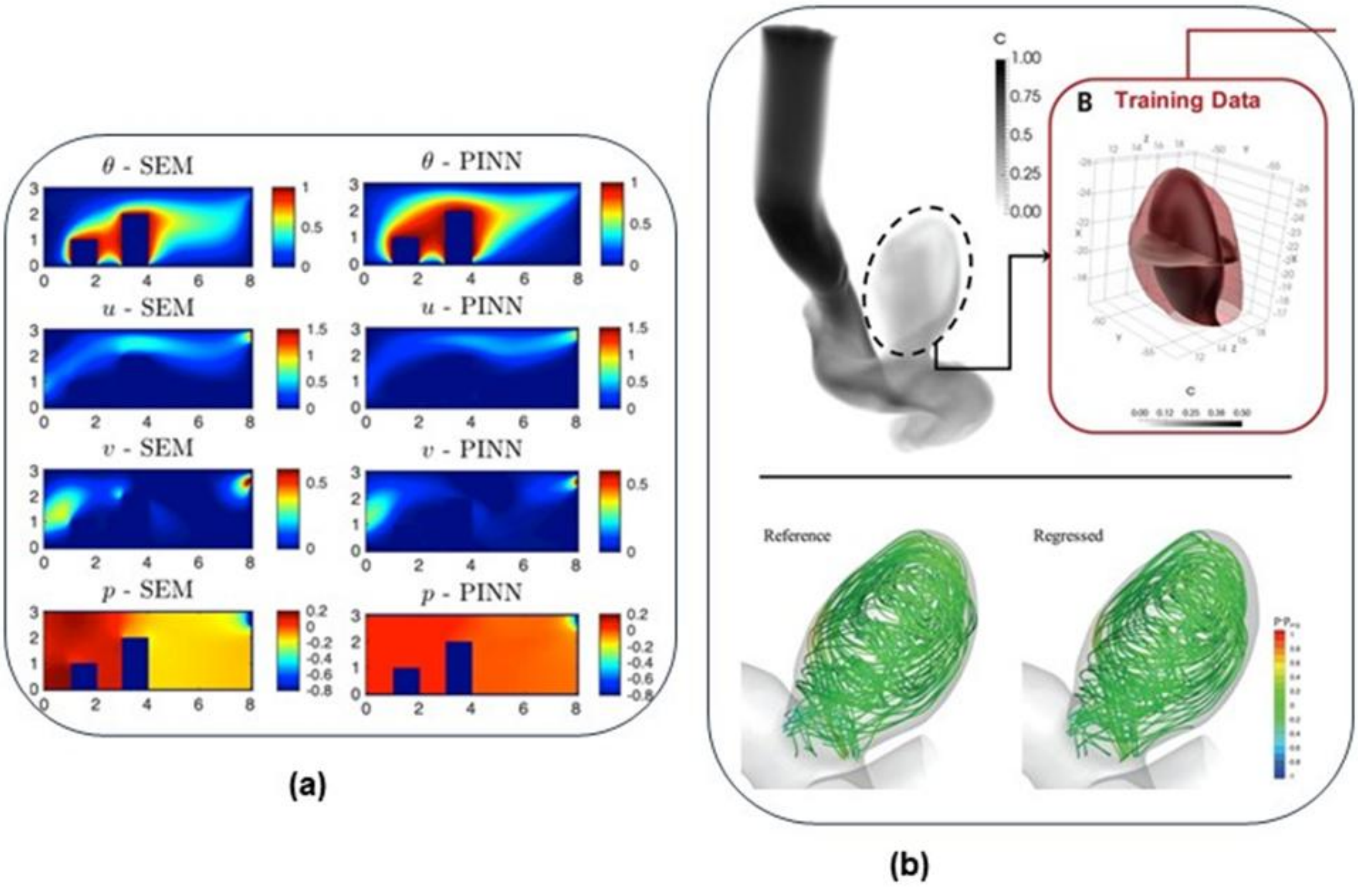

**Figure 8.** Hidden Boundary Detection a) Comparing PINN-derived h, u, v, and p to reference simulation findings using the spectral element approach for forced convection in enclosures. b) Quantitative hemodynamic prediction in a three-dimensional intracranial aneurysm.

Flow field in microfluidic devices is frequently studied using Particle Image Velocimetry (PIV) or Lagrangian Particle Tracking. Shin et al. (2025) [186]demonstrated that PINNs can recuperate flow structures and underlying channel limitations even from low-density particle clouds, exceeding standard interpolation approaches. As illustrated in Fig. 9, PINN modeling of pressure and velocity fields with error visualization. It is observed that PINN can predict and give very accurate field with error accuracy less than 1.

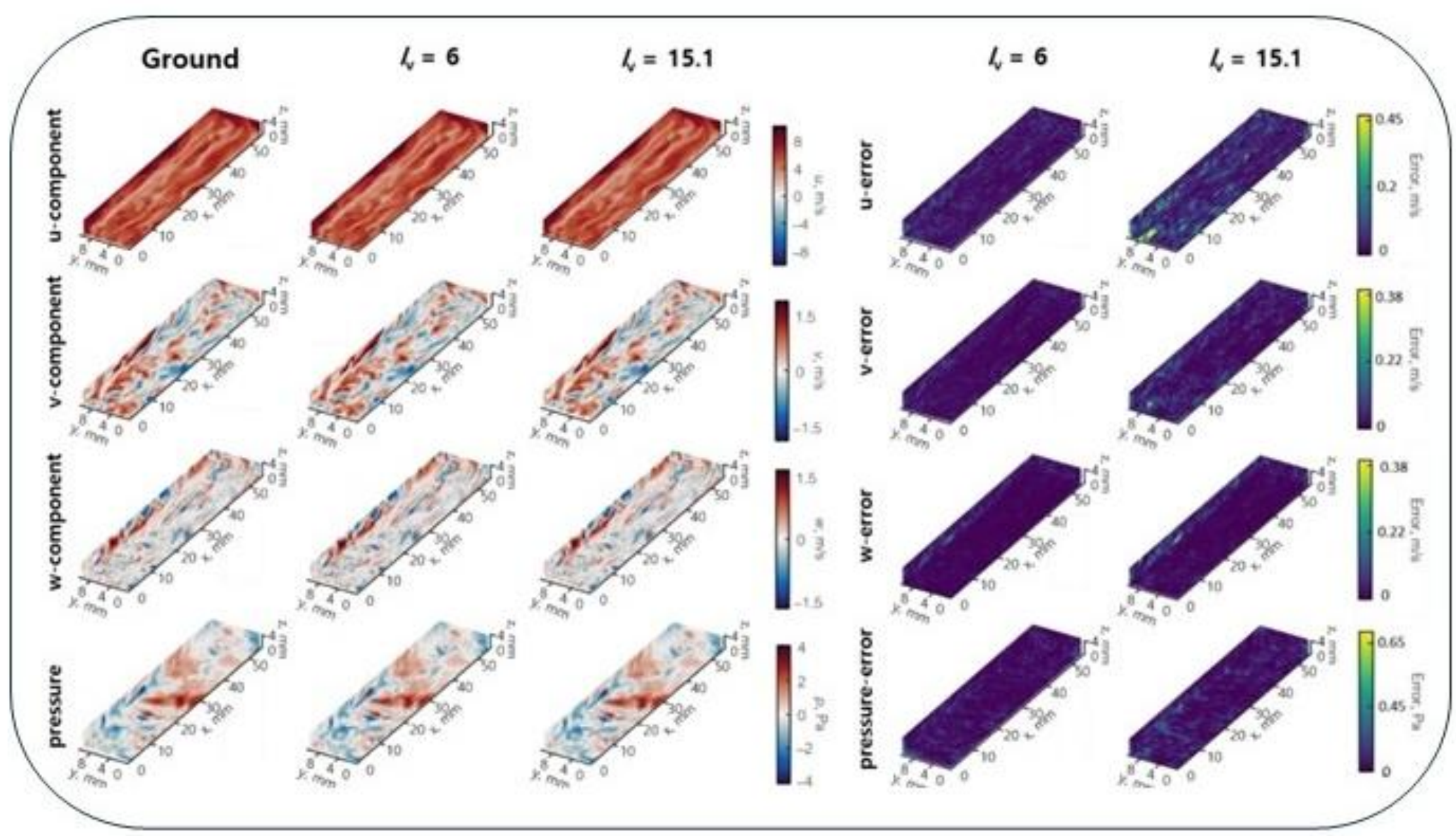

**Figure 9.** Velocity and pressure contours at different inter-particle spacings, along with corresponding error plots at the 24th timestep.

Furthermore, Wassing et al. (2025) [187]have contributed to rapid the development of transonic environments through their work on advanced shock-capturing and surface roughness parameterization, allowing PINNs to model real-world degradation and high-speed flow discontinuities with unprecedented sharpness. Collectively, these studies represent an evolution from data-driven models to hybrid frameworks capable of real-time, physically grounded asset monitoring and flow reconstruction. Fig. 10 depicts an overview of the resulting pressure field compared to the reference solution at five different angles of attack. At the lowest angle of attack $\alpha \approx 0^o$, the model accurately approximates totally subsonic flow. For $\alpha \approx 1^o$, the prediction is very accurate overall, but since this shock is very weak, it is not captured by the PINN. At $\alpha \approx 2^o$ The PINN can resolve the shock, but it is still smoothed out. The stronger shocks are accurately estimated at $\alpha \geq 3^o$.

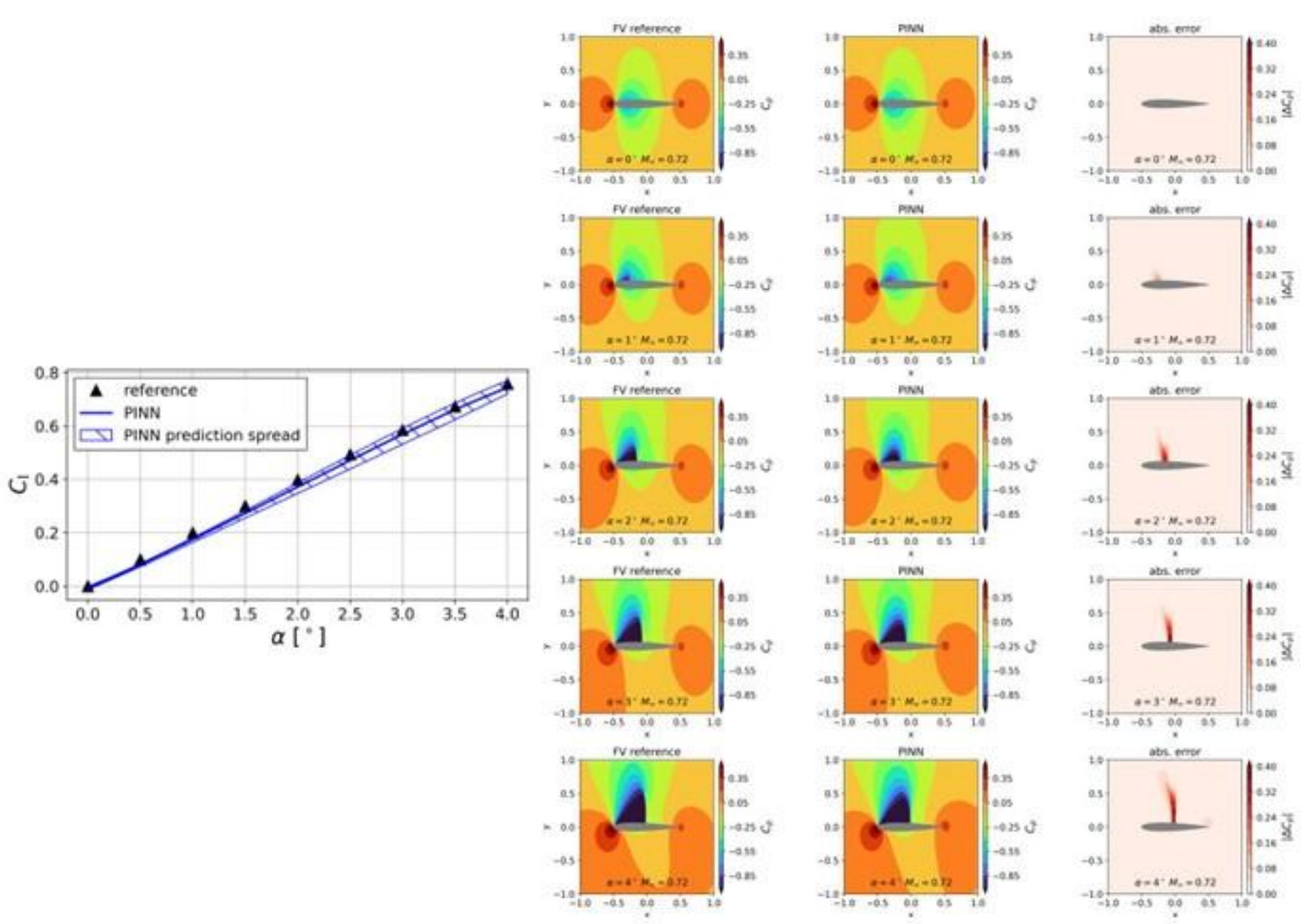

**Figure 10.** Comparing results between PINN and reference simulation at different angle of attack.

## 4.2.2 Viscosity and permeability prediction

Inverse design involves detecting material and transport parameters such as fluid viscosity or porous media permeability via indirect observations. Li et al. (2022) [188]investigate the reconstruction of thermal fluid fields for nanofluids using physics-informed Deep learning. It has been employing a multi-objective loss function that constrains the neural network to adhere to the linked conservation laws of mass, momentum, and energy. Unlike standard CFD, this approach deals with the nanofluid as a single-phase or two-phase mixture. Physical properties, such as effective dynamic viscosity ($\mu_{nf}$) and thermal conductivity ($k_{nf}$), are often treated as functions of the nanoparticle volume fraction (Ø). The governing equations encoded in the PINN loss function for steady-state incompressible nanofluid flow are written as follows:

$$\nabla . u = 0 \quad (28)$$

$$\rho_{nf} \ (u.\nabla)u = -\nabla p + \mu_{nf} \ \nabla^2 u + (\rho\beta)_{nf} \ (T - T_{ref})g \quad (29)$$

$$\left(\rho C_p\right)_{nf}(u.\nabla)\,T \;=\; k_{nf}\;\nabla^2 T \tag{30}$$

The PINN can reconstruct high-resolution velocity and temperature fields from sparse or noisy experimental data (such surface temperature readings) by minimizing the residuals of these equations at collocation locations. Because it can concurrently infer unknown parameters, like the localized heat transfer coefficient or the effective distribution of nanoparticles, this framework is especially powerful for nano fluids. This allows for a deeper examination of the enhancement mechanisms of convective heat transfer without requiring high-density sensor grids.

PINNs have been used to replace empirical constitutive relations like porosity in related fluid flows (like transverse Liquid Composite Moulding), with neural networks trained jointly with PDE residuals. This allows for the real-time extraction of permeability relations from flow rate measurements. This illustrates in Fig. 11 how PINNs can infer permeability or other transport coefficients without depending on traditional empirical models. PINNs have significantly replaced traditional empirical power-law equations in fluid flow simulation through porous media. Classical models, such as the Carman-Kozeny equation or basic power-law fits (Fig. 11a), offer a baseline for forecasting permeability and compaction stress, but they frequently have trouble with complex boundary conditions and non-uniform Fiber Volume Fraction (FVF) distributions. PINNs integrate the governing partial differential equations, like the diffusion-like equation directly into the loss function of a Feed-forward Neural Network (FNN), as shown in the given framework (Fig. 11b). This method enables the model to confine its predictions using hard constraints for beginning conditions and constitutive relations, two basic physical principles (Lee et al, 2025)[189].

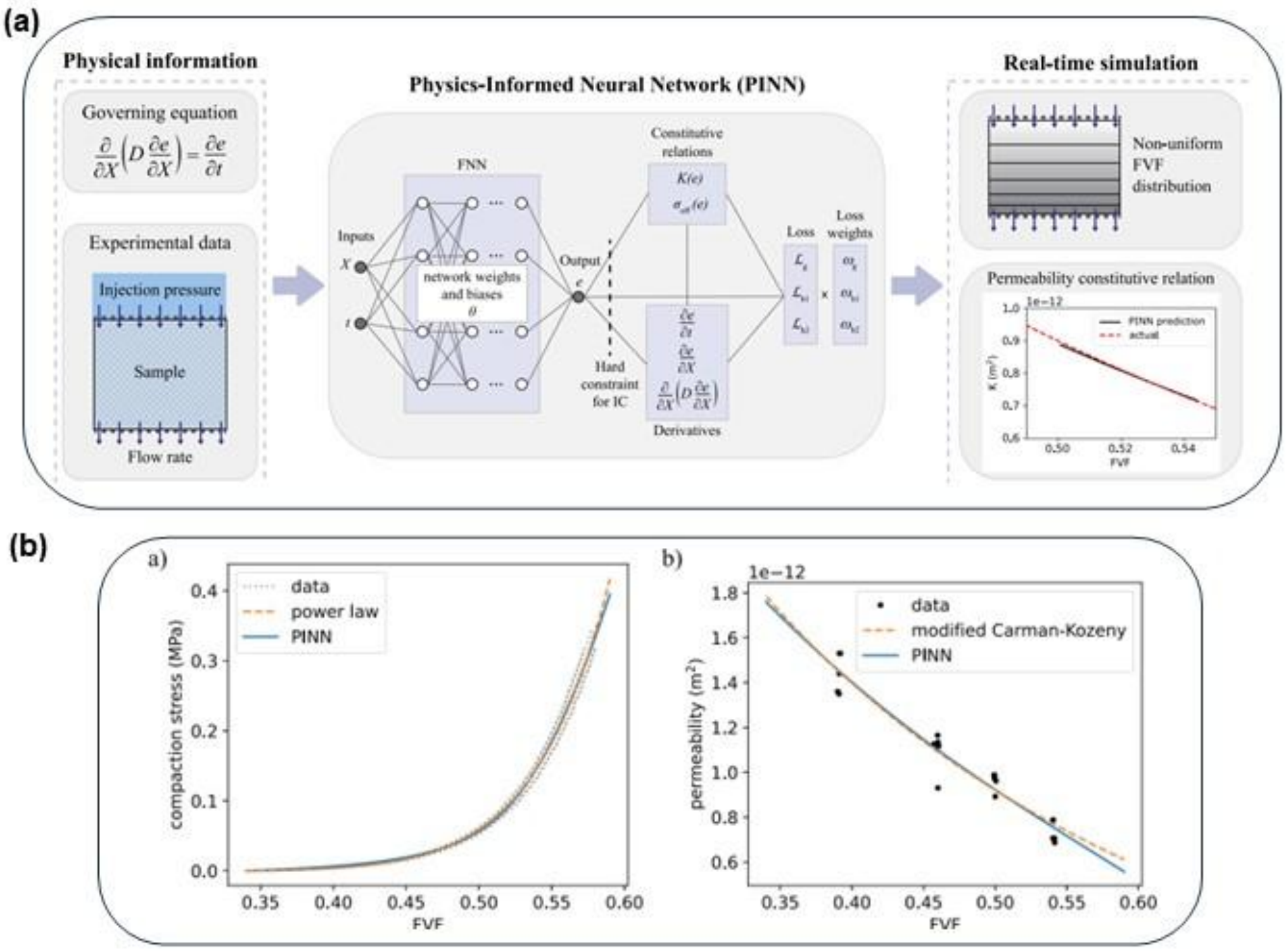

**Figure 11.** Characterization of porous media using PINNs. (a) Diagrammatic workflow showing how experimental data and the governing diffusion equation are coupled within a deep learning framework to forecast a non-uniform FVF distribution. (b) Verification of PINN predictions for permeability and compaction stress using experimental data and traditional empirical models (power law and modified Carman-Kozeny).

Recently, permeability inversion has been specifically targeted in heterogeneous porous flow utilizing self-adaptive PINNs. Using indirect pressure and velocity observations, models directly forecast spatially changing permeability, stabilizing training using adaptive loss weighting. These methods can be applied to microfluidic cases in which surface conditions or channel structure affect transport characteristics[190].

As a result, the literature on fluid dynamics demonstrates that PINNs can effectively infer parameters from sparse data, such as constitutive model parameters or unknown transport coefficients. This capacity is crucial for predicting viscosity or permeability in nanofluidic and microfluidic environments. Reviews emphasize PINN's superiority in field reconstruction and parameter identification[191].

## 4.3 Inverse Design in Astrophysics

### 4.3.1 Astronomy

Magnetic sounding combined with Juno mission data is used to constrain Jupiter's internal boundaries. Traditional interior reconstructions, assuming zero electrical conductivity and spherical harmonic representation, are limited by noise amplification at small scales. This study presents new time-independent reconstructions of Jupiter's internal magnetic field using PINNs, based on the first 33 (PINN33) and 50 (PINN50) Juno orbits. The method effectively identifies local magnetic structures and accommodates weak ambient electric currents. Our models avoid noise amplification with depth and produce a smooth picture of the planet's interior without explicit regularization. The results suggest that the dynamo boundary lies at a fractional radius of 0.8, where the magnetic field forms longitudinal bands, and strong local features such as the Great Blue Spot appear to be anchored in adjacent regions of opposite magnetic polarity.[192]

We present a PINNs extrapolation method that uses machine learning techniques to reconstruct coronal magnetic fields. The traditional neural network architecture is improved by introducing a quasi-output layer to maintain physical constraints during the extrapolation process. Second-order optimization methods are employed for training, offering greater efficiency than the first-order techniques commonly used in classical machine learning. The proposed method is evaluated using the well-known semi-analytical model developed by Low and Lou, achieving high accuracy across multiple metrics. Finally, the approach is validated on observed magnetogram data from active solar regions.[193]

PI-AstroDeconv is a physics-informed semi-supervised learning (PINN) method designed to remove beam effects in astronomical telescope observations. It employs an encoder–decoder architecture that integrates the telescope's point-spread function (PSF) as prior information and uses Fast Fourier Transform–based accelerated convolution. The model effectively eliminates beam distortions, handles multiple PSFs, and tolerates moderate measurement errors. It greatly enhances the efficiency and precision of astronomical image deconvolution without relying on annotated data. The method was validated using the Square Kilometre Array Science Data Challenge 3a datasets and compared with the CLEAN deconvolution approach. Results show that PI-AstroDeconv restores fine image details, reduces blurring, and accurately recovers the true neutral hydrogen power spectrum.[194]

### 4.3.2 Cosmology

We use the network-generated solutions to perform statistical analyses and estimate the parameters of each model using observational data: Hubble parameter measurements from cosmic chronometers, Type Ia supernova data from the Pantheon compilation, and baryon acoustic oscillation observations. The results for all models are consistent with similar estimates reported in the literature using numerical solvers. We assess the accuracy of the trained networks by comparing their solutions to analytical ones when available, or to high-precision numerical solutions otherwise. The estimated errors remain below 1% within the parameter space regions corresponding to the 95% confidence intervals derived from the data. These results were made possible through methodological improvements in solving differential equations using artificial neural networks.[146]

We use PINNs to modularly solve the Boltzmann equations governing dark matter (DM) freeze-in in alternative cosmologies. Through inverse PINNs and a single observational point—the measured relic density—we infer key theoretical parameters, including power-law cosmologies inspired by braneworld scenarios and particle interaction cross sections. The Universe's expansion is modeled using a smooth, switch-like function that reproduces the Hubble law at late times. A distinct correlation is observed between the power-law exponent and interaction strength: negative exponents require smaller cross sections, while positive exponents demand larger ones to match observations. Bayesian methods are employed to quantify the epistemic uncertainty of the inferred parameters.[195]

We present a neural network–based data inversion method to reduce structured contamination and large-scale systematic effects in Planck High Frequency Instrument (HFI) sky maps. The method exploits spatiotemporal couplings of contamination sources to learn optimized low-dimensional representations for robust data inversion and mapping. Physics-informed constraints, transfer learning, and data augmentation are integrated to enhance performance. The approach is validated on simulated 545 GHz and real 857 GHz HFI data, showing up to an order-of-magnitude improvement in contamination removal. These methods are being incorporated into the new SRoll3 algorithm, with 857 GHz detector maps released to the community.[196]

We apply cosmology-informed neural networks (CINNs) to identify deviations from the Lambda Cold Dark Matter (ACDM) model in dark matter power spectra. By integrating partial differential equations (PDEs) into the training process, CINNs are designed to differentiate between the $\Lambda$CDM model and alternative dark matter and dark energy scenarios. The network is trained to classify power spectra in the range of 0.01 $hMpc^{-1}$ to 2.5 $hMpc^{-1}$, achieving a classification accuracy of about 97%. CINNs show high sensitivity to subtle deviations from the ACDM model, making them an effective tool for constraining cosmological parameters and deepening our understanding of dark matter and dark energy.
Future extensions may use larger datasets to further improve model accuracy and performance in cosmological research.[197]

We analyze the Barrow–Tsallis Holographic Dark Energy (BTHDE) model using both the classical MCMC and Bayesian PINN frameworks with various cosmological datasets, including CMB, BAO, CC, and Pantheon+ supernova data. Key parameters ($H_0$, $q$, $\Delta$, $\alpha$, $\beta$, $\Sigma m\nu$) are constrained, with PINN providing tighter bounds than MCMC, especially for $\beta$ and $\Sigma m\nu$. The inferred $H_0$ values lie between Planck 2018 and SH0ES (R22), reducing the Hubble tension to $1.3\sigma$–$2.1\sigma$. Combining CMB and late-time probes yields $\Sigma m\nu < 0.114$ eV and $H_0 = 70.6 \pm 1.35$ km/s/Mpc. PINN ensures physical consistency through embedded differential constraints. The BTHDE model effectively addresses cosmological tensions and explores modified entropy scenarios. This work demonstrates the synergy between machine learning and Bayesian inference in cosmological modeling.[198]

Non-Gaussian distributions in cosmology are analyzed using Markov Chain Monte Carlo (MCMC) methods, as the Fisher matrix applies only to Gaussian cases. The Metropolis–Hastings algorithm samples the posterior after a burn-in phase, with convergence tested by the Gelman–Rubin criterion. We explore its convergence through analogies with Hamiltonian systems in thermal equilibrium, quantifying virialization, equipartition, and thermalization. These criteria are compared with Gelman–Rubin results using a toy model and a dark energy model from supernova data. The study identifies broader, physically motivated convergence measures with clear equilibrium targets. PINNs are employed to accelerate sampling and enhance convergence.[199]

### 4.3.3 Space Weather

Data-driven ionospheric modeling has struggled to comply with physical laws. To address this, we propose an ionospheric inversion model, PINN-SAMI3, based on the physical SAMI3 framework. It integrates SAMI3 governing equations into a neural network to reconstruct temporal and spatial plasma distributions. The goal is to assess the integration of physical models with machine learning for ionospheric modeling. PINN-SAMI3 enforces physical constraints through continuity, momentum, and temperature equations in magnetic dipole coordinates. Simulations show that with sparse ion density data, electron densities, ion velocities, and ion temperatures can be retrieved. The impact of E × B velocity terms is also analyzed during quiet and storm conditions. PINN-SAMI3 achieves accurate inversion results and supports AI-based space weather forecasting.[200]

Data-driven ionospheric modeling does not fully comply with physical laws. PINN-SAMI3 [1], an advanced data assimilation framework, integrates the governing equations of the complete physical model SAMI3 with neural networks to create a physics-informed ionospheric inversion model. We compare PINN-SAMI3 with traditional data assimilation methods, EKF [2] and 4D-Var. The results show that while PINN-SAMI3 outperforms EKF, its accuracy is still lower than 4D-Var. This framework marks an important step toward developing hybrid space weather forecasting models that combine physical modeling with neural networks.[201]

PINNs are used to reconstruct full magnetohydrodynamic (MHD) solutions from partial samples, simulating the space-time environment around spacecraft observations. One-dimensional MHD benchmarks—Sod, Ryu-Jones, and Brio-Wu shock tubes—provide plasma state variables along linear trajectories. The simulated spacecraft data serve as boundary constraints for the PINN, which embeds the full set of 1D MHD equations in its loss function. Results show that PINNs can reconstruct complete one-dimensional solutions even with Gaussian noise. The transformer-based PINN architecture does not scale effectively to higher dimensions. Nonetheless, PINNs show strong potential for reconstructing magnetic structures and plasma dynamics from satellite data in near-Earth space.[202]

The near-Earth space environment has been contaminated by the rapid growth of artificial objects, particularly satellite constellation projects. This increase makes accurate estimation and prediction of dynamical parameters crucial for orbit determination and improved space situational awareness (SSA). As a proof of concept, we developed a physics-informed deep neural network (PINN-D) to estimate key near-Earth parameters such as the Earth's gravitational constant (μE) and the J2 coefficient of the gravitational potential. Unlike conventional DNNs, PINN-D penalizes predictions that violate physical laws. Using normalized Keplerian orbital elements, trigonometric functions, and Gauss's Variational Equations (GVEs) as constraints, the model accurately estimates these parameters. The approach is extended to model the satellite drag coefficient under an exponential atmospheric density model, predicting it within 0.5% of its true value.[203]

We perform 3D magnetic field extrapolations using the nonlinear force-free NF2 code based on PINNs and compute the vector potential to ensure divergence-free solutions. Vector magnetograms from the Solar Dynamics Observatory's Helioseismic and Magnetic Imager are analyzed at 12-minute intervals to study the magnetic evolution of AR 13664 and its major flares. A decrease in calculated relative free magnetic energy corresponds to solar flares in about 90% of cases, and all X-class flares show this correlation. Regions of enhanced and depleted magnetic energy align with brightness increases in extreme ultraviolet observations. A detailed analysis of the X3.9 flare on May 10 links the interaction between separated magnetic fields to major flare events. We provide a comprehensive dataset of the magnetic evolution of AR 13664 for future research.[204]

### 4.3.4 Gravitational Waves

Gravitational Waves (GWs) were predicted as waves propagating in spacetime by the theory of General Relativity (GR) as early as 1916. The search for GWs continued for decades, spanning one generation of resonant bar gravitational antennas and two generations of interferometric detectors. Gravitational-wave signals are produced by changes in mass-energy distribution at the quadrupole (or higher multipole) level. The most likely events that can produce GWs are the coalescences of compact objects like Black Holes (BHs) or Neutron Stars (NSs), which are extremely dense objects with a mass similar to or larger than the solar mass (1 M⊙ ≈ 1.989 *10^30 kg), enclosed within a radius of the order of tens of km. In particular, they can also be found in a binary system, orbiting each other. These orbital configurations are strong GWs emitters, and the outgoing flux of GWs causes a loss of energy from the system, leading to an inspiral phase that continues until a plunge phase when the gravitational potential becomes too strong, leading to the final merger of the objects into a single one. This sequence produces the characteristic chirping waveform shown in Fig.1.d, which remains the only type of GW signal detected so far.[205]

In parallel, machine learning (ML) has become an influential tool in astrophysics, including GW research. The LIGO and Virgo collaborations have applied ML algorithms to tackle complex challenges in data analysis and detector characterization. Since LIGO/Virgo data are inherently non-stationary and non-Gaussian, ML techniques can improve data quality and classification. For instance, ML models can help distinguish between genuine astrophysical signals and transient detector noise by analyzing input from thousands of instrumental and environmental monitors. They are also effective for identifying the origins of noise through the classification of transient events. Some projects even use citizen science to generate training datasets.[206]

By contrast, traditional gravity field reconstruction methods rely heavily on mathematical modeling and numerical computations rooted in physical principles. The spherical harmonics method, in particular, has long been the standard in astrodynamics and satellite orbital studies. It represents deviations from a perfectly spherical gravity field by superimposing harmonic functions onto the central gravitational term. However, when applied inside the Brillouin sphere of a small body, spherical harmonics encounter limitations due to inaccuracies in regions within the effective radius. Consequently, while this method performs well for nearly spherical asteroids such as Bennu, it is less effective for modeling gravity fields in irregular or heterogeneous bodies.

Recently, PINNs have been applied to reconstruct gravity fields within the surface regions of small bodies. A self-adaptive PINN model, incorporating auxiliary-point data augmentation, residual-based refinement, and adaptive weighting, has shown superior accuracy over previous methods and offers a pathway to extending reconstruction to regions farther from the body.[207] Additionally, PINNs have been used to solve wave propagation and full waveform inversions (FWIs) and it was shown that PINNs are able to accurately solve the forward and inverse modeling of the acoustic wave equation in complex media. PINNs are meshless and possess impressive generalization capabilities. Given only a set of two early-time snapshots, they can predict the wavefield solution much later in time. Also, with minimal observed data only at the surface, and without any training data within the computational domain (other than the early-time snapshots) PINNs yield excellent results for seismic inversions. This shows that the PINN formalism can be implemented independent of other numerical solvers that require more complex implementation.[152]

### 4.3.5 Stellar

One of the most crucial steps to understanding the solar and heliospheric activities would be by studying the magnetic field properties on the solar surface. Solar surface magnetic activity, driven by dynamo processes operating in the convection zone, manifests as active regions whose emergence and transport govern the evolution of the Sun's large-scale magnetic field and the solar cycle. Surface flux transport (SFT) models describe the advection and diffusion of the radial magnetic field on the photosphere and play a central role in constraining solar dynamo theories, extrapolating coronal and heliospheric magnetic fields, and forecasting future solar activity. While traditional SFT approaches employ numerical solvers, spherical harmonics, and data assimilation techniques, recent advances have explored machine-learning methods to address the multiscale and computational challenges inherent to these systems[208]. In this context, physics-informed neural networks provide a physically constrained, mesh-free framework for modeling solar surface flows and magnetic flux evolution, extending their growing application to problems in solar and heliospheric physics.

In this study[208], a PINNs model was developed, and its results were benchmarked against an analytical solution obtained using the Runge–Kutta implicit–explicit (RK-IMEX) scheme. The RK-IMEX method is a second-order accurate approach in both space and time, in which the advective term of the governing PDE is evolved explicitly using an upwind Godunov-type scheme, while the diffusive term is treated implicitly[209],[210]. Fig.1.e presents a detailed comparison between the PINN and RK-IMEX solutions using both local and global diagnostics, including the L1 error norm as a function of spatial resolution, the latitudinal magnetic field profile at the diffusion timescale $\tau_{(d)}$, and the total magnetic flux integrated over the southern hemisphere. The results demonstrate distinct convergence characteristics, highlighting the strong grid dependence of conventional numerical solvers in contrast to the grid-independent accuracy of the PINN framework (Figure 12). [208]

### 4.3.6 Astrophysics

The extraction of scientific knowledge from raw astronomical observations fundamentally relies on the resolution of ill-posed inverse problems, a challenge that has intensified as astronomy enters the Big Data regime with instruments like the Atacama Large Millimeter/submillimeter Array (ALMA), the Square Kilometer Array (SKA), and the TOLIMAN space telescope. Mathematically, these problems are formulated as finding a model m (physical parameters) given a set of measurements b, connected by a forward operator G such that $G(m) \approx b$. In radio interferometry, for instance, this involves reconstructing the true sky brightness distribution from incomplete Fourier space samplings (visibilities), a task historically handled by algorithms like CLEAN which iteratively deconvolve the instrument's point spread function (PSF). However, traditional algorithms struggle to cope with the increasing volume, velocity, and complexity of data (approaching exabyte scales for SKA), and often fail to satisfy Hadamard conditions of uniqueness and stability without extensive, computationally expensive regularization.

The necessity of employing Deep Learning (DL) and physics-informed pipelines in this domain arises from the limitations of classical solvers in high-dimensional spaces. DL architectures, such as Convolutional Autoencoders (CAE) and Residual Neural Networks (ResNets), can effectively learn the complex, non-linear inverse mapping from "dirty" observations to "clean" physical models by incorporating priors directly into the training process through realistic simulations. These methods provide a mechanism to regularize ill-conditioned problems by learning hierarchical feature representations, ranging from spatial morphology in interferometric cubes to time-series periodicity in astrometry, thereby surpassing the sensitivity limits of traditional compressed sensing or matched filtering approaches.

In the study[211], they developed DL-based pipelines to address the inverse problems of source detection, characterization, and signal recovery for ALMA, SKA, and TOLIMAN, achieving the following key results:

1. **ALMA Reconstruction and Characterization:** A pipeline combining a Convolutional Autoencoder (Blobs Finder) and a Recurrent Neural Network (Deep GRU) was developed to process dirty ALMA data cubes. The method achieved a source detection rate of 92.3% on a simulated test set with zero false positives after spectral focusing, detecting source positions with sub-pixel accuracy (10^(-3) pixel error). Crucially, when compared to the standard tCLEAN algorithm, the DL approach demonstrated a computational speed-up factor of 200 and superior image reconstruction quality, characterized by lower residuals and higher structural similarity.

2. **SKA Source Detection:** In the context of the SKA Data Challenge 2, a revised DL pipeline utilizing 3D ResNets for segmentation and classification outperformed widely used traditional algorithms (e.g., SoFiA) and other competing DL solutions. The pipeline achieved the highest score in the challenge, maintaining high reliability and completeness even at the bright end of the flux distribution where other DL models struggled.

3. **TOLIMAN Signal Detection:** For the problem of detecting exoplanets via astrometric wobbles, an unsupervised CAE was employed to compress image time series into a latent space analyzed by a Lomb-Scargle periodogram. This method successfully recovered periodic astrometric signals with amplitudes as small as 10^(-6) pixels, a regime where state-of-the-art sparsity-based methods (PCA and Wasserstein Dictionary Learning) failed.

The fig.1.f demonstrates the capability of the Deep Learning approach (Blobs Finder) to solve the inverse deconvolution problem compared to the traditional tCLEAN algorithm. The DL model recovers the target sky model with significantly higher fidelity and fewer artifacts than the classical physics-based solver.[211]

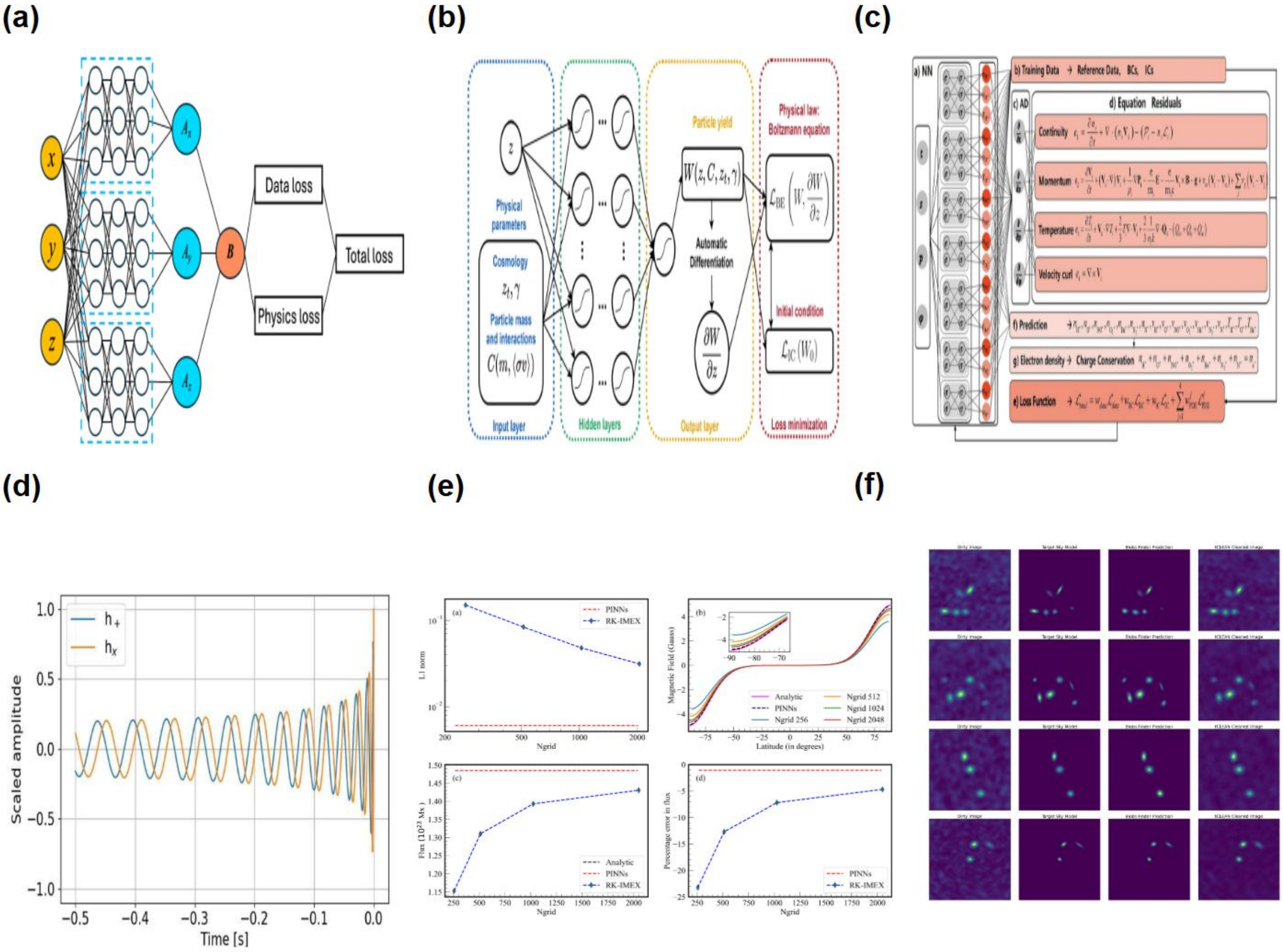

**Figure 12.** Some of the Inverse Design PINNs architectures in Various specializations, and comparing the PINN model with other models. **(a)** PINN architectures In Astronomy.[192] **(b)** PINN architectures in Cosmology.[195] **(c)** PINN architectures in Space Weather.[200] **(d)** Newtonian model of the chirping waveform emitted by a system of two merging equal masses m = 30 M⊙. Both polarizations are shown.[205] **(e)** Evolved magnetic field for β = 10. (a) L1 norm calculated for the PINNs and RK-IMEX scheme compared to the analytic solution. (b) Solution for the time stamp t = τd for different numbers of grid points. The PINN solution (blue) and analytical s.[147] **(f)** Examples of Dirty Images (first column), target Sky Models (second), Blobs Finder's reconstructions (third), and tCLEAN reconstructions with niter = 200 (fourth). The Deep Learning reconstruction (Blobs Finder) separates the sources from the ins.[211]

## 4.4 Biomedical Imaging and Tissue Characterization

PINNs overcome the limitations that face traditional inverse problem solvers due to ill-conditioned systems, or signals that carry much noise. The inclusion of bio-thermal laws into the network's architecture allows the models to work backward from sometimes scarce clinical data to reconstruct internal thermal diffusivity and inversely design targeted therapies.

### 4.4.1 Inverse thermal diffusivity in cancer detection

Infrared thermography is a technique that captures abnormal surface temperature distributions due to the elevated metabolic heat of subcutaneous tumors. Such technique is used for the non-invasive early detection of breast cancer through the translation of the obtained 2D surface thermograms into 3D tumor characteristics, which requires solving an Inverse Heat Conduction Problem.[212] Using the Pennes BHTE, one can relate metabolic heat generation and blood perfusion to surface skin temperatures. However, classical numerical methods used to invert the Pennes equation are ill-conditioned and often yield non-unique solutions due to the multi-layered heterogeneity of human tissue, which is where PINNs can be used.[213, 214]

The introduction of PINNs has paved the way for more advanced architectures that impose physical limitations on neural networks to achieve an accurate result. For instance, in 2026, Ding et al. presented a Physics-Informed Hierarchical Neural Operator (PIHNO) which replace traditional networks that learn a single solution with a neural operator that learn the solution mapping across infinite dimensional spaces. They were able to apply PIHNO in cancer detection by fully reconstructing the unsteady spatiotemporal heat conduction process and precisely inverting the internal heat source (the tumor) using only discrete temperature measurements from a single time slice.[215] The PIHNO was benchmarked against three numerical experiments, and the results showed of the algorithm outperformed the numerical methods, showing highly accurate, real-time mapping of transient thermal diffusivity without being affected much by noise.

Another application of PINNs in inverse design is the estimation of biological and treatment parameters from noisy clinical data. PINNs have been employed to solve inverse problems related to tumor-immune dynamics, approximating complex ordinary differential equations that govern cancer progression based on sparse clinical measurements.[216] For instance, therapeutic applications that use Magnetic Fluid Hyperthermia (MFH) were reported to have used PINNs in such treatments. Traditional MFH works by infusing superparamagnetic nanoparticles to heat tumors locally. PINNs are used in the prediction of the required nanoparticle dosage and spatial distribution based on the target therapeutic temperature required to induce tumor necrosis, which is often known. This is done by using the Pennes BHTE and mass conservation laws into the neural network's loss function, which allows the algorithm to work backward from a prescribed optimal temperature field to calculate the exact nanoparticle distribution and magnetic field intensity required. This eliminates the numerical solutions which consider the complex dynamic effects of blood perfusion and the heterogeneous mass transfer of the nanoparticles within the tumor environment.[217] A similar application was demonstrated by.[216] who demonstrated how PINNs can be utilized to solve a similar inverse parameter discovery problem that aimed to determine the biological parameters governing a patient's specific cancer progression. They used synthetic clinical data that described the interaction between different patient cells, and were able to establish a prediction of future tumor growth and a simulated efficacy of certain types of therapies.

### 4.4.2 Reconstructing biophysical tissue parameters from MRI or optical signals

Magnetic resonance imaging (MRI) is a medical technique used in the diagnostic radiology of many living organisms which, despite being useful, cannot provide quantitative results that are comparable across different clinical institutions or research centers. The introduction of quantitative Magnetic resonance imaging (qMRI) addressed this limitation by providing measurements for specific physical parameters related to the nuclear spin of water protons (which are the main drivers for MRI scans), and proton density. These parameters, which include the longitudinal and transverse relaxation times $T_1$ and $T_2$ respectively, can be used to infer biological information about a living brain (such as myelination, and iron content).[218]

As qMRI works by attempting to fit qualitative weighted images to quantitative parameter maps that represent biophysical and microstructural processes at each voxel, it can benefit from PINNs to drastically improve its efficiency.[219] Traditional parameter mapping is based on pattern recognition algorithms, along with nonlinear iterative optimization schemes and dictionary-matching, which are very prone to noise and have high computational costs. PINNs can replace these mathematical solvers with deep learning models that integrate biophysical constraints onto the system to ensure an accurate, and biologically plausible outputs that are efficient.[219, 220] This was demonstrated by Barzaghi et al. who designed a PINN-based model called Myo-DINO that does fast magnetic resonance (MR) parameter mapping in lower limb muscles in order to check for neuromuscular disorders. The researchers utilized Multi-Echo Spin Echo to take 2165 slices from a total of 232 subjects to be used as input for the dataset in their model. Extended Phase Graph theory was incorporated in the loss function of the neural network, which helped the Myo-DINO model yield high reconstruction similarity ot reference mape with a structural similarity index > 0.92, and a noise-signal ratio that peaked at > 30 dB. The model showed minimal root mean square errors (RMSE) < 0.038 which demonstrates the accuracy of this model. The researchers were able to use this data to predict the Fat Fraction (FF) content, water-T2 (wT2), and B1 field maps.[220]

In 2023, a group of researchers successfully developed a PINN-based framework that performs cine-MR image registration for heart deformations using Fourier feature mappings. Called warpPINN, their model was tested on synthetic examples along with a real-world cine-MRI benchmark dataset of 15 healthy subjects. The researchers were able to successfully predict local metrics of heart deformation including physiological strain estimations in radial and circumferential directions, as well as cardiac landmark tracking.[221] The results of the proposed algorithm were benchmarked against a public, standard test from a 2011 medical imaging challenge called STACOM'11 which works by showing the error in the distance between manually tracked points on the heart muscle versus ones generated through an algorithm. The warpPINN algorithm was the most accurate compared to 3 other models, which was demonstrated by achieving an ultimate median error of 2.91 mm compared to 3.17 mm for the second best algorithm (called UPF.) This was deemed as a very accurate algorithm for the diagnosis of heart failure. A similar approach for prostate intervention was introduced by Min et al., who combined MR and 3D intraoperative transrectal ultrasound (TRUS) images for registration during prostate biopsy.[222] Instead of using Fourier feature mapping, this approach used PointNet as a feature extractor because it is invariant to permutations, which anatomical feature extraction generalizable to patients with different geometries. With a training dataset from 75 subjects, this algorithm resulted in low registration error values that reached $1.90 \pm 0.59$ mm for all nodes (Figure 13).

In addition to image registration of biomechanics, PINNs have applications in the reconstruction of electrical properties of tissues from tomography fields. For instance, in 2024, a group of researchers developed PIFON-EPT, a Physics-Informed Fourier Networks for Electrical Properties Tomography that predicts the electrical properties (relative permittivity and electric conductivity) of tissues from noisy B1 transmit fields. The algorithm is implemented through two Physics-Informed Fourier Networks on a phantom object: One network is responsible for denoising and completing the transmit field, while a second one estimates the object's electrical properties.[223] This PIFON-EPT algorithm was able to reconstruct the electrical properties of the object with $\leq 5\%$ error when using only 20% of the noisy measured fields as inputs. In another work by Schote et al., researchers were able to employ SH-Net, a physics-informed algorithm for computing the spherical harmonic coefficients to guarantee a spatially smooth field map estimate. The method works in low field MRI regime ($B_0$-field ≈ 50 mT), which has growing applications in intensive care units and facilities with limited resources. By correcting for the noise and image distortions that arise inherently in such low-field environments due to $B_0$ inhomogeneities, the algorithm was able to outperform purely model-based methods, improving the Peak Signal-to-Noise Ratio (PSNR) by up to 11.7% and reducing the Root Mean Square Error (RMSE) by up to 86.3%.[224]

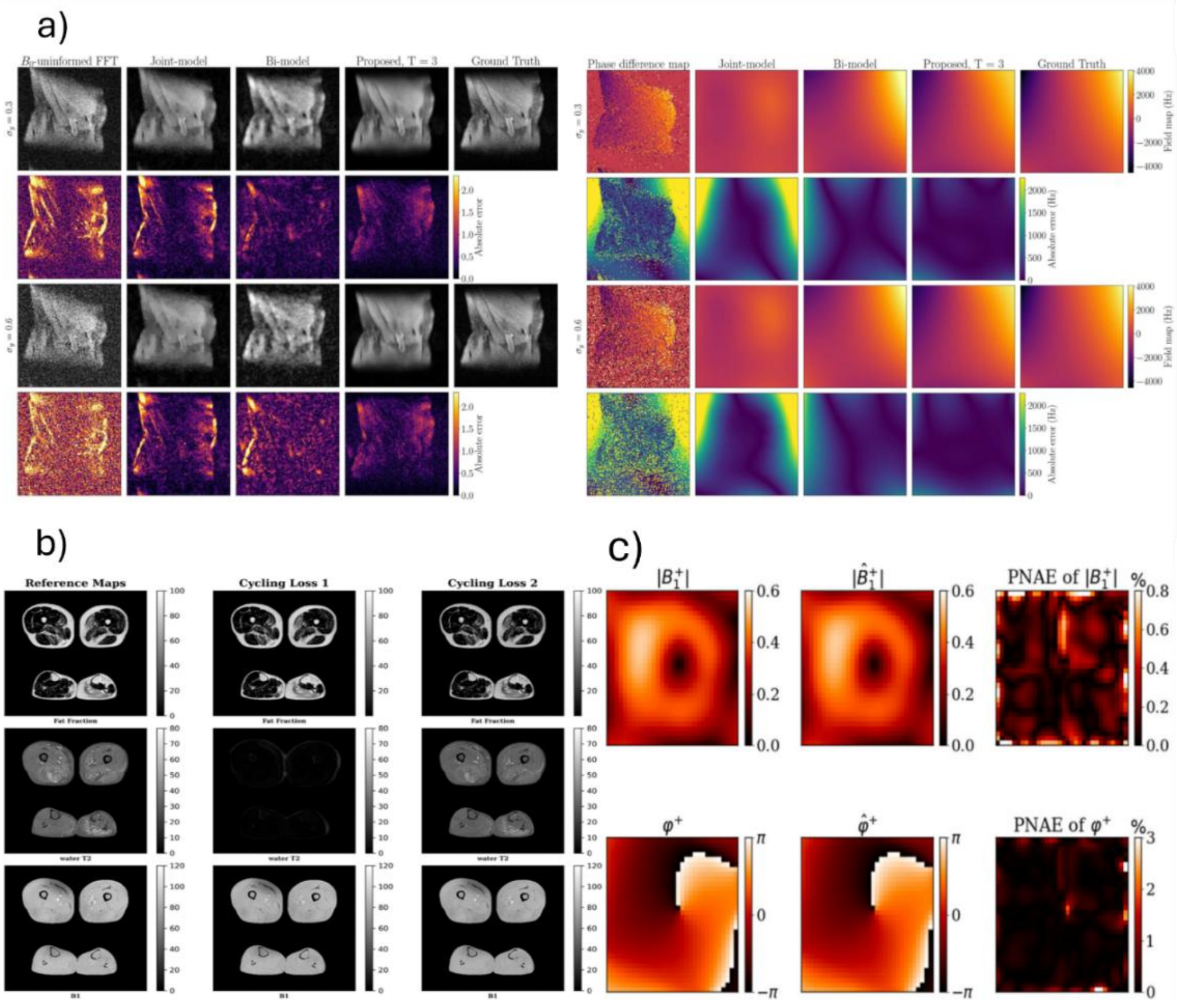

**Figure 13.** Clinical applications and multiparametric map reconstructions utilizing PINNs. a) Joint estimation of $B_0$-field maps and structural images in low-field MRI[224]. b) Fast, high resolution quantitative mapping of Fat Fraction and water-$T_2$ in lower limb muscles for the assessment of neuromuscular disorders using Extended Phase Graph (EPG) theory within the loss function[220]. c) MR-based Electrical Property Tomography (EPT) showing how the PIFON algorithm reconstructs spatial maps of tissues' electrical conductivity and relative permittivity from noise $B_1^+$ fields[223].

# 5. Data-Driven Equation Discovery

## 5.1 Advantages of using PINNs

For centuries, from Newton's laws to Maxwell's equations and the Navier–Stokes equations, scientific progress has relied on expressing natural phenomena through compact mathematical equations that provided predictivity, mechanistic understanding, and the ability to generalize beyond observed data. However, many modern scientific challenges, such as climate dynamics, plasma physics, biological systems, and complex materials, are too high-dimensional, nonlinear, or poorly understood to be described easily from first principles alone.

At the same time, we are living in an era of unprecedented data availability. High-resolution simulations, satellite observations, laboratory diagnostics, and sensor networks generate vast datasets describing complex systems. Traditional modeling approaches often struggle to fully benefit from this data, and they may fail to generalize outside their training distribution. Data-driven equation discovery offers a middle ground. Rather than learning input–output mappings, it seeks to uncover the underlying governing equations directly from data. Overall, equation discovery is an active research area bridging physics and machine learning. Methods such as GP, SINDy, PINNs, and their hybrids are being applied across every field.[225]

For example, data-driven equation discovery is used in Predicting the dynamics of energetic electrons in Earth's radiation belts using incomplete observational data, which is essential for both satellite protection and understanding space plasma processes. Traditional reduced models rely on diffusion equations derived from the quasi-linear approximation. In this work, a PINNs framework is trained and validated using data from the Van Allen Probes. The approach enables the extraction of physical insight by solving the inverse problem of determining model parameters from data. Results reveal that the dynamics of "killer electrons" are better represented by a drift-diffusion equation rather than a simple diffusion model. The proposed drift and diffusion coefficients offer a simpler yet more accurate parameterization compared to existing models.[226]

## 5.2 Data-Driven Equation Discovery in Nanophotonics

In the equation-discovery paradigm, the neural network is not merely asked to solve a known PDE but to simultaneously identify the unknown parameters or even the functional form of the governing equation from data, a task that integrates forward simulation and parameter inversion into a single optimization. The physics-informed loss serves a dual purpose: it constrains the field prediction to remain consistent with the postulated equation structure, while the unknown equation coefficients are promoted to trainable variables whose converged values constitute the discovered physical model. In electromagnetic and photonic systems, this capability is especially valuable because the effective nonlinear coefficients, higher-order dispersion terms, gain and loss parameters, and constitutive relations of real fibers, amplifiers, and nanostructured media are difficult to characterize directly from component-level measurements and are often uncertain or spatially distributed in ways that resist conventional fitting. The subsections below trace the progression of PINN-based discovery from the canonical nonlinear Schrödinger equation and its higher-order variants, through complex fiber amplification and generation systems, to the more exploratory domain of EM constitutive parameter identification, where the boundary between inverse scattering and equation discovery becomes intentionally obscured

### 5.2.1 Data-Driven Discovery in Nonlinear Optics: NLSE and Beyond

The nonlinear Schrödinger equation (NLSE) and its higher-order generalizations occupy a central position in photonic equation discovery because their solutions, solitons, breathers, and rogue waves, are analytically characterized sufficiently well to provide ground-truth validation. However, the physical coefficients of real fibers deviate from handbook values owing to fabrication variability, aging, and nonlinear cross-effects, making autonomous coefficient identification from propagating pulse measurements a practically important problem. Fang et al. [227]demonstrated the first comprehensive PINN-based study of simultaneous solving and parameter identification for the high-order NLSE, constructing a network that embeds the full equation as a residual constraint while treating the nonlinear and dispersive coefficients as additional trainable scalar variables; training the network on rogue-wave trajectory datasets, the framework recovers five classes of femtosecond soliton solutions, one-soliton, two-soliton, rogue-wave, W-soliton, and M-soliton, alongside the unknown coefficients, with one-soliton and W/M-soliton prediction errors systematically smaller than two-soliton errors due to the greater complexity of two-particle interaction dynamics, and providing an architectural sensitivity study correlating layer count, neuron number, and collocation point density with discovery accuracy that serves as a practical design guide for subsequent NLSE discovery works.

Building directly on this foundation, Zhou and Yan[228], narrowed the focus to the dimensionless third-order NLSE, the Hirota equation, which adds third-order dispersion and self-steepening to the standard NLSE, and performed coefficient discovery from bright soliton, breather, and rogue-wave training data by augmenting the PDE residual with a parameter regularization term,

$$L_{total} = L_{PDE}\,(\theta, c_1, c_2) + \lambda_{data} L_{data}(\theta) + \lambda_{reg} L_{reg}\,(c_1, c_2) \quad (31)$$

where $c_1$, $c_2$ are the unknown third-order dispersion and self-steepening coefficients and $\lambda_{data}$, $\lambda_{reg}$ are weighting factors; by adding 2% Gaussian noise to the training waveforms, Zhou and Yan demonstrated robust coefficient identification in the presence of measurement-level noise, a practically critical test that established noise tolerance as a reachable standard for photonic PINN discovery tasks. Subsequently, Zhao et al. [229]extended the discovery to the nonlocal NLS equation, a variant arising from the Manakov system under parity-time symmetry that relates the two coupled field components through a spatial reflection rather than through independent evolution equations. By exploiting this nonlocal symmetry constraint, which reduces the degrees of freedom of the problem, the authors showed that a simpler network with fewer physical constraints suffices for accurate coefficient identification compared to the local coupled NLS case, recovering four distinct two-soliton collision patterns and correctly identifying nonlinear coefficients at multiple noise levels, thereby establishing that the symmetry structure of the governing equation directly informs optimal PINN discovery architecture choices. Chuprov et al. [230]tackled the substantially more complex multimode NLSE (MMNLSE) governing simultaneous propagation and nonlinear coupling of multiple spatial modes in multimode fibers, a system that couples dozens of field envelopes through cross-phase modulation, four-wave mixing, and inter-modal dispersion terms; recognizing that the wide range of magnitude across the equation's dispersion, nonlinearity, and gain terms causes severe gradient imbalance during PINN training, they proposed a scaling transformation specifically for the zeroth-order dispersion coefficient that normalizes the relative contribution of each physical term and allows the PINN optimizer to advance simultaneously on all components, validating the resulting framework against the split-step Fourier method for fiber lengths reaching several hundred meters and demonstrating that physics-informed discovery can be extended to multimode regimes of practical relevance for spatial-division multiplexing.

### 5.2.2 Fiber System Parameter Discovery

The discovery capabilities developed for the NLSE family become practically transformative when applied to more complex, experimentally accessible systems encountered in real fiber-optic transmission and generation infrastructure, where the governing rate equations or propagation models contain multiple coupled parameters that cannot be isolated through standard characterization procedures. The erbium-doped fiber amplifier (EDFA) is a canonical example: its gain and noise figure depend on the absorption cross-sections, stimulated emission cross-sections, saturation powers, and background loss coefficients that are nominally specified by manufacturers but deviate from tabulated values in deployed amplifiers owing to temperature, pump aging, and splice variations. Jiang et al. [231]directly addressed this practical identification problem by formulating a two-stage PIML workflow: in the first stage, a PINN inverse model embeds the EDFA rate equations as physics constraints and optimizes the absorption, gain, saturation, and background loss coefficients jointly with the network weights using only a few measured input-output signal power pairs as data observations; in the second stage, the identified coefficients initialize a forward PINN-based gain estimation model that predicts the amplifier gain spectrum across any operating condition. Experimental validation on a laboratory-constructed EDFA system yielded a mean absolute gain estimation error of 0.127 dB with a standard deviation of 0.065 dB, which is consistent with or better than the estimates obtained by substituting published manufacturer parameter values, demonstrating that autonomous in-situ amplifier characterization through physics-informed discovery is both achievable and practically meaningful.

A closely related challenge arises in ultrafast fiber laser systems, where higher-order nonlinear and dispersive effects that are negligible in long-haul transmission become dominant at femtosecond pulse durations and must be identified from experimental mode-locked pulse waveforms to enable accurate pulse-shaping and stability prediction. Sun et al. [232]approached this problem by designing a physics-informed reversible neural network (PIRNN) whose layer architecture explicitly mirrors the alternating linear and nonlinear propagation steps of the split-step Fourier method (SSFM), with each linear layer implementing a parameterized group delay operator and each nonlinear layer implementing a parameterized nonlinear phase shift; by making the propagation steps explicitly reversible, the PIRNN can run forward to predict pulse evolution and backward to invert for the governing higher-order dispersion and nonlinear coefficients from experimental pulse measurements governed by the generalized Ginzburg-Landau equation, providing a physically transparent discovery architecture whose structure encodes prior knowledge about the operator-splitting nature of fiber propagation. In a different fiber-optic context, Wu et al. [233]targeted electro-optic frequency comb generation, where the governing equation involves coupled nonlinear parametric amplification and phase modulation terms whose coefficients determine the comb bandwidth and spectral flatness. They identified that standard PINN training suffers from gradient imbalance between the propagation physics and the frequency-domain measurement data, and proposed a modified PINN incorporating a power neuron, a specialized activation unit with a learnable exponent, and a local gradient boosting scheme that adaptively amplifies gradients in spectral regions where the predicted comb diverges most from measurements, reducing the mean squared error by up to 82.4% for inverse estimation and 42.0% for forward prediction compared to standard PINNs, and demonstrating that problem-specific architectural modifications driven by understanding of the governing physics yield more significant accuracy improvements than generic hyperparameter tuning.

At the transmission system level, Song et al. [234]developed SRS-Net, a universal physics-informed framework that unifies three operationally distinct tasks within a single coherent architecture: forward prediction of the multichannel signal power evolution along a Raman-amplified fiber, inverse identification of the fiber's attenuation and Raman gain coefficient profiles from measured channel powers, and optimization of the Raman pump power allocation to flatten the channel power ripple, all governed by the stimulated Raman scattering (SRS) propagation equation embedded as the shared physics constraint. The framework uses separate branch networks for the forward and inverse subproblems, with the identified fiber parameters becoming trainable variables that are shared between branches and optimized end-to-end, allowing the measurement data from any channel combination to simultaneously constrain both the physical model and the forward prediction. Experimental validation on a C+L-band multichannel fiber-optic system demonstrated full-field power prediction with a maximum error below 0.3 dB across all channels, and Raman pump optimization achieved a reduction in channel power ripple from 6.5 to 2.2 dB, which is one of the most experimentally comprehensive PINN demonstrations in fiber-optic system-level applications, establishing the framework as a practical tool for deployed network monitoring and optimization

### 5.2.3 EM Constitutive Parameter and Field Equation Discovery

For recovering material distributions from observed field data, two of these contributions extend naturally into the equation-discovery paradigm by identifying constitutive parameters without assuming any particular spatial functional form for the material distribution, which is the defining characteristic that distinguishes equation discovery from conventional constrained inversion. Hu et al. [173]demonstrate that unknown permittivity distributions can be recovered from multi-frequency scattered-field data using an unsupervised PINN without imposing any parametric spatial model, the recovered $\varepsilon_r$ map emerges as a free field variable constrained only by Maxwell's equations and the scattering observations, representing a form of constitutive-law discovery from electromagnetic measurements in the same sense that sparse regression methods discover active terms in governing equations from trajectory data. Zeng et al. [176]similarly recover the spatial relative permittivity distribution in the Poisson equation framework from sparse electric field measurements in 2D and 3D configurations, using a sigmoid-based material model that allows the permittivity to vary continuously across the domain and is optimized entirely by physics-guided data fitting without structural assumptions, interpretable as simultaneous forward-field reconstruction and constitutive-parameter discovery from sparse observations. It should be noted that full equation-structure discovery in the electromagnetic domain, identifying which interaction terms in a generalized Maxwell system are physically active in a given medium, analogous to SINDy-class sparse regression for dynamical systems, has not yet been demonstrated in the available reference corpus and represents a compelling open direction connecting PINN-based photonic discovery with the broader scientific machine learning literature. Table 3 overview recent PINNs equation discovery applications.

**Table 3.** PINNs for Equation Discovery

| Ref | Year | Application | Input | Output | Architecture | Parameter Size | Dataset | Governing Loss | Accuracy |
|---|---|---|---|---|---|---|---|---|---|
| [227] | 2021 | Data-driven discovery of high-order NLSE parameters | (x, t) coordinates | Complex field Q, discovered coefficients $\lambda_i$ | FCNN, 7 hidden layers, 30-40 neurons | - | Synthetic soliton/rogue wave data from exact solutions | Initial/boundary data MSE + PDE residual | Relative L2 errors: one-soliton 9.125e-3, two-soliton 1.634e-2, rogue wave 2.556e-2 |
| [228] | 2021 | Coefficient discovery for Hirota equation | (x, t) coordinates | Complex field Q, discovered $c_1$, $c_2$ | FCNN | - | Synthetic bright soliton, breather, rogue-wave data with 2% noise | $L_{total} = L_{PDE} + \lambda_{data} L_{data} + \lambda_{reg} L_{reg}$ | - |
| [229] | 2025 | Parameter identification for nonlocal NLS equation | (x, t) coordinates | Complex field q, discovered nonlinear coefficients | PINN with parameter regularization | - | Synthetic one- and two-soliton solutions | PDE residual + data loss + parameter regularization | - |
| [230] | 2024 | Solving and parameter discovery for multimode NLSE (MMNLSE) | (z, T) coordinates | Complex envelopes $A_p$ for 3 modes | PINN with 6 ResNet blocks, 150 neurons/block | - | 240,000 collocation points | PDE residual (MMNLSE) + initial condition | MSE below 8e-2 for linear 300m case |
| [231] | 2025 | Parameter identification and gain estimation for EDFA | Propagation distance z | Pump and signal powers, identified physical parameters | PINN, 2 hidden layers, 200 neurons | - | 5 sets of input-output power pairs from experiment | $\omega_p$ $Loss_{PHY}$ + $\omega_c$ $Loss_{CON}$ (simplified Giles model) | Gain MAE 0.127 dB, STD 0.065 dB |
| [232] | 2024 | Reverse exploration of higher-order effects in fiber laser | Initial field (reconstructed from experiment) or white noise | Predicted spectra, ACTs, discovered coefficients (β_n, $\gamma_{2m+1}$) | PIRNN with LinearONet and NonlinearONet | - | Experimental spectra and ACTs from mode-locked fiber laser | $(A_{exp} - A_{pred})^2/A_{pred}^2 + (I_{exp} - I_{pred})^2/A_{pred}^2$ | MSE 2.89e-9; coefficient uncertainties reported |
| [233] | 2024 | Inverse parameters estimation for electro-optic frequency combs | (z, t) coordinates | Complex envelope A, identified $\beta_2$, $\beta_3$, $\gamma$ | Modified PINN with power neuron and local gradient boosting | - | Simulation data from solving NLSE | MSE of initial, observed, PDE residual + gradient constraint | Relative power error reduction: 43.0% and 82.4% |
| [234] | 2024 | Universal solver for stimulated Raman scattering (SRS) in fiber-optic systems | (z, t) coordinates (and pump powers for optimization) | Complex field $A_k$, identified fiber parameters, optimized pump powers | SRS-Net: unified PINN framework with separate branches | - | Simulated (SSFM) and experimental (C+L-band WDM) data | Data loss (initial/boundary or measured) + PDE residual | Max error <0.3 dB for power prediction; power ripple reduced from 6.5 to 2.2 dB |

## 5.3 Data-Driven Equation Discovery in Astronomy

Despite their promise, PINNs face notable challenges in data-driven equation discovery in astronomy. Astronomical data are often sparse, noisy, and irregular, making it difficult to reliably infer governing equations without careful regularization and uncertainty handling. Additionally, the highly nonlinear and multiscale nature of astrophysical systems can lead to unstable or slow training, especially when physical priors or boundary conditions are uncertain. Nevertheless, ongoing advances in adaptive loss strategies, uncertainty quantification, and hybrid symbolic–neural approaches are enhancing the robustness and interpretability of PINNs. With increasingly rich observational datasets and growing computational power, PINNs are expected to play an important role in uncovering hidden physical laws and bridging observational data with first-principles models in astronomy.

This paper shows that the ionosphere significantly affects the quality of satellite navigation and shortwave communication. Current AI models often lack physical interpretability, limiting insights into ionospheric dynamics. To overcome this, we applied three hybrid methods combining neural networks with PDEs: PDE-Net2, PINNs, and SINDy. These models were tested for their ability to reconstruct ionospheric total electron content (TEC) data. Among them, PDE-Net2 showed superior accuracy and efficiency in TEC reconstruction. Its extracted PDEs revealed dominant longitudinal convection and notable latitudinal diffusion effects. Longitudinal electron transport is mainly influenced by meridional winds and solar radiation variations. Meanwhile, the diffusion process indicates nonlinear coupling between the magnetosphere and ionosphere. Overall, PDE-Net2 provides both high performance and enhanced physical interpretability for ionospheric modeling (Figure 14).[235]

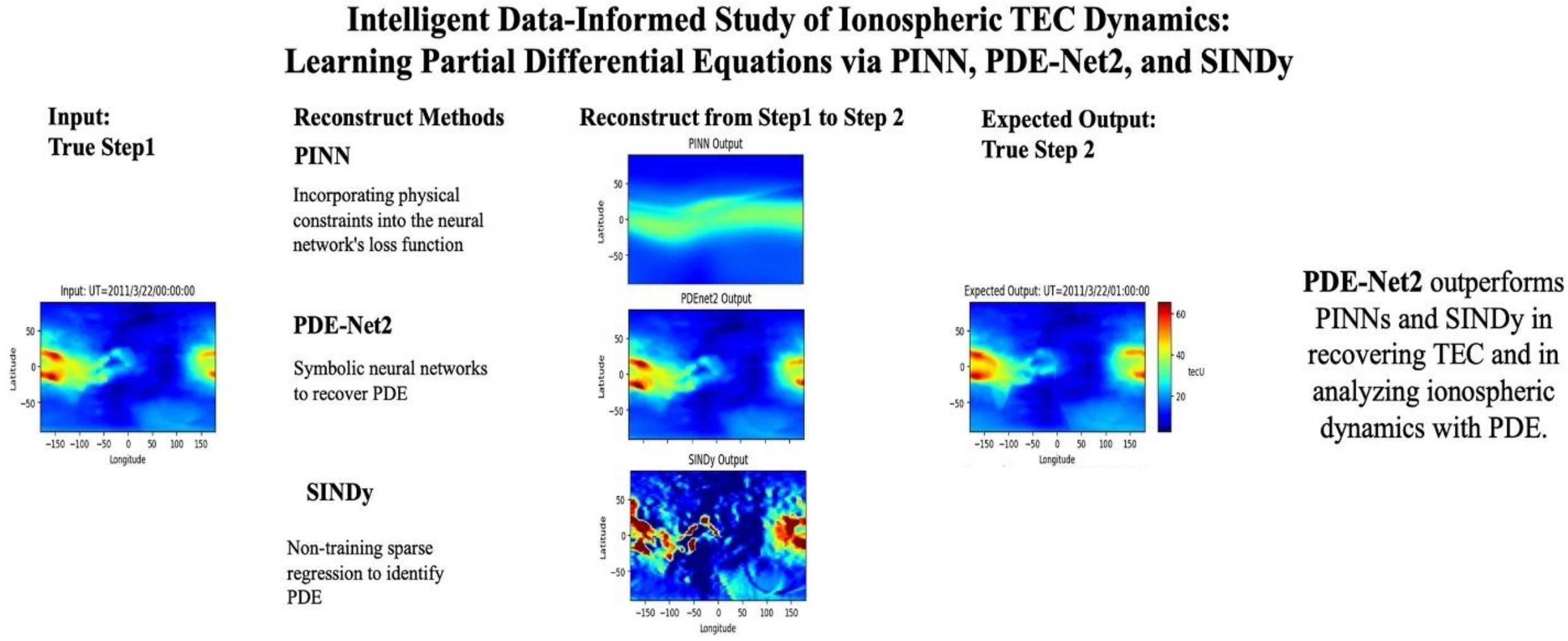

**Figure 14.** Comparison of PINN, PDE-Net2, and SINDy for reconstructing ionospheric Total Electron Content (TEC) dynamics from Step 1 to Step 2. The results show that PDE-Net2 achieves the closest agreement with the true TEC distribution, outperforming PINN and SINDy in capturing ionospheric structures.[235]

## 5.4 Data-Driven Equation Discovery in Biomedical Engineering

In the biomedical field, some effort has been dedicated to using PINNs for inferring equations for modeling physiological fluids and tissues. This is because standard fluid movement equations such as Navier-Stokes equations do not characterize blood hemodynamics accurately, since it behaves as a complex non-Newtonian fluid in the microvasculature. To address this, recent frameworks have combined PINNs with symbolic regression to discover some missing mathematical terms that govern such non-Newtonian flows. By analyzing raw flow data, this hybrid architecture sifts through libraries of mathematical operators to identify the precise algebraic terms needed to correct the model misspecification, which results in accurate cardiovascular modeling.[236]

Moreover, equation discovery is also being used in epidemiology model predictions. Traditional compartmental models (such as SIR or SEIR models) used to forecast infectious disease spread rely on ordinary differential equations where the exact transmission, and recovery rates are changing and are often unavailable to infer from public health records. Physics-informed architectures paired with functional interpolation have proved to be able to mathematically discover these such epidemiological parameters and time-varying dynamics. Such approach was empirically demonstrated and resulted in highly accurate models that showed accurate parameter predictions with low computation time.[237]

# 6. Multi-Physics and Coupled Domains

The application of PINNs to multi-physics systems represents one of the most dynamic frontiers in scientific machine learning, as it enables monolithic treatment of coupled PDEs that have traditionally required staggered partitioned solvers and intricate interface handling. In engineering contexts, this capability is particularly valuable when tight feedback exists between subsystems, when disparate scales must be resolved simultaneously, or when inverse identification of coupling parameters is desired from sparse, modality-wise measurements.[238-240]

## 6.1 Solving Coupled PDEs with PINNs

### 6.1.1 Thermo-optical simulations

In integrated silicon photonics, thermo-optic phase shifters are central components for modulators, interferometers, and reconfigurable photonic circuits used in data communications, neuromorphic photonics, and sensing. These devices exploit the temperature dependence of the refractive index of silicon to modulate the optical phase of guided modes by localized Joule heating.[241] The underlying multi-physics problem involves steady or transient heat conduction in a layered stack, thermo-optic coupling of material indices, and electromagnetic wave propagation governed by Helmholtz or Maxwell equations. A typical formulation couples the thermal and optical fields through

$$\rho C_p \frac{\partial T}{\partial t} - \nabla \cdot (\kappa \nabla T) = Q_{\mathrm{source}}(\mathbf{x}) \tag{32}$$

and

$$n(\mathbf{x}, T) = n_0(\mathbf{x}) + \frac{\mathrm{d}n}{\mathrm{d}T}\,(T(\mathbf{x}) - T_0), \tag{33}$$

where $\rho$ is density, $C_p$ heat capacity, $\kappa$ thermal conductivity, $Q_{\mathrm{source}}$ Joule heating, and $\mathrm{d}n/\mathrm{d}T \approx 1.8 \times 10^{-4}$ for crystalline silicon near telecommunication wavelengths. The optical field $E(\mathbf{x})$ satisfies the scalar Helmholtz equation for a given polarization,

$$\nabla^2 E + k_0^2 n^2(\mathbf{x}, T) E = 0, \tag{34}$$

or the full-vector Maxwell system in more general settings.

Solving this coupled problem with finite element or finite difference time domain (FDTD) methods typically requires separate meshes and solvers for thermal and electromagnetic fields, along with iterative coupling schemes to converge to a self-consistent solution.[242] For design optimization, this must be repeated many times as heater geometries, metal traces, or cladding configurations are varied. PINN-based formulations directly approximate both temperature $T(\mathbf{x})$ and electric field $E(\mathbf{x})$ using a multi-output network, with a composite loss that enforces the heat equation, thermo-optic relation, Helmholtz equation, and boundary conditions on the same set of collocation points.[243]

Recent studies have demonstrated that such thermo-optic PINNs can accurately predict stationary temperature distributions, effective index shifts, and accumulated phase in Mach–Zehnder interferometers, while simultaneously supporting inverse design tasks.[92] For example, given a target phase-shift profile or $\pi$-phase length at a prescribed drive power, the heating distribution $Q_{\mathrm{source}}$ and heater geometry can be treated as design variables and optimized within the PINN training loop, thereby reducing thermal cross-talk and power consumption in densely integrated photonic meshes. The mesh-free representation is particularly advantageous for free-form or non-standard heater layouts, where conventional topology optimization must repeatedly rebuild and adapt meshes. Extensions to transient heating allow the same framework to optimize switching speeds and dynamic phase stability, key metrics for high-speed reconfigurable photonics and neuromorphic signal processing.

### 6.1.2 Bio-mechano-electrical systems

Cardiac function emerges from the tight coupling between electrical excitation, active contraction of myocardial tissue, and blood flow within the cardiac chambers.[244] High-fidelity models often rely on bidomain or monodomain formulations for electrophysiology, large-deformation hyperelasticity for myocardium, and Navier–Stokes equations for intracardiac hemodynamics, leading to a multi-scale, stiff, and computationally intensive simulation problem. The bidomain equations can be written as

$$\chi C_m \frac{\partial V_m}{\partial t} + I_{\text{ion}}(V_m, \mathbf{w}) - \nabla \cdot (\boldsymbol{\Sigma}_i \nabla V_m) - \nabla \cdot (\boldsymbol{\Sigma}_i \nabla \phi_e) = I_{\text{stim}}, \tag{35}$$

$$\nabla \cdot ((\boldsymbol{\Sigma}_i + \boldsymbol{\Sigma}_e) \nabla \phi_e) + \nabla \cdot (\boldsymbol{\Sigma}_i \nabla V_m) = 0, \tag{36}$$

where $V_m$ is the transmembrane potential, $\phi_e$ extracellular potential, $\boldsymbol{\Sigma}_{i,e}$ anisotropic conductivity tensors, $I_{\text{ion}}$ ionic currents parameterized by gating variables $\mathbf{w}$, and $I_{\text{stim}}$ external stimulation. Mechanical equilibrium of the myocardium is governed by

$$\nabla \cdot \mathbf{P}(\mathbf{F}, V_m) = \mathbf{0}, \tag{37}$$

with Piola–Kirchhoff stress $\mathbf{P} = \mathbf{P}_{\text{passive}}(\mathbf{F}) + \mathbf{P}_{\text{active}}(V_m)$ and deformation gradient $\mathbf{F}$.

Direct simulation of this coupled system via FEM demands fine spatial resolution to capture steep electrical wavefronts and small time steps to resolve ionic transients, rendering it impractical for routine patient-specific modelling or real-time clinical decision support.[245] PINNs have been proposed as an alternative framework for solving inverse problems in this setting,[246] including identification of patient-specific conductivity tensors, material stiffness parameters, or scar locations from sparse body-surface potential mapping (BSPM), intracardiac electrograms, and MRI-derived displacement fields. By embedding the bidomain equations and hyperelastic equilibrium in the loss function, the PINN acts as a physics-based regularizer that infers unobserved fields consistent with both measurements and governing laws.

A notable difficulty in this context is the stiffness induced by ionic models, where gating variables evolve on sub-millisecond scales, while mechanical deformation and hemodynamics evolve on orders of magnitude slower time scales. Stiff-PINN formulations have therefore been introduced, employing adaptive temporal sampling, time-windowing, and specialized loss weighting to ensure that rapid depolarization fronts and plateau phases are faithfully captured without overwhelming the optimization with slow modes. Hybrid PINNs further partition the problem by integrating stiff ODE subsystems (e.g. ionic currents) using classical solvers inside the training loop, while the spatial PDEs are represented by neural networks, thus combining numerical stability with representational flexibility.[247] Beyond cardiology, similar strategies have been applied to dielectric elastomer actuators, where electromechanical coupling produces large, nonlinear deformations and instabilities such as snap-through under applied voltage. PINN models combining incompressible neo-Hookean elasticity with Maxwell stress have been shown to reproduce experimental inflation curves and to support inverse identification of material parameters and optimal geometries for energy harvesting and soft robotics.[248]

## 6.2 Domain Decomposition and Modular PINNs

### 6.2.1 XPINNs, FBPINNs for complex geometries

When applied to domains with complex geometries or highly heterogeneous solution features, single-network PINNs often exhibit degraded convergence and accuracy due to spectral bias and the difficulty of representing localized phenomena with a global set of parameters. address this limitation by decomposing the computational domain into overlapping or non-overlapping subdomains, each represented by a dedicated neural network with potentially distinct architecture, depth, and activation functions.[50] The training loss comprises a sum of subdomain residuals and interface penalties enforcing continuity of state and fluxes,

$$\mathcal{L} = \sum_i \mathcal{L}_{\mathrm{PDE}}^{(i)} + \sum_{(i,j)} \mathcal{L}_{\mathrm{interface}}^{(ij)}, \tag{38}$$

with interface terms of the form

$$\mathcal{L}_{\mathrm{interface}}^{(ij)} = \frac{1}{N_{\Gamma_{ij}}} \sum_{k=1}^{N_{\Gamma_{ij}}} \left( \| u_{i,\theta_i} - u_{j,\theta_j} \|^2 + \alpha \| \mathbf{n} \cdot \nabla u_{i,\theta_i} - \mathbf{n} \cdot \nabla u_{j,\theta_j} \|^2 \right)_{\mathbf{x} \in \Gamma_{ij}}, \tag{39}$$

where $\Gamma_{ij}$ denotes the interface between subdomains and $\alpha$ a weighting parameter. This construction allows localized high-capacity networks to focus on regions with shocks, boundary layers, or sharp material interfaces, while simpler regions are treated with shallower, more efficient networks, leading to substantial gains in accuracy and computational efficiency for canonical flow and transport benchmarks.

Finite-basis PINNs (FBPINNs) further combine domain decomposition with local spectral expansions by expressing the solution as a sum of compactly supported basis functions whose coefficients are predicted by neural networks.[249] In this setting, high-frequency components of the solution are captured by the local basis, while the network predominantly learns smooth coefficient fields, thereby mitigating spectral bias and improving convergence for multi-scale problems, including two-phase flow with capillarity and high-Reynolds-number turbulence. Reported error reductions of several orders of magnitude with respect to standard PINNs highlight the potential of these architectures for large-scale engineering simulations.[250]

### 6.2.2 Coupling between different physics (e.g., electromagnetism + mechanics)

Domain decomposition also provides a natural route to modular multi-physics PINN architectures, where different physical subdomains are governed by distinct operators and may be solved by different numerical technologies. In an aeroelastic wing simulation, for example, the external fluid domain is described by compressible Navier–Stokes equations, while the structural domain is governed by linear or nonlinear elasticity.[251] XPINN-style modularization assigns a fluid PINN to the aerodynamic subdomain and a solid PINN (or even a classical FEM solver) to the structural subdomain, with interface conditions enforcing kinematic continuity and dynamic equilibrium,

$$\mathbf{v}_{\text{fluid}} = \frac{\partial \mathbf{u}_{\text{solid}}}{\partial t}, \sigma_{\text{fluid}}\mathbf{n} + \sigma_{\text{solid}}\mathbf{n} = \mathbf{0} \text{ on } \Gamma_{\text{FSI}}. \tag{40}$$

This modular set-up promotes code reuse, as pre-trained fluid models can be coupled to new structural models with limited additional training, and it facilitates hybrid solvers in which PINNs operate in complex or poorly meshed regions while established FEM/FVM solvers handle regular subdomains.

Such modular approaches are not limited to fluid–structure interaction. In micro- and nano-systems, they can be used to couple electromagnetic PINNs in optical regions with mechanical PINNs in deformable supports, or to connect electrochemical PINNs in battery electrodes with thermal PINNs in cooling structures, enabling multi-domain digital twins with consistent cross-domain coupling.[252, 253] From an engineering standpoint, this modularity aligns with conventional practice of splitting large systems into discipline-specific solvers, but now within a differentiable and end-to-end trainable framework that is well suited for gradient-based design optimization and uncertainty quantification.

# 7. Benchmarking, Generalization, and Limitations

A rigorous evaluation of PINNs demands systematic benchmarking against established numerical methods such as FEM and FDTD, alongside a critical assessment of their generalization properties, optimization pathologies, and uncertainty quantification capabilities. In engineering terms, the key questions concern when PINNs are competitive or superior to classical solvers, how they scale with problem size and complexity, and what methodological safeguards are required for reliable deployment in safety-critical contexts.[254]

## 7.1 Accuracy vs. Classical Solvers

Numerical studies consistently demonstrate that, for well-posed forward problems with simple geometries and linear physics, FEM and FDTD achieve machine-precision accuracy with significantly lower computational cost than PINNs. Relative $L^2$ errors as low as $10^{-10}$~$10^{-12}$ are routine for benchmark elliptic and parabolic problems, whereas standard PINNs often plateau at $10^{-3}$~$10^{-5}$, even after extensive training. This discrepancy can be traced to the fundamentally different numerical engines: classical methods rely on mature linear and nonlinear solvers with well-understood convergence theory, while PINNs use stochastic gradient descent or quasi-Newton optimization in non-convex landscapes, where convergence to the global minimum is not guaranteed.[14]

From a performance perspective, solving a single Poisson or heat equation instance via PINNs is typically several orders of magnitude slower than FEM, with reported slowdowns around $10^4$~$10^5$ depending on resolution and implementation. The principal advantage of PINNs emerges when a single trained network is used as a surrogate for many-query scenarios: once trained, evaluation costs are negligible and solutions can be produced in milliseconds, enabling real-time control, embedded inference, or inner-loop evaluations in optimization. PINNs also circumvent the mesh-generation bottleneck in complex or evolving geometries, though accurate enforcement of boundary conditions on intricate CAD surfaces still requires careful construction of distance functions, parameterizations, or immersed-boundary formulations.

Recent work has examined whether carefully engineered PINNs can match or even surpass FEM in specific regimes, for example by employing high-order architectures, NTK-guided initialization, or gradient-free training schemes. While promising results have been reported in certain low-dimensional problems, the broader consensus remains that classical solvers are superior for routine forward simulations, whereas PINNs derive their competitive edge from their ability to unify forward and inverse problems, assimilate data, and operate in high-dimensional or poorly meshed spaces.[255] Table 4 summarizes the key features of PINNs, FEM, and FDTD.

**Table 4.** Comparative Benchmarking of PINNs vs. Classical Solvers (FEM and FDTD)

| Feature | FEM | FDTD | PINNs |
|---|---|---|---|
| **Accuracy** | **High (Gold Standard)**<br>Excellent for complex geometries; error decreases predictably with mesh refinement. | **High**<br>Standard for broadband/transient problems; accuracy depends on grid density (staircasing error). | **Moderate to High**<br>Often struggles with high-frequency/multiscale features (spectral bias). Can match FDTD with "hybrid" training but generally lower precision. |

| Solution Time (Training) | **Fast to Moderate** Solves linear systems directly or iteratively. Scale: Minutes/Hours. | **Fast** Explicit time-stepping is efficient for large domains. Scale: Minutes/Hours. | **Very Slow** Training a network takes orders of magnitude longer (hours/days) to converge than FEM/FDTD run. |
|---|---|---|---|
| **Inference Time** | **Slow** Re-evaluating at a new point or parameter usually requires a full re-simulation or costly interpolation. | **Slow** Must re-run the time-stepping loop for new parameters or inputs. | **Extremely Fast (Real-Time)** Once trained, predicting fields at any point takes milliseconds. Ideal for real-time applications. |
| **Mesh Required** | **High (Unstructured)** Requires complex mesh generation (tetrahedral or hexahedral). Quality affects stability. | **High (Structured)** Requires a structured grid (Yee grid). complex shapes cause "staircasing" errors. | **None (Mesh-Free)** Uses random "collocation points." No mesh generation needed, handling complex geometry easily. |
| **Inverse Problems** | **Difficult / Expensive** Requires iterative loops (adjoint method), running the forward solver hundreds of times. | **Difficult / Expensive** Same as FEM; computationally prohibitive for many-parameter searches. | **Excellent / Efficient** Seamlessly integrates data and physics. Parameters are learned alongside the solution without a separate loop. |

## 7.2 Transfer Learning and Few-shot Learning in PINNs

The high training cost of PINNs motivates strategies that exploit transferability of learned representations across related parameter regimes or geometries. In engineering design, it is common to explore families of solutions parametrized by Reynolds number, material properties, or device dimensions. Training a separate PINNs from random initialization for each parameter setting is inefficient and often unnecessary, because the underlying PDE operator is unchanged and the solution manifold evolves smoothly in parameter space.[256]

Transfer learning approaches pre-train a PINN on a base configuration and then fine-tune it for new parameter values, for example by initializing a network for $Re = 200$ with weights obtained from a network trained at $Re = 100$. Numerical studies report reductions in training time of 50–80% while achieving comparable accuracy, as optimization starts from a representation that already encodes the relevant physics. Similar strategies have been applied in geometry transfer, where PINNs trained on flows over canonical shapes (such as cylinders) are adapted to related geometries (e.g. ellipses or airfoils) with modest additional training. This few-shot capability is particularly valuable in design loops, where geometries and parameters are perturbed iteratively and fast convergence is essential for tractable optimization.[257] More general meta-learning and operator-learning approaches aim to train models that can generalize across broader families of PDE instances by encoding the solution map itself, but these directions are still under active development and their interplay with physics-informed regularization remains a topic of ongoing research.[258]

## 7.3 Optimization Challenges: Local Minima, Convergence, and Stiffness

Training PINNs entails solving a high-dimensional, non-convex optimization problem with multiple competing loss terms, and a variety of failure modes have been documented. A central issue is optimization stiffness: the combined loss

$$\mathcal{L} = w_r \mathcal{L}_{\text{PDE}} + w_b \mathcal{L}_{\text{BC/IC}} + w_d \mathcal{L}_{\text{data}} \tag{41}$$

may be dominated by one component if the corresponding gradients are much larger in magnitude than the others. In such cases, gradient updates disproportionately reduce boundary-condition or data misfit while leaving the PDE residual largely unchanged, resulting in solutions that fit measurements but violate conservation laws or stability constraints.[259]

NTK-based analysis has further shown that the convergence rate of different Fourier modes of the solution is tied to eigenvalues of the NTK matrix, with high-frequency components associated with small eigenvalues and consequently slow convergence. This explains the observed spectral bias of PINNs: they tend to learn smooth, low-frequency approximations and struggle to capture shocks, boundary layers, or small-scale turbulence without specialized treatment.[260]

Proposed remedies include adaptive loss weighting methods that dynamically adjust $w_r$, $w_b$, $w_d$ based on gradient norms or training progress, curriculum strategies that gradually shift collocation sampling towards regions with high residuals or steep gradients, and re-parameterizations of the solution that precondition the optimization landscape. Additionally, hybrid training schemes employing second-order optimizers, natural gradient methods, or gradient-free solvers have been explored to escape shallow minima and improve convergence robustness, although at increased per-iteration cost.[261]

## 7.4 Interpretability and Uncertainty Quantification

For engineering deployment, particularly in safety-critical domains such as aerospace or medicine, it is insufficient to provide point predictions without quantified uncertainty or interpretable mechanisms. Bayesian PINNs address this by placing prior distributions over network weights or directly over the solution fields and inferring posterior distributions conditioned on data and physics constraints. The resulting predictive distributions yield both mean and variance estimates for state variables and parameters, allowing assessment of confidence in regions with sparse data or extrapolation beyond the training domain.[262]

In practice, Bayesian training via Markov chain Monte Carlo (MCMC) or variational inference is computationally demanding, but recent developments in multi-level sampling and hybrid MCMC–PINN frameworks have improved scalability for moderate problem sizes. Complementary efforts seek to enhance interpretability by combining PINNs with equation-discovery techniques, enabling extraction of explicit PDE forms or reduced-order models from trained networks. These approaches are especially promising in biomedical and materials science applications where mechanistic understanding and regulatory transparency are as important as predictive accuracy.[263]

## 7.5 Cross-Disciplinary Opportunities and Future Directions

The cross-domain nature of PIML has led to transfer of techniques originally developed in one field to others with analogous mathematical structures. Shock-capturing strategies from compressible aerodynamics, such as localized artificial viscosity and adaptive sampling near discontinuities, have been adapted to nanophotonics to sharpen field discontinuities at high-index interfaces and to electrophysiology to better resolve steep activation fronts. Multi-scale network hierarchies used in cosmological simulations to capture galaxy formation within large-scale structures inspire similar architectures in materials science, where grain-boundary evolution and microstructural features must be resolved within macroscopic components.[264]

Data assimilation methodologies, particularly four-dimensional variational (4D-Var) methods widely used in numerical weather prediction, are being reinterpreted in the PINNs framework, allowing time-continuous fusion of sensor data with PDE models for real-time process monitoring and digital twins in industrial settings. This convergence suggests that future PIML architectures will rely increasingly on principled combinations of domain knowledge, optimization theory, and probabilistic modelling rather than purely heuristic loss designs.[265]

## 7.6 Unified PINN Frameworks for Multi-Domain Scientific Discovery

The maturation of physics-informed learning has been accelerated by open-source libraries and industrial frameworks that encapsulate best practices, common PDE templates, and hardware-optimized implementations. DeepXDE, built atop TensorFlow and PyTorch, provides a flexible interface for specifying PDEs, boundary conditions, and geometry via constructive solid geometry, and it supports both forward and inverse problems across a wide range of applications. NVIDIA Modulus targets large-scale industrial use cases, with native support for multi-GPU execution, integration with CAD formats, and dedicated modules for turbulence, heat transfer, and structural mechanics, effectively positioning PINNs as components within larger digital twin environments.[266]

These frameworks not only reduce entry barriers for domain experts but also facilitate systematic benchmarking and reproducibility, enabling the community to compare alternative architectures, loss designs, and training strategies on standardized test cases. As additional modules emerge for specialized domains such as nanophotonics, power systems, and biomechanics, they will support increasingly complex multi-domain workflows, including co-simulation with legacy solvers and integration into engineering design pipelines.

## 7.7 Hardware-accelerated PINNs (TPUs, neuromorphic computing)

The computational intensity of PINNs that stemming from repeated evaluation of networks and their derivatives over large collocation sets has motivated exploration of specialized hardware beyond conventional GPUs. Tensor Processing Units (TPUs), originally optimized for low-precision deep learning workloads, offer high-throughput matrix operations that can significantly reduce training time for large PINNs models, provided that numerical precision is managed carefully. Mixed-precision strategies, in which bulk computations are performed in reduced precision and critical residual evaluations or accumulation in higher precision, have been investigated to balance speed and accuracy, particularly in stiff PDE problems that demand tight residual tolerances.[267]

Neuromorphic platforms such as Intel's Loihi represent a more radical departure from von Neumann architectures by exploiting spiking neural networks and event-driven computation. Prototype studies have shown that certain PDEs, notably diffusion-type equations, can be mapped onto the dynamics of spiking neuron ensembles, yielding orders-of-magnitude improvements in energy efficiency relative to GPU-based implementations. While full-fledged PINNs implementations on neuromorphic hardware remain in early stages, the alignment between time-dependent PDEs and intrinsic temporal dynamics of spiking networks suggests a promising route toward ultra-low-power "edge physics" applications, including autonomous drones, implantable medical devices, and distributed sensor networks.[268]

# 8. Conclusion and Outlook

Physics-informed machine learning has transitioned from a conceptual framework to a rapidly expanding pillar of computational science, fundamentally reshaping how engineers and scientists approach modelling, inference, and design in complex physical systems. By embedding governing equations directly into the training objective, PINNs alleviate the data hunger, lack of physical consistency, and limited extrapolation capability that constrain purely data-driven deep learning models, especially in domains where high-fidelity observations are sparse, expensive, or partially inaccessible.

The present body of work demonstrates that PINNs are not intended as wholesale replacements for FEM or FVM, but rather as complementary tools that excel in specific regimes: inverse problems with sparse or noisy measurements, high-dimensional PDEs that defy mesh-based discretization and tightly coupled multi-physics systems for which monolithic formulations are desirable. Across nanophotonics, fluid mechanics, astronomy, and biomedical engineering, successful case studies have highlighted their capability to reconstruct hidden fields from limited observations, identify effective parameters, and enable new classes of digital twins that integrate data and physics in a unified computational graph.

At the same time, critical limitations remain. Accuracy for standard forward problems lags behind high-order classical solvers, training is prone to optimization pathologies and spectral bias, and hardware requirements can be substantial for three-dimensional transient problems. Addressing these limitations will require coordinated advances along several axes. First, algorithmic innovations such as XPINNs, FBPINNs, conservative formulations, NTK-guided preconditioning, and adaptive loss-balancing must be systematically integrated and benchmarked to improve robustness and accuracy across representative engineering problems, rather than only on idealized test cases.[269] Second, multi-fidelity and hybrid solvers that couple PINNs tightly with FEM/FVM, reduced-order models, or operator-learning networks offer a promising path toward leveraging the strengths of each paradigm, particularly in industrial environments with established legacy codes.[270]

Third, uncertainty quantification and interpretability must be elevated from add-on features to integral components of physics-informed frameworks, enabling rigorous risk assessment and regulatory acceptance in domains such as medical device design, structural health monitoring, and aerospace certification. Bayesian PINNs, physics-guided neural operators, and equation-discovery approaches provide important building blocks in this direction, but further effort is needed to make them scalable and user-friendly for domain practitioners.[271] Finally, the convergence of algorithm and hardware from exascale TPU clusters to energy-efficient neuromorphic devices will determine whether PIML can deliver real-time predictive capabilities at the edge, a key requirement for autonomous systems, adaptive manufacturing, and in situ monitoring of critical infrastructures.[272]

Looking ahead, a plausible trajectory is the emergence of unified PIML ecosystems that integrate domain-decomposition PINNs, operator-learning surrogates, uncertainty-aware inference, and multi-physics coupling within modular, open frameworks such as DeepXDE and NVIDIA Modulus. In such ecosystems, engineers would specify governing physics, design variables, and data sources at a high level, while the underlying tooling orchestrates solver selection, training, and deployment across heterogeneous hardware platforms. Achieving this vision will require sustained cross-disciplinary collaboration among applied mathematicians, domain experts, and computer architects, as well as rigorous community standards for benchmarking, validation, and reproducibility. If these challenges are met, physics-informed machine learning is poised to become a cornerstone of next-generation computational engineering, accelerating discovery and enabling predictive capabilities over scales and complexities that remain beyond the reach of current methods.[269]

# Acknowledgement

The authors would like to acknowledge and thank Egypt Scholars organization and specially the Advanced Labs for organizing the team-work to produce this paper.

## References

[1] L. Quartapelle, *Numerical solution of the incompressible Navier-Stokes equations*. Birkhäuser, 2013.

[2] R. Fitzpatrick, *Maxwell's Equations and the Principles of Electromagnetism*. Jones & Bartlett Publishers, 2008.

[3] C. Nagaiah, K. Kunisch, and G. Plank, "Numerical solution for optimal control of the reaction-diffusion equations in cardiac electrophysiology," *Computational Optimization and Applications,* vol. 49, no. 1, pp. 149-178, 2011.

[4] W. Huang, "Practical aspects of formulation and solution of moving mesh partial differential equations," *Journal of Computational Physics,* vol. 171, no. 2, pp. 753-775, 2001.

[5] P. F. Fischer, "Scaling limits for PDE-based simulation," in *22nd AIAA Computational Fluid Dynamics Conference*, 2015, p. 3049.

[6] D. B. Davidson, *Computational electromagnetics for RF and microwave engineering*. Cambridge University Press, 2010.

[7] B. Labbé, R. Hérault, and C. Chatelain, "Learning deep neural networks for high dimensional output problems," in *2009 International Conference on Machine Learning and Applications*, 2009: IEEE, pp. 63-68.

[8] D. P. Solomatine and A. Ostfeld, "Data-driven modelling: some past experiences and new approaches," *Journal of hydroinformatics,* vol. 10, no. 1, pp. 3-22, 2008.

[9] G. E. Karniadakis, I. G. Kevrekidis, L. Lu, P. Perdikaris, S. Wang, and L. Yang, "Physics-informed machine learning," *Nature Reviews Physics,* vol. 3, no. 6, pp. 422-440, 2021.

[10] B. Amos, "Differentiable optimization-based modeling for machine learning," 2019.

[11] W. T. Riley, "A new era of clinical research methods in a data-rich environment," in *Oncology informatics*: Elsevier, 2016, pp. 343-355.

[12] D. Pfenniger, F. Combes, and L. Martinet, "Is dark matter in spiral galaxies cold gas? I. Observational constraints and dynamical clues about galaxy evolution," *Astronomy and Astrophysics, Vol. 285, p. 79-93 (1994),* vol. 285, pp. 79-93, 1994.

[13] X. W. Klapa Antonion, M. Raissi, and L. Joshie, "Machine learning through physics–informed neural networks: Progress and challenges," *Academic Journal of Science and Technology,* vol. 9, no. 1, p. 2024, 2024.

[14] O. A. M. Abdelraouf, A. M. A. Ahmed, E. Eldele, and A. A. Omar, "From maxwell's equations to artificial intelligence: The evolution of physics-guided AI in nanophotonics and electromagnetics," *Optics & Laser Technology,* vol. 192, p. 113828, 2025/12/01/ 2025, doi: https://doi.org/10.1016/j.optlastec.2025.113828.

[15] J. Gurieva, E. Vasiliev, and L. Smirnov, "Application of conservation laws to the learning of physics-informed neural networks," *Procedia Computer Science,* vol. 212, pp. 464-473, 2022.

[16] Y. Yang and Y. Mesri, "Learning by neural networks under physical constraints for simulation in fluid mechanics," *Computers & Fluids,* vol. 248, p. 105632, 2022.

[17] I. I. Abdulaal, A. W. Elsayed, and O. A. Abdelraouf, "Terahertz Quasi-BIC Metasurfaces for Ultra-Sensitive Biosensing and High-Speed Wireless Communications," *arXiv preprint arXiv:2510.00357,* 2025.

[18] O. M. Halawa, E. Ahmed, M. M. Abdelrazek, Y. M. Nagy, and O. A. M. Abdelraouf, "Recent Progress in Nanophotonics for Green Energy, Medicine, Healthcare, and Optical Computing Applications," *Materials,* vol. 19, no. 8, p. 1660, 2026. [Online]. Available: https://www.mdpi.com/1996-1944/19/8/1660.

[19] H. Liu *et al.*, "High-order photonic cavity modes enabled 3D structural colors," *ACS nano,* vol. 16, no. 5, pp. 8244-8252, 2022.

[20] O. A. Abdelraouf *et al.*, "Recent advances in tunable metasurfaces: materials, design, and applications," *ACS nano,* vol. 16, no. 9, pp. 13339-13369, 2022.

[21] O. A. Abdelraouf *et al.*, "All-Optical Switching of Structural Color with a Fabry–Pérot Cavity," *Advanced Photonics Research,* vol. 4, no. 11, p. 2300209, 2023.

[22] S. Jana *et al.*, "Aperiodic Bragg reflectors for tunable high-purity structural color based on phase change material," *Nano Letters,* vol. 24, no. 13, pp. 3922-3929, 2024.
[23] T. S. Almurayziq, O. A. Abdelraouf, A. Shaker, M. Salem, M. T. Alshammari, and M. T. Qureshi, "Nanophotonic Engineering of CZTSSe Thin-Film Solar Cells for Near-Unity Absorption and Enhanced Efficiency," *Photonics and Nanostructures-Fundamentals and Applications,* p. 101528, 2026.
[24] D. Khodair, A. Saeed, A. Shaker, M. Abouelatta, O. A. Abdelraouf, and S. EL-Rabaie, "A review on tandem solar cells based on Perovskite/Si: 2-T versus 4-T configurations," *Solar Energy,* vol. 300, p. 113815, 2025.
[25] N. Atef, S. S. Emara, D. S. Eissa, A. El-Sayed, O. A. Abdelraouf, and N. K. Allam, "Well-dispersed Au nanoparticles prepared via magnetron sputtering on TiO2 nanotubes with unprecedentedly high activity for water splitting," *Electrochemical Science Advances,* vol. 1, no. 1, p. e2000004, 2021.
[26] O. A. Abdelraouf and N. K. Allam, "Nanostructuring for enhanced absorption and carrier collection in CZTS-based solar cells: coupled optical and electrical modeling," *Optical Materials,* vol. 54, pp. 84-88, 2016.
[27] O. A. Abdelraouf and N. K. Allam, "Towards nanostructured perovskite solar cells with enhanced efficiency: Coupled optical and electrical modeling," *Solar Energy,* vol. 137, pp. 364-370, 2016.
[28] O. A. Abdelraouf, M. I. Abdelrahaman, and N. K. Allam, "Plasmonic scattering nanostructures for efficient light trapping in flat czts solar cells," in *Metamaterials XI*, 2017, vol. 10227: SPIE, pp. 90-98.
[29] O. A. Abdelraouf, H. A. Ali, and N. K. Allam, "Optimizing absorption and scattering cross section of metal nanostructures for enhancing light coupling inside perovskite solar cells," in *2017 Conference on Lasers and Electro-Optics Europe & European Quantum Electronics Conference (CLEO/Europe-EQEC)*, 2017: IEEE, pp. 1-1.
[30] O. A. Abdelraouf, A. Shaker, and N. K. Allam, "Using all dielectric and plasmonic cross grating metasurface for enhancing efficiency of CZTS solar cells," in *Nanophotonics VII*, 2018, vol. 10672: SPIE, pp. 246-255.
[31] O. A. Abdelraouf, A. Shaker, and N. K. Allam, "All dielectric and plasmonic cross-grating metasurface for efficient perovskite solar cells," in *Metamaterials Xi*, 2018, vol. 10671: SPIE, pp. 104-112.
[32] O. A. Abdelraouf, A. Shaker, and N. K. Allam, "Enhancing light absorption inside CZTS solar cells using plasmonic and dielectric wire grating metasurface," in *Metamaterials XI*, 2018, vol. 10671: SPIE, pp. 165-174.
[33] O. A. Abdelraouf, A. Shaker, and N. K. Allam, "Design methodology for selecting optimum plasmonic scattering nanostructures inside CZTS solar cells," in *Photonics for Solar Energy Systems VII*, 2018, vol. 10688: SPIE, pp. 24-32.
[34] O. A. Abdelraouf, A. Shaker, and N. K. Allam, "Design of optimum back contact plasmonic nanostructures for enhancing light coupling in CZTS solar cells," in *Photonics for Solar Energy Systems VII*, 2018, vol. 10688: SPIE, pp. 33-41.
[35] O. A. Abdelraouf, A. Shaker, and N. K. Allam, "Plasmonic nanoscatter antireflective coating for efficient CZTS solar cells," in *Photonics for Solar Energy Systems VII*, 2018, vol. 10688: SPIE, pp. 15-23.
[36] O. A. Abdelraouf, A. Shaker, and N. K. Allam, "Novel design of plasmonic and dielectric antireflection coatings to enhance the efficiency of perovskite solar cells," *Solar Energy,* vol. 174, pp. 803-814, 2018.
[37] O. A. Abdelraouf, A. Shaker, and N. K. Allam, "Front dielectric and back plasmonic wire grating for efficient light trapping in perovskite solar cells," *Optical materials,* vol. 86, pp. 311-317, 2018.
[38] P. R. Wiecha, A. Arbouet, C. Girard, and O. L. Muskens, "Deep learning in nano-photonics: inverse design and beyond," *Photon. Res.,* vol. 9, no. 5, pp. B182-B200, 2021.

[39] V. Roquemen-Echeverri and C. Mosquera-Lopez, "Recent advancements and applications of physics-informed machine learning in biomedical research," *Current Opinion in Biomedical Engineering,* vol. 35, p. 100612, 2025.
[40] A. D. Jagtap and G. E. Karniadakis, "Extended physics-informed neural networks (XPINNs): A generalized space-time domain decomposition based deep learning framework for nonlinear partial differential equations," *Communications in Computational Physics,* vol. 28, no. 5, 2020.
[41] D. Ceccarelli, "Bayesian physics-informed neural networks for inverse uncertainty quantification problems in cardiac electrophysiology," 2019.
[42] H. Wang, L. Lu, S. Song, and G. Huang, "Learning specialized activation functions for physics-informed neural networks," *arXiv preprint arXiv:2308.04073,* 2023.
[43] G. M. Gunay, "Effect of Activation Functions on Performance of Physics Informed Neural Networks: Beam Deflection Case Study," *Available at SSRN 5642997,* 2025.
[44] V. Sitzmann, J. Martel, A. Bergman, D. Lindell, and G. Wetzstein, "Implicit neural representations with periodic activation functions," *Advances in neural information processing systems,* vol. 33, pp. 7462-7473, 2020.
[45] K. Luo *et al.*, "Physics-informed neural networks for PDE problems: A comprehensive review," *Artificial Intelligence Review,* vol. 58, no. 10, p. 323, 2025.
[46] S. L. Brunton, J. L. Proctor, and J. N. Kutz, "Sparse identification of nonlinear dynamics with control (SINDYc)," *IFAC-PapersOnLine,* vol. 49, no. 18, pp. 710-715, 2016.
[47] Z. Chen, V. Badrinarayanan, C.-Y. Lee, and A. Rabinovich, "Gradnorm: Gradient normalization for adaptive loss balancing in deep multitask networks," in *International conference on machine learning*, 2018: PMLR, pp. 794-803.
[48] A. A. Heydari, C. A. Thompson, and A. Mehmood, "Softadapt: Techniques for adaptive loss weighting of neural networks with multi-part loss functions," *arXiv preprint arXiv:1912.12355,* 2019.
[49] A. Jacot, F. Gabriel, and C. Hongler, "Neural tangent kernel: Convergence and generalization in neural networks," *Advances in neural information processing systems,* vol. 31, 2018.
[50] Z. Hu, A. D. Jagtap, G. E. Karniadakis, and K. Kawaguchi, "When do extended physics-informed neural networks (XPINNs) improve generalization?," *arXiv preprint arXiv:2109.09444,* 2021.
[51] D. Hansen, D. C. Maddix, S. Alizadeh, G. Gupta, and M. W. Mahoney, "Learning physical models that can respect conservation laws," in *International Conference on Machine Learning*, 2023: PMLR, pp. 12469-12510.
[52] O. A. M. Abdelraouf, "NanoPhotoNet-Inverse: AI-driven inverse design of dual-resonance multi-layer metasurface for enhanced single photon emission," *Optics Communications,* vol. 610, p. 133152, 2026/08/01/ 2026, doi: https://doi.org/10.1016/j.optcom.2026.133152.
[53] O. A. M. Abdelraouf, "NanoPhotoNet-NL: AI-Assisted Design of Nonlinear Multilayer Metasurfaces for Broadband Tunable Deep Ultraviolet Emission," *Optics Communications,* p. 132526, 2025/10/03/ 2025, doi: https://doi.org/10.1016/j.optcom.2025.132526.
[54] O. A. Abdelraouf, A. Mousa, and M. Ragab, "NanoPhotoNet: AI-Enhanced Design Tool for Reconfigurable and High-Performance Multi-Layer Metasurfaces," *Photonics and Nanostructures-Fundamentals and Applications,* p. 101379, 2025.
[55] P. Zhang, Y. Hu, Y. Jin, S. Deng, X. Wu, and J. Chen, "A Maxwell's Equations Based Deep Learning Method for Time Domain Electromagnetic Simulations," *IEEE Journal on Multiscale and Multiphysics Computational Techniques,* vol. 6, pp. 35-40, 2021, doi: 10.1109/JMMCT.2021.3057793.
[56] H. Qin, T. Zhang, H. Bao, Z. Yu, and D. Ding, "Physics-Informed Neural Network for Solving Three-Dimensional Maxwell's Equations," in *2025 International Conference on Microwave and Millimeter Wave Technology (ICMMT)*, 19-22 May 2025 2025, pp. 1-3, doi: 10.1109/ICMMT65948.2025.11188574.

[57] G. G. Shaviner, H. Chandravamsi, S. Pisnoy, Z. Chen, and S. H. Frankel, "PINNs for solving unsteady Maxwell’s equations: convergence issues and comparative assessment with compact schemes," *Neural Computing and Applications,* vol. 37, no. 29, pp. 24103-24122, 2025/10/01 2025, doi: 10.1007/s00521-025-11554-2.

[58] S. Oh and S. K. Hong, "Physics-Informed Neural Modeling of 2D Transient Electromagnetic Fields," *Applied Sciences,* vol. 15, no. 23, p. 12612, 2025. [Online]. Available: https://www.mdpi.com/2076-3417/15/23/12612.

[59] M. Nohra and S. Dufour, "Physics-Informed Neural Networks for the Numerical Modeling of Steady-State and Transient Electromagnetic Problems with Discontinuous Media," arXiv, arXiv:2406.04380 [physics.comp-ph], 2024/06/06 2024. [Online]. Available: https://arxiv.org/abs/2406.04380

[60] H. Zhang *et al.*, "FE-PIRBN: Feature-enhanced physics-informed radial basis neural networks for solving high-frequency electromagnetic scattering problems," *Journal of Computational Physics,* vol. 527, p. 113798, 2025/04/15/ 2025, doi: https://doi.org/10.1016/j.jcp.2025.113798.

[61] A. Khan and D. A. Lowther, "Physics Informed Neural Networks for Electromagnetic Analysis," *IEEE Transactions on Magnetics,* vol. 58, no. 9, pp. 1-4, 2022, doi: 10.1109/TMAG.2022.3161814.

[62] Z. Gong, Y. Chu, and S. Yang, "Physics-Informed Neural Networks for Solving 2-D Magnetostatic Fields," *IEEE Transactions on Magnetics,* vol. 59, no. 11, pp. 1-5, 2023, doi: 10.1109/TMAG.2023.3281863.

[63] A. Beltrán-Pulido, I. Bilionis, and D. Aliprantis, "Physics-Informed Neural Networks for Solving Parametric Magnetostatic Problems," *IEEE Transactions on Energy Conversion,* vol. 37, no. 4, pp. 2678-2689, 2022, doi: 10.1109/TEC.2022.3180295.

[64] Y. Su, S. Zeng, X. Wu, Y. Huang, and J. Chen, "Physics-Informed Graph Neural Network for Electromagnetic Simulations," in *2023 XXXVth General Assembly and Scientific Symposium of the International Union of Radio Science (URSI GASS)*, 19-26 Aug. 2023 2023, pp. 1-3, doi: 10.23919/URSIGASS57860.2023.10265621.

[65] J. Lim and D. Psaltis, "MaxwellNet: Physics-driven deep neural network training based on Maxwell’s equations," *APL Photonics,* vol. 7, no. 1, 2022, doi: 10.1063/5.0071616.

[66] C. Gigli, A. Saba, A. B. Ayoub, and D. Psaltis, "Predicting nonlinear optical scattering with physics-driven neural networks," *APL Photonics,* vol. 8, no. 2, 2023, doi: 10.1063/5.0119186.

[67] H. Liu *et al.*, "Physics-Informed Deep Model for Fast Time-Domain Electromagnetic Simulation and Inversion," *IEEE Transactions on Antennas and Propagation,* vol. 72, no. 10, pp. 7807-7820, 2024, doi: 10.1109/TAP.2024.3433389.

[68] V. Medvedev, A. Erdmann, and A. Rosskopf, "Physics-informed deep learning for 3D modeling of light diffraction from optical metasurfaces," *Optics Express,* vol. 33, no. 1, pp. 1371-1384, 2025.

[69] W. Xu, Q. Zhong, M. Wang, Z. Wei, Z. Wang, and X. Cheng, "High precision, full-vector optical mode solving in waveguides via fourth-order derivative physics-informed neural networks," *Optics Express,* vol. 33, no. 18, pp. 38317-38328, 2025/09/08 2025, doi: 10.1364/OE.571156.

[70] J. Ding *et al.*, "Virtual draw of microstructured optical fiber based on physics-informed neural networks," *Optics Express,* vol. 32, no. 6, pp. 9316-9331, 2024/03/11 2024, doi: 10.1364/OE.518238.

[71] O. A. Abdelraouf, "Amplified Directional Photoluminescence from CIS Quantum Dots and hBN Quantum Emitters using Tunable BIC Metasurfaces," *arXiv preprint arXiv:2510.10470,* 2025.

[72] O. A. M. Abdelraouf, "Electrically tunable photon-pair generation in nanostructured NbOCl2 for quantum communications," *Optics & Laser Technology,* vol. 192, p. 113517, 2025/12/01/ 2025, doi: https://doi.org/10.1016/j.optlastec.2025.113517.

[73] O. A. Abdelraouf, "Phase-matched Deep Ultraviolet Chiral Bound States in the Continuum Metalens," *arXiv preprint arXiv:2509.15904,* 2025.
[74] O. A. Abdelraouf *et al.*, "Tunable transmissive THG in silicon metasurface enabled by phase change material," in *CLEO: QELS_Fundamental Science*, 2021: Optica Publishing Group, p. FTh4K. 3.
[75] O. A. Abdelraouf *et al.*, "Multistate tuning of third harmonic generation in fano-resonant hybrid dielectric metasurfaces," *Advanced Functional Materials,* vol. 31, no. 48, p. 2104627, 2021.
[76] O. A. Abdelraouf, A. P. Anthur, X. R. Wang, Q. J. Wang, and H. Liu, "Modal phase-matched bound states in the continuum for enhancing third harmonic generation of deep ultraviolet emission," *ACS nano,* vol. 18, no. 5, pp. 4388-4397, 2024.
[77] O. A. Abdelraouf, M. Wu, and H. Liu, "Hybrid Metasurfaces Enabling Focused Tunable Amplified Photoluminescence Through Dual Bound States in the Continuum," *Advanced Functional Materials,* p. 2505165, 2025.
[78] O. A. M. Abdelraouf, T. Wang, and Z. Wang, "Recent Developments in Deep-Ultraviolet Flat Optics," *Nanophotonics,* vol. 15, no. 10, p. e70136, 2026, doi: https://doi.org/10.1002/nap2.70136.
[79] A. Ghosh, M. Elhamod, J. Bu, W.-C. Lee, A. Karpatne, and V. A. Podolskiy, "Physics-Informed Machine Learning for Optical Modes in Composites," *Advanced Photonics Research,* vol. 3, no. 11, p. 2200073, 2022, doi: https://doi.org/10.1002/adpr.202200073.
[80] F. Davoodi, "Active Physics-Informed Deep Learning: Surrogate Modeling for Nonplanar Wavefront Excitation of Topological Nanophotonic Devices," *Nano Letters,* vol. 25, no. 2, pp. 768-775, 2025/01/15 2025, doi: 10.1021/acs.nanolett.4c05120.
[81] J. Gu *et al.*, "NeurOLight: a physics-agnostic neural operator enabling parametric photonic device simulation," presented at the Proceedings of the 36th International Conference on Neural Information Processing Systems, New Orleans, LA, USA, 2022.
[82] Z. Fang and J. Zhan, "Deep Physical Informed Neural Networks for Metamaterial Design," *IEEE Access,* vol. 8, pp. 24506-24513, 2020, doi: 10.1109/ACCESS.2019.2963375.
[83] Y. Wang and S. Zhang, "Multi-receptive-field physics-informed neural network for complex electromagnetic media," *Opt. Mater. Express,* vol. 14, no. 11, pp. 2740-2754, 2024/11/01 2024, doi: 10.1364/OME.533643.
[84] L. Armbruster, V. Medvedev, and A. Rosskopf, *Physics-Informed PointNets for Modeling Electromagnetic Scattering from All-Dielectric Metasurfaces with Inclined Nanopillars*. 2025.
[85] l. Xiao *et al.*, *Forward electromagnetic modeling and inverse design of etch cylindrical holes using physics-informed neural network*. 2025, p. 23.
[86] O. Elsheikh, A. Shaaban, A. Arafa, A. S. E. Yahia, L. Gomaa, and N. Gad, "Predicting Fundamental Transverse Electric Mode of Slab Waveguide Based on Physics-Informed Neural Networks," *Egyptian Journal of Pure and Applied Science,* vol. 61, no. 1, pp. 1-10, 2023, doi: 10.21608/ejaps.2023.181263.1047.
[87] M. G. Mahmoud, A. S. Hares, M. F. O. Hameed, M. S. El-Azab, and S. S. A. Obayya, "AI-driven photonics: Unleashing the power of AI to disrupt the future of photonics," *APL Photonics,* vol. 9, no. 8, 2024, doi: 10.1063/5.0220766.
[88] M. R. Khan, C. L. Zekios, S. Bhardwaj, and S. V. Georgakopoulos, "A Physics-Informed Neural Network-Based Waveguide Eigenanalysis," *IEEE Access,* vol. 12, pp. 120777-120787, 2024, doi: 10.1109/ACCESS.2024.3452160.
[89] H. S. Ünal and A. C. Durgun, "A physics-aware neural network for effective refractive index prediction of photonic waveguides," *Optical and Quantum Electronics,* vol. 57, no. 1, p. 107, 2025/01/11 2025, doi: 10.1007/s11082-024-08009-8.
[90] A. S. Hares, M. S. El-Azab, and S. S. A. Obayya, "A stable physics-guided neural networks approach for electromagnetic problems," *Journal of Computational Physics,* vol. 548, p. 114568, 2026/03/01/ 2026, doi: https://doi.org/10.1016/j.jcp.2025.114568.

[91] L. Huang, L. Chen, and R. Bai, "Physics-Informed Extreme Learning Machine Applied for Eigenmode Analysis of Waveguides and Transmission Lines," *International Journal of RF and Microwave Computer-Aided Engineering,* vol. 2025, no. 1, p. 6233356, 2025, doi: https://doi.org/10.1155/mmce/6233356.

[92] W. Gao *et al.*, "High-Precision Large-Scale Optical Phased Array Calibration Using Physics-Informed Neural Network With Transfer Learning," *J. Lightwave Technol.,* vol. 43, no. 22, pp. 10203-10209, 2025/11/15 2025. [Online]. Available: https://opg.optica.org/jlt/abstract.cfm?URI=jlt-43-22-10203.

[93] P. Ma, H. Yang, Z. Gao, D. S. Boning, and J. Gu, "PIC2O-Sim: A physics-inspired causality-aware dynamic convolutional neural operator for ultra-fast photonic device time-domain simulation," *APL Photonics,* vol. 10, no. 3, 2025, doi: 10.1063/5.0242728.

[94] I. Teofilovic and F. Da Ros, *Physics-informed Data-driven Model for Predicting Thermal Crosstalk in a Photonic Integrated Circuit*. 2025.

[95] S. Qi and C. D. Sarris, "Physics-Informed Neural Networks for Multiphysics Simulations: Application to Coupled Electromagnetic-Thermal Modeling," in *2023 IEEE/MTT-S International Microwave Symposium - IMS 2023*, 11-16 June 2023 2023, pp. 166-169, doi: 10.1109/IMS37964.2023.10188015.

[96] Q. Zhang, Y. Sui, S. Rothe, and J. Czarske, "Physics-informed mode decomposition neural network for structured light in multimode fibers," in *2023 IEEE Photonics Conference (IPC)*, 12-16 Nov. 2023 2023, pp. 1-2, doi: 10.1109/IPC57732.2023.10360643.

[97] X. Luo, M. Zhang, Z. Wang, X. Jiang, Y. Song, and D. Wang, "PINN-BPM: An Enhanced Physics-Informed Neural Network Framework of Solving Helmholtz Equation for Light Field Propagation in Optical Fiber," *Journal of Lightwave Technology,* vol. 43, no. 23, pp. 10380-10401, 2025, doi: 10.1109/JLT.2025.3615984.

[98] Y. Zang, Z. Yu, K. Xu, M. Chen, S. Yang, and H. Chen, "Universal Fiber Models based on PINN Neural Network," in *2020 Asia Communications and Photonics Conference (ACP) and International Conference on Information Photonics and Optical Communications (IPOC)*, 24-27 Oct. 2020 2020, pp. 1-3.

[99] M. Liu and G. Liu, "Smoothed Particle Hydrodynamics (SPH): an Overview and Recent Developments," *Archives of Computational Methods in Engineering,* vol. 17, pp. 25-76, 2010.

[100] A. Jameson and A. Katz, "Meshless methods for computational fluid dynamics," 2009.

[101] G. E. Karniadakis and S. J. Sherwin, "Spectral/hp Element Methods for Computational Fluid Dynamics," 2005.

[102] A. N. Brooks and T. J. R. Hughes, "Streamline upwind/Petrov-Galerkin formulations for convection dominated flows with particular emphasis on the incompressible Navier-Stokes equations," *Computer Methods in Applied Mechanics and Engineering,* vol. 32, pp. 199-259, 1990.

[103] A. Abdelsamie *et al.*, "Comparing LES and URANS results with a reference DNS of the transitional airflow in a patient-specific larynx geometry during exhalation," *Computers & Fluids,* 2023.

[104] P. Sharma, W. T. Chung, B. Akoush, and M. Ihme, "A Review of Physics-Informed Machine Learning in Fluid Mechanics," *Energies,* 2023.

[105] M. Raissi, P. Perdikaris, and G. E. Karniadakis, "Physics-informed neural networks: A deep learning framework for solving forward and inverse problems involving nonlinear partial differential equations," *J. Comput. Phys.,* vol. 378, pp. 686-707, 2019.

[106] R. Laubscher, "Simulation of multi-species flow and heat transfer using physics-informed neural networks," 2021.

[107] Y. Weng and D. Zhou, "Multiscale Physics-Informed Neural Networks for Stiff Chemical Kinetics," *The journal of physical chemistry. A,* 2022.

[108] W. Ji, W. Qiu, Z. Shi, S. Pan, and S. Deng, "Stiff-PINN: Physics-Informed Neural Network for Stiff Chemical Kinetics," *The journal of physical chemistry. A,* 2020.

[109] M. Aliakbari, M. Mahmoudi, P. Vadasz, and A. Arzani, "Predicting high-fidelity multiphysics data from low-fidelity fluid flow and transport solvers using physics-informed neural networks," *International Journal of Heat and Fluid Flow,* 2022.
[110] R. Qiu *et al.*, "Physics-informed neural networks for phase-field method in two-phase flow," *Physics of Fluids,* 2022.
[111] H. Wang, Y. Liu, and S. Wang, "Dense velocity reconstruction from particle image velocimetry/particle tracking velocimetry using a physics-informed neural network," *Physics of Fluids,* 2022.
[112] H. Eivazi, M. Tahani, P. Schlatter, and R. Vinuesa, "Physics-informed neural networks for solving Reynolds-averaged Navier-Stokes equations," *ArXiv,* vol. abs/2107.10711, 2021.
[113] S. Cai, Z. Wang, F. Fuest, Y. J. Jeon, C. Gray, and G. E. Karniadakis, "Flow over an espresso cup: inferring 3-D velocity and pressure fields from tomographic background oriented Schlieren via physics-informed neural networks," *Journal of Fluid Mechanics,* vol. 915, 2021.
[114] X. Jin, S. Cai, H. Li, and G. E. Karniadakis, "NSFnets (Navier-Stokes flow nets): Physics-informed neural networks for the incompressible Navier-Stokes equations," *J. Comput. Phys.,* vol. 426, p. 109951, 2020.
[115] M. Raissi, A. Yazdani, and G. E. Karniadakis, "Hidden fluid mechanics: Learning velocity and pressure fields from flow visualizations," *Science,* vol. 367, pp. 1026 - 1030, 2020.
[116] M. Raissi, Z. Wang, M. S. Triantafyllou, and G. E. Karniadakis, "Deep learning of vortex-induced vibrations," *Journal of Fluid Mechanics,* vol. 861, pp. 119 - 137, 2018.
[117] K. Duraisamy, G. Iaccarino, and H. Xiao, "Turbulence Modeling in the Age of Data," *Annual Review of Fluid Mechanics,* 2018.
[118] J.-X. Wang, J.-L. Wu, and H. Xiao, "Physics-informed machine learning approach for reconstructing Reynolds stress modeling discrepancies based on DNS data," *arXiv: Fluid Dynamics,* vol. 2, p. 034603, 2016.
[119] A. P. Singh, S. Medida, and K. Duraisamy, "Machine Learning-augmented Predictive Modeling of Turbulent Separated Flows over Airfoils," *ArXiv,* vol. abs/1608.03990, 2016.
[120] H. Xiao, J.-L. Wu, J.-X. Wang, R. Sun, and C. J. Roy, "Quantifying and reducing model-form uncertainties in Reynolds-averaged Navier-Stokes simulations: A data-driven, physics-informed Bayesian approach," *J. Comput. Phys.,* vol. 324, pp. 115-136, 2015.
[121] E. J. Parish and K. Duraisamy, "A paradigm for data-driven predictive modeling using field inversion and machine learning," *J. Comput. Phys.,* vol. 305, pp. 758-774, 2016.
[122] J. Ling, A. Kurzawski, and J. A. Templeton, "Reynolds averaged turbulence modelling using deep neural networks with embedded invariance," *Journal of Fluid Mechanics,* vol. 807, pp. 155 - 166, 2016.
[123] J. Ling, R. E. Jones, and J. A. Templeton, "Machine learning strategies for systems with invariance properties," *J. Comput. Phys.,* vol. 318, pp. 22-35, 2016.
[124] J. A. Templeton, "Evaluation of machine learning algorithms for prediction of regions of high Reynolds averaged Navier Stokes uncertainty," *Physics of Fluids,* vol. 27, p. 085103, 2015.
[125] M. Bode *et al.*, "Using physics-informed enhanced super-resolution generative adversarial networks for subfilter modeling in turbulent reactive flows," 2021.
[126] R. Maulik and O. San, "A neural network approach for the blind deconvolution of turbulent flows," *Journal of Fluid Mechanics,* vol. 831, pp. 151 - 181, 2017.
[127] A. D. Beck, D. G. Flad, and C.-D. Munz, "Deep Neural Networks for Data-Driven Turbulence Models," *J. Comput. Phys.,* vol. 398, 2018.
[128] C. J. Lapeyre, A. Misdariis, N. Cazard, D. Veynante, and T. Poinsot, "Training convolutional neural networks to estimate turbulent sub-grid scale reaction rates," *Combustion and Flame,* 2018.
[129] M. d. A. Cruz, R. L. Thompson, L. Sampaio, and R. D. A. Bacchi, "The use of the Reynolds force vector in a physics informed machine learning approach for predictive turbulence modeling," *Computers & Fluids,* 2019.

[130] G. Berkooz, P. Holmes, and J. L. Lumley, "The Proper Orthogonal Decomposition in the Analysis of Turbulent Flows," *Annual Review of Fluid Mechanics,* vol. 25, pp. 539-575, 1993.
[131] K. E. Meyer, J. M. Pedersen, and O. Ozcan, "A turbulent jet in crossflow analysed with proper orthogonal decomposition," *Journal of Fluid Mechanics,* vol. 583, pp. 199 - 227, 2007.
[132] H. F. S. Lui and W. R. Wolf, "Construction of reduced-order models for fluid flows using deep feedforward neural networks," *Journal of Fluid Mechanics,* vol. 872, pp. 963 - 994, 2019.
[133] D. S. Karim S. Ahmed, Dylan SitarskiAnupam SharmaAnupam SharmaPaul DurbinPaul Durbin, "Enhanced RANS Modeling of Separation Induced Transition Flows Using Field Inversion," presented at the 13th International Symposium on Turbulence and Shear Flow Phenomena (TSFP13)At: Montreal, Canada.
[134] K. Ahmed, S. Menon, D. Sitarski, A. Sharma, and P. Durbin, "Data-Enhanced RANS Modeling of Unsteady Transitional Flows," *AIAA AVIATION FORUM AND ASCEND 2025,* 2025.
[135] Z. Mao, A. D. Jagtap, and G. E. Karniadakis, "Physics-informed neural networks for high-speed flows," *Computer Methods in Applied Mechanics and Engineering,* vol. 360, p. 112789, 2020/03/01/ 2020, doi: https://doi.org/10.1016/j.cma.2019.112789.
[136] J. Li, X. Wu, and Z. Li, "Physics-informed neural network approach for solving vortex-induced vibration problems of cylinder," *Applied Ocean Research,* 2025.
[137] K. Cong, G. Li, Y. Sun, and H. Ren, "Using physics-informed derivative networks to solve the forward problem of a free-convective boundary layer problem," *Scientific Reports,* vol. 15, 2025.
[138] P. Stefanou, J. F. Urbán, and J. A. Pons, "Solving the pulsar equation using physics-informed neural networks," *Monthly Notices of the Royal Astronomical Society,* vol. 526, no. 1, pp. 1504-1511, 2023.
[139] A. S. Cornell, S. R. Herbst, A. M. Ncube, and H. Noshad, "Solving the Regge-Wheeler and Teukolsky equations: supervised versus unsupervised physics-informed neural networks," *arXiv preprint arXiv:2402.11343,* 2024.
[140] D. Dahlbüdding, K. Molaverdikhani, B. Ercolano, and T. Grassi, "Approximating Rayleigh scattering in exoplanetary atmospheres using physics-informed neural networks," *Monthly Notices of the Royal Astronomical Society,* vol. 533, no. 3, pp. 3475-3483, 2024.
[141] L. G. Bachar, A. T. Chantada, S. J. Landau, C. G. Scóccola, and P. Protopapas, "Evolution of linear matter perturbations with error-bounded bundle physics-informed neural networks," *arXiv preprint arXiv:2508.08443,* 2025.
[142] A. K. Mishra and E. Tolley, "SPINN: Advancing Cosmological Simulations of Fuzzy Dark Matter with Physics Informed Neural Networks," (in English), *Astrophysical Journal,* Article vol. 988, no. 1, 2025, Art no. 114, doi: 10.3847/1538-4357/ade43e.
[143] Z. Dai, B. Moews, R. Vilalta, and R. Davé, "Physics-informed neural networks in the recreation of hydrodynamic simulations from dark matter," (in English), *Monthly Notices of the Royal Astronomical Society,* Article vol. 527, no. 2, pp. 3381-3394, 2024, doi: 10.1093/mnras/stad3394.
[144] P. Flores, O. Graf, P. Protopapas, and K. Pichara, "Improved Uncertainty Quantification in Physics-Informed Neural Networks Using Error Bounds and Solution Bundles," *arXiv preprint arXiv:2505.06459,* 2025.
[145] D. Korber, M. Bianco, E. Tolley, and J. P. Kneib, "PINION: physics-informed neural network for accelerating radiative transfer simulations for cosmic reionization," (in English), *Monthly Notices of the Royal Astronomical Society,* Article vol. 521, no. 1, pp. 902-915, 2023, doi: 10.1093/mnras/stad615.

[146] A. T. Chantada, S. J. Landau, P. Protopapas, C. G. Scóccola, and C. Garraffo, "Cosmology-informed neural networks to solve the background dynamics of the Universe," *Physical Review D,* vol. 107, no. 6, p. 063523, 2023.

[147] J. J. Athalathil, B. Vaidya, S. Kundu, V. Upendran, and M. C. M. Cheung, "Surface Flux Transport Modeling Using Physics-informed Neural Networks," (in English), *Astrophysical Journal,* Article vol. 975, no. 2, 2024, Art no. 258, doi: 10.3847/1538-4357/ad7d91.

[148] N. Costa, F. S. Barros, J. J. Lima, R. F. Pinto, and A. Restivo, "Leveraging Physics-Informed Neural Networks as Solar Wind Forecasting Models," in *Proceedings of the 32nd European Symposium on Artificial Neural Networks, Computational Intelligence and Machine Learning (ESANN'2024), Bruges (Belgium), page in this volume, Louvain-La-Neuve, Belgium*, 2024, p. i6doc.

[149] H. Baty and V. Vigon, "Modelling solar coronal magnetic fields with physics-informed neural networks," *Monthly Notices of the Royal Astronomical Society,* vol. 527, no. 2, pp. 2575-2584, 2024.

[150] X. Lian, S. Liu, X. Cao, H. Wang, W. Deng, and X. Ning, "Agile control of test mass based on PINN-DDPG for drag-free satellite," *ISA Transactions,* vol. 157, pp. 306-317, 2025/02/01/ 2025, doi: https://doi.org/10.1016/j.isatra.2024.11.049.

[151] J. Martin and H. Schaub, "Physics-informed neural networks for gravity field modeling of the Earth and Moon," *Celestial Mechanics and Dynamical Astronomy,* vol. 134, no. 2, p. 13, 2022.

[152] M. Rasht-Behesht, C. Huber, K. Shukla, and G. E. Karniadakis, "Physics-informed neural networks (PINNs) for wave propagation and full waveform inversions," *Journal of Geophysical Research: Solid Earth,* vol. 127, no. 5, p. e2021JB023120, 2022.

[153] J. Li, M. Jian, Y.-S. Ting, and G. M. Green, "Differentiable Stellar Atmospheres with Physics-Informed Neural Networks," *arXiv preprint arXiv:2507.06357,* 2025.

[154] M. Raissi, P. Perdikaris, and G. E. Karniadakis, "Physics informed deep learning (part i): Data-driven solutions of nonlinear partial differential equations," *arXiv preprint arXiv:1711.10561,* 2017.

[155] W.-B. Tay, M. Damodaran, Z.-D. Teh, and R. Halder, "Investigation of applying physics informed neural networks (PINN) and variants on 2d aerodynamics problems," in *Fluids Engineering Division Summer Meeting*, 2020, vol. 83730: American Society of Mechanical Engineers, p. V003T05A055.

[156] L. Liu *et al.*, "Discontinuity computing using physics-informed neural networks," *Journal of Scientific Computing,* vol. 98, no. 1, p. 22, 2024.

[157] J. Vandegriff, K. Wagstaff, G. Ho, and J. Plauger, "Forecasting space weather: Predicting interplanetary shocks using neural networks," *Advances in Space Research,* vol. 36, no. 12, pp. 2323-2327, 2005.

[158] D. Sudar, B. Vršnak, and M. Dumbović, "Predicting coronal mass ejections transit times to Earth with neural network," *Monthly Notices of the Royal Astronomical Society,* vol. 456, no. 2, pp. 1542-1548, 2015.

[159] S. Moschou, E. Hicks, R. Parekh, D. Mathew, S. Majumdar, and N. Vlahakis, "Physics-informed neural networks for modeling astrophysical shocks," *Machine Learning: Science and Technology,* vol. 4, no. 3, p. 035032, 2023.

[160] Z. Anni, F. Davood, and H. Xiao, "Physics-informed neural networks for physiological signal processing and modeling: a narrative review," *Physiological Measurement,* vol. 46, pp. 07TR02 , pmid = 40680769 , publisher = Institute of Physics, 7 2025, doi: 10.1088/1361-6579/ADF1D3.

[161] S. Mohammad, B. Hessam, and L. Kaveh, "Physics-Informed Neural Networks for Brain Hemodynamic Predictions Using Medical Imaging," *IEEE Transactions on Medical Imaging,* vol. 41, pp. 2285 - 2303, 2022, doi: 10.1109/TMI.2022.3161653 , issue = 9.

[162] B. Bivas, D. Soumen, and C. Satyasaran, "Deep learning based solution of nonlinear partial differential equations arising in the process of arterial blood flow," *Mathematics and Computers in Simulation,* vol. 217, pp. 21 - 36, 2024, doi: 10.1016/j.matcom.2023.10.011.

[163] K. Georgios, Y. Yibo, H. Eileen, R. W. Walter, A. D. John, and P. Paris, "Machine learning in cardiovascular flows modeling: Predicting arterial blood pressure from non-invasive 4D flow MRI data using physics-informed neural networks," *Computer Methods in Applied Mechanics and Engineering,* vol. 358, pp. 112623 , publisher = North-Holland, 1 2020, doi: 10.1016/J.CMA.2019.112623.

[164] A. M. Philipp, F. Wolfgang, T. Stefan, G. Isabell, and G. Michael, "Modeling of 3D Blood Flows with Physics-Informed Neural Networks: Comparison of Network Architectures," *Fluids,* vol. 8, 2023, doi: 10.3390/fluids8020046 , issue = 2.

[165] Z. Xuelan *et al.*, "Physics-informed neural networks (PINNs) for 4D hemodynamics prediction: An investigation of optimal framework based on vascular morphology," *Computers in Biology and Medicine,* vol. 164, 2023, doi: 10.1016/j.compbiomed.2023.107287.

[166] F. Fredrik Eikeland *et al.*, "Machine learning augmented reduced-order models for FFR-prediction," *Computer Methods in Applied Mechanics and Engineering,* vol. 384, 2021, doi: 10.1016/j.cma.2021.113892.

[167] M. Olzhas, Z. Yong, S. M. Aigerim, Z. Vasilios, N. Eddie Yin Kwee, and N. Aidossov, "Physics-informed neural network for fast prediction of temperature distributions in cancerous breasts as a potential efficient portable AI-based diagnostic tool," *Computer Methods and Programs in Biomedicine,* vol. 242, 2023, doi: 10.1016/j.cmpb.2023.107834.

[168] A. P. Danang, A. B. Maharani, I. Nur Fadhilah, I. Ruwaidiah, and M. M. Norizan, "Physical restriction neural networks with restarting strategy for solving mathematical model of thermal heat equation for early diagnose breast cancer," *Results in Applied Mathematics,* vol. 19, 2023, doi: 10.1016/j.rinam.2023.100384.

[169] B. Brett, A. O. Chad, K. Jason, R. K. Taufiquar, M. G. Eddie, and J. G. Nicholas, "Physics-Informed Neural Networks for the Heat Equation with Source Term under Various Boundary Conditions," *Algorithms,* vol. 16, 2023, doi: 10.3390/a16090428 , issue = 9.

[170] R. Farnaz, D. Weizhong, D. Shayan, and B. Aniruddha, "A physics-informed neural network method for thermal analysis in laser-irradiated 3D skin tissues with embedded vasculature, tumor and gold nanorods," *International Journal of Heat and Mass Transfer,* vol. 245, 2025, doi: 10.1016/j.ijheatmasstransfer.2025.126980.

[171] Y. Chen, L. Lu, G. E. Karniadakis, and L. Dal Negro, "Physics-informed neural networks for inverse problems in nano-optics and metamaterials," *Opt. Express,* vol. 28, no. 8, pp. 11618-11633, 2020/04/13 2020, doi: 10.1364/OE.384875.

[172] Y. Chen and L. Dal Negro, "Physics-informed neural networks for imaging and parameter retrieval of photonic nanostructures from near-field data," *APL Photonics,* vol. 7, no. 1, 2022.

[173] Y. D. Hu, X. H. Wang, H. Zhou, L. Wang, and B. Z. Wang, "A More General Electromagnetic Inverse Scattering Method Based on Physics-Informed Neural Network," *IEEE Transactions on Geoscience and Remote Sensing,* vol. 61, pp. 1-9, 2023, doi: 10.1109/TGRS.2023.3301455.

[174] Y. D. Hu, X. H. Wang, H. Zhou, and L. Wang, "A Priori Knowledge-Based Physics-Informed Neural Networks for Electromagnetic Inverse Scattering," *IEEE Transactions on Geoscience and Remote Sensing,* vol. 62, pp. 1-9, 2024, doi: 10.1109/TGRS.2024.3371528.

[175] S. Piao, H. Gu, A. Wang, and P. Qin, "A Domain-Adaptive Physics-Informed Neural Network for Inverse Problems of Maxwell's Equations in Heterogeneous Media," *IEEE Antennas and Wireless Propagation Letters,* vol. 23, no. 10, pp. 2905-2909, 2024, doi: 10.1109/LAWP.2024.3413851.

[176] X. Zeng, S. Zhang, C. Ren, and T. Shao, "Physics informed neural networks for electric field distribution characteristics analysis," *Journal of Physics D: Applied Physics,* vol. 56, no. 16, p. 165202, 2023/03/14 2023, doi: 10.1088/1361-6463/acbec3.

[177] M. Baldan, P. D. Barba, and D. A. Lowther, "Physics-Informed Neural Networks for Inverse Electromagnetic Problems," *IEEE Transactions on Magnetics,* vol. 59, no. 5, pp. 1-5, 2023, doi: 10.1109/TMAG.2023.3247023.
[178] M. Zhelyeznyakov *et al.*, "Large area optimization of meta-lens via data-free machine learning," *Communications Engineering,* vol. 2, no. 1, p. 60, 2023/08/21 2023, doi: 10.1038/s44172-023-00107-x.
[179] Y. Q. Pan, R. Wang, and B. Z. Wang, "Physics-Informed Neural Networks With Embedded Analytical Models: Inverse Design of Multilayer Dielectric-Loaded Rectangular Waveguide Devices," *IEEE Transactions on Microwave Theory and Techniques,* vol. 72, no. 7, pp. 3993-4005, 2024, doi: 10.1109/TMTT.2023.3343028.
[180] Y. Torabi, A. Ekhteraei, and M. Khajezadeh, "Physics-Informed Graph Neural Network for Inverse Design of Integrated Photonic Biosensors," in "arXiv," 2602.19082, 2026/02/22 2026. [Online]. Available: https://arxiv.org/abs/2602.19082
[181] X. Luo, M. Zhang, Z. Wang, X. Jiang, Y. Song, and D. Wang, "Hybrid physics-informed and data-driven mode solver for optical fiber design," *Advanced Photonics Nexus,* vol. 4, no. 6, p. 066016, 2025. [Online]. Available: https://doi.org/10.1117/1.APN.4.6.066016.
[182] B. Liang *et al.*, "Physics-Guided Neural-Network-Based Inverse Design of a Photonic–Plasmonic Nanodevice for Superfocusing," *ACS Applied Materials & Interfaces,* vol. 14, no. 23, pp. 27397-27404, 2022/06/15 2022, doi: 10.1021/acsami.2c05083.
[183] Y. Pan, H. Liu, Z. Cheng, Z. Wei, X. Huang, and C. Yu, "Supercell-based metasurfaces for arbitrary polarization beam splitting: physics-informed U-Net design with high extinction ratio," *Optics Express,* vol. 33, no. 19, pp. 39960-39976, 2025/09/22 2025, doi: 10.1364/OE.561950.
[184] L. Lu, R. Pestourie, W. Yao, Z. Wang, F. Verdugo, and S. G. Johnson, "Physics-informed neural networks with hard constraints for inverse design," *ArXiv,* vol. abs/2102.04626, 2021.
[185] V. S. K. Malineni and S. Rajendran, "Physics-Informed Neural Network Approaches for Sparse Data Flow Reconstruction of Unsteady Flow Around Complex Geometries," *ArXiv,* vol. abs/2508.01314, 2025.
[186] K. Zhou *et al.*, "Benchmarking Data Assimilation Algorithms For 3D Lagrangian Particle Tracking," *Proceedings of the International Symposium on the Application of Laser and Imaging Techniques to Fluid Mechanics,* 2024.
[187] S. Wassing, S. Langer, and P. Bekemeyer, "Physics-informed neural networks for inviscid transonic flows around an airfoil," *Physics of Fluids,* 2024.
[188] Y. Li, T. Liu, and Y. Xie, "Thermal fluid fields reconstruction for nanofluids convection based on physics-informed deep learning," *Scientific Reports,* vol. 12, 2022.
[189] J. Lee, M. Duhovic, D. May, T. Allen, and P. Kelly, "Physics-informed neural networks for real-time simulation of transverse Liquid Composite Moulding processes and permeability measurements," *Composites Part A: Applied Science and Manufacturing,* 2025.
[190] M. A. Aziz, T. Strauss, M. Mohebujjaman, and T. Khan, "Self-adaptive physics-informed neural network for forward and inverse problems in heterogeneous porous flow," *ArXiv,* vol. abs/2512.14610, 2025.
[191] J. Sa, B. Jeon, Y. Seo, J. Lim, and S. Q. Choi, "Physics-informed neural networks for complex fluids: opportunities and limitations," *Korea-Australia Rheology Journal,* vol. 37, pp. 469 - 492, 2025.
[192] P. W. Livermore, L. Wu, L. Chen, and S. de Ridder, "Reconstructions of Jupiter's magnetic field using physics-informed neural networks," *Monthly Notices of the Royal Astronomical Society,* vol. 533, no. 4, pp. 4058-4067, 2024.
[193] Y. Zhang, L. Xu, and Y. Yan, "Physics-informed Neural Network for Force-free Magnetic Field Extrapolation," *Research in Astronomy and Astrophysics,* vol. 24, no. 10, p. 105010, 2024.
[194] S. Ni *et al.*, "Application of Physics-Informed Neural Networks in Removing Telescope Beam Effects," *The Astrophysical Journal,* vol. 990, no. 2, p. 122, 2025.

[195] M. P. Bento, H. B. Câmara, and J. F. Seabra, "Unraveling particle dark matter with Physics-Informed Neural Networks," (in English), *Physics Letters, Section B: Nuclear, Elementary Particle and High-Energy Physics,* Article vol. 868, 2025, Art no. 139690, doi: 10.1016/j.physletb.2025.139690.
[196] M. Lopez-Radcenco, J. M. Delouis, and L. Vibert, "SRoll3: A neural network approach to reduce large-scale systematic effects in the Planck High-Frequency Instrument maps," (in English), *Astronomy and Astrophysics,* Article vol. 651, 2021, Art no. A65, doi: 10.1051/0004-6361/202040152.
[197] A. Mathivanan, N. Kumar, P. Kumar, A. Nivesh, S. Gupta, and G. Thakur, "Detection of Deviations from Lambda-CDM in Dark Matter Power Spectrum Using Cosmology Informed Neural Networks," in *2024 3rd Edition of IEEE Delhi Section Flagship Conference (DELCON)*, 2024: IEEE, pp. 1-5.
[198] M. Yarahmadi and A. Salehi, "A Bayesian PINN Framework for Barrow-Tsallis Holographic Dark Energy with Neutrinos: Toward a Resolution of the Hubble Tension," *arXiv preprint arXiv:2506.02235,* 2025.
[199] L. Röver, H. von Campe, M. P. Herzog, R. M. Kuntz, and B. M. Schaefer, "Partition function approach to non-Gaussian likelihoods: physically motivated convergence criteria for Markov chains," (in English), *Monthly Notices of the Royal Astronomical Society,* Article vol. 526, no. 1, pp. 473-482, 2023, doi: 10.1093/mnras/stad2726.
[200] J. Ma, H. Fu, J. D. Huba, and Y. Jin, "A Novel Ionospheric Inversion Model: PINN-SAMI3 (Physics Informed Neural Network Based on SAMI3)," (in English), *Space Weather,* Article vol. 22, no. 4, 2024, Art no. e2023SW003823, doi: 10.1029/2023SW003823.
[201] J. Ma, Y. Liu, and H. Fu, "Comparison of PINN-SAMI3 and 4D-Var Data Assimilation Methods for Ionospheric Prediction," 2024: Institute of Electrical and Electronics Engineers Inc., doi: 10.1109/ISAPE62431.2024.10840758. [Online]. Available: https://www.scopus.com/inward/record.uri?eid=2-s2.0-85218337451&doi=10.1109%2FISAPE62431.2024.10840758&partnerID=40&md5=d5da83281367e0d54d370c314cbfe66b
[202] C. Bard and J. C. Dorelli, "Neural Network Reconstruction of Plasma Space-Time," *Front. Astron. Space Sci.,* vol. 8, p. 732275, 2021/09/01/ 2021, doi: 10.3389/fspas.2021.732275.
[203] S. N. Paul and P. Mehta, "Development of methodologies for modeling and estimation of drag parameters using physics-informed neural network," 2022.
[204] R. Jarolim, A. M. Veroni, S. Purkhart, P. Zhan, and M. Rempel, "Magnetic Field Evolution of the Solar Active Region 13664," (in English), *Astrophysical Journal Letters,* Article vol. 976, no. 1, 2024, Art no. L12, doi: 10.3847/2041-8213/ad8914.
[205] M. Scialpi, "A machine learning approach to parameter inference in gravitational-wave signal analysis," 2023.
[206] E. Cuoco *et al.*, "Enhancing gravitational-wave science with machine learning," *Machine Learning: Science and Technology,* vol. 2, no. 1, p. 011002, 2020.
[207] G. Ai, J. Yin, and L. Cui, "Learning Gravity Fields of Small Bodies: Self-adaptive Physics-informed Neural Networks," *The Astronomical Journal,* vol. 168, no. 6, p. 242, 2024.
[208] J. J. Athalathil, B. Vaidya, S. Kundu, V. Upendran, and M. C. Cheung, "Surface Flux Transport Modeling Using Physics-informed Neural Networks," *The Astrophysical Journal,* vol. 975, no. 2, p. 258, 2024.
[209] L. Pareschi and G. Russo, "Implicit–explicit Runge–Kutta schemes and applications to hyperbolic systems with relaxation," *Journal of Scientific computing,* vol. 25, no. 1, pp. 129-155, 2005.
[210] S. Kundu, B. Vaidya, and A. Mignone, "Numerical modeling and physical interplay of stochastic turbulent acceleration for nonthermal emission processes," *The Astrophysical Journal,* vol. 921, no. 1, p. 74, 2021.
[211] M. D. Veneri, "Resolution of Inverse Problems in Astrophysics through Deep Learning," Università degli Studi di Napoli Federico II.

[212] M. Aigerim, Z. Yong, Y. K. N. Eddie, Z. Vasilios, F. Sai Cheong, and M. Olzhas, "Early detection of the breast cancer using infrared technology – A comprehensive review," *Thermal Science and Engineering Progress,* vol. 27, pp. 101142 , publisher = Elsevier, 1 2022, doi: 10.1016/j.tsep.2021.101142.
[213] L. Chongyuan *et al.*, "The application and development of Pennes' biological heat transfer model in human comfort and thermal therapy," *International Communications in Heat and Mass Transfer,* vol. 169, pp. 109724 , publisher = Pergamon, 12 2025, doi: 10.1016/j.icheatmasstransfer.2025.109724.
[214] P. Bhawani Sankar, S. N. Srigitha, A. Pankaj, B. M. kumar, K. Syed, and M. Neeladri, "Hybrid AI models for predicting heat distribution in complex tissue structures with bioheat transfer simulation," *Journal of Thermal Biology,* vol. 129, pp. 104122 , pmid = 40311397 , publisher = Pergamon, 4 2025, doi: 10.1016/j.jtherbio.2025.104122.
[215] D. Shan, T. Yongfu, Q. Lang, M. Hongxiang, and Y. Rui, "Physics-informed hierarchical neural operator for solving inverse problem of unsteady heat conduction," *International Journal of Heat and Mass Transfer,* vol. 258, pp. 128335 , publisher = Pergamon, 5 2026, doi: 10.1016/j.ijheatmasstransfer.2026.128335.
[216] M. Nicolás, C. Aníbal, and M. Gastón, "Assessing the performance of physics-informed neural networks for tumor growth prediction under noisy and sparse data conditions," *Computational Biology and Chemistry,* vol. 122, pp. 108915 , publisher = Elsevier, 6 2026, doi: 10.1016/j.compbiolchem.2026.108915.
[217] R. Pratik, G. Ranjan, and B. Nirmalendu, "A critical review on magnetic fluid hyperthermia treatment for cancers: Challenges and pathways from computation to clinic," *Journal of Thermal Biology,* vol. 133, pp. 104284 , publisher = Pergamon, 10 2025, doi: 10.1016/j.jtherbio.2025.104284.
[218] N. Weiskopf, L. J. Edwards, G. Helms, S. Mohammadi, and E. Kirilina, "Quantitative magnetic resonance imaging of brain anatomy and in vivo histology," *Nature Reviews Physics,* vol. 3, no. 8, pp. 570-588, 2021/08/01 2021, doi: 10.1038/s42254-021-00326-1.
[219] B. Florian, M. Lucas, S. Julius, W. Qi, S. S. Klaus, and H. Rahel, "Flexible and cost-effective deep learning for accelerated multi-parametric relaxometry using phase-cycled bSSFP," *Scientific Reports,* vol. 15, 2025, doi: 10.1038/s41598-025-88579-z , issue = 1.
[220] B. Leonardo *et al.*, "Myo-regressor Deep Informed Neural NetwOrk (Myo-DINO) for fast MR parameters mapping in neuromuscular disorders," *Computer Methods and Programs in Biomedicine,* vol. 256, 2024, doi: 10.1016/j.cmpb.2024.108399.
[221] L. Pablo Arratia, M. Hernan, A. U. Sergio, E. H. Daniel, and C. Francisco Sahli, "WarpPINN: Cine-MR image registration with physics-informed neural networks," *Medical Image Analysis,* vol. 89, 2023, doi: 10.1016/j.media.2023.102925.
[222] M. Zhe *et al.*, *Non-rigid Medical Image Registration using Physics-informed Neural Networks* (Lecture Notes in Computer Science). 2023, pp. 601 - 613.
[223] Y. Xinling *et al.*, "PIFON-EPT: MR-Based Electrical Property Tomography Using Physics-Informed Fourier Networks," *IEEE Journal on Multiscale and Multiphysics Computational Techniques,* vol. 9, pp. 49 - 60, 2024, doi: 10.1109/JMMCT.2023.3345798.
[224] S. David, W. Lukas, K. Christoph, H. R. Georg, S. Oliver, and K. Andreas, "Joint B0 and Image Reconstruction in Low-Field MRI by Physics-Informed Deep-Learning," *IEEE Transactions on Biomedical Engineering,* vol. 71, pp. 2842 - 2853, 2024, doi: 10.1109/TBME.2024.3396223 , issue = 10.
[225] G. Camps-Valls *et al.*, "Discovering causal relations and equations from data," *Physics Reports,* vol. 1044, pp. 1-68, 2023.
[226] E. Camporeale, G. J. Wilkie, A. Drozdov, and J. Bortnik, "Machine-learning based discovery of missing physical processes in radiation belt modeling," *arXiv preprint arXiv:2107.14322,* 2021.

[227] Y. Fang, G.-Z. Wu, Y.-Y. Wang, and C.-Q. Dai, "Data-driven femtosecond optical soliton excitations and parameters discovery of the high-order NLSE using the PINN," *Nonlinear Dynamics,* vol. 105, no. 1, pp. 603-616, 2021/07/01 2021, doi: 10.1007/s11071-021-06550-9.

[228] Z. Zhou and Z. Yan, "Deep learning neural networks for the third-order nonlinear Schrödinger equation: bright solitons, breathers, and rogue waves," *Communications in Theoretical Physics,* vol. 73, no. 10, p. 105006, 2021/09/03 2021, doi: 10.1088/1572-9494/ac1cd9.

[229] N. Zhao, Y. Chen, L. Cheng, and J. Chen, "Data-driven soliton solutions and parameter identification of the nonlocal nonlinear Schrödinger equation using the physics-informed neural network algorithm with parameter regularization," *Nonlinear Dynamics,* vol. 113, no. 8, pp. 8801-8817, 2025/04/01 2025, doi: 10.1007/s11071-024-10562-6.

[230] I. A. Chuprov, J. Gao, D. S. Efremenko, F. A. Buzaev, and V. V. Zemlyakov, "Solution of the Multimode Nonlinear Schrödinger Equation Using Physics-Informed Neural Networks," *Doklady Mathematics,* vol. 110, no. 1, pp. S15-S24, 2024/12/01 2024, doi: 10.1134/S1064562424602105.

[231] X. Jiang, J. Dong, Y. Song, J. Li, M. Zhang, and D. Wang, "Physics-Informed Machine Learning for EDFA: Parameter Identification and Gain Estimation," *Journal of Lightwave Technology,* vol. 43, no. 11, pp. 5040-5054, 2025, doi: 10.1109/JLT.2025.3548285.

[232] J. Sun, Y. Shu, Y. Ge, J. Li, and W. Chen, "Reversely Exploring Higher-Order Effects in a Fiber Laser through Physics-Informed Recursive Neural Network," *ACS Photonics,* vol. 11, no. 10, pp. 4306-4315, 2024/10/16 2024, doi: 10.1021/acsphotonics.4c01235.

[233] Y. Wu, X. Zou, J. Hu, and H. Yang, "Inverse Parameters Estimation for Electro-Optic Frequency Combs Generation via Modified Physics-Informed Neural Network," in *2024 Asia Communications and Photonics Conference (ACP) and International Conference on Information Photonics and Optical Communications (IPOC)*, 2-5 Nov. 2024 2024, pp. 1-4, doi: 10.1109/ACP/IPOC63121.2024.10810095.

[234] Y. Song *et al.*, "SRS-Net: a universal framework for solving stimulated Raman scattering in nonlinear fiber-optic systems by physics-informed deep learning," *Communications Engineering,* vol. 3, no. 1, p. 109, 2024/08/06 2024, doi: 10.1038/s44172-024-00253-w.

[235] K. Zhang, Z. Chen, J. Wang, and C. Zhou, "Intelligent data-informed study of ionospheric TEC dynamics: learning partial differential equations via PINN, PDE-Net2, and SINDy," (in English), *Intelligence and Robotics,* Article vol. 5, no. 3, pp. 488-504, 2025, doi: 10.20517/ir.2025.25.

[236] Z. Zongren, M. Xuhui, and K. George Em, "Correcting model misspecification in physics-informed neural networks (PINNs)," *Journal of Computational Physics,* vol. 505, p. 112918, 2024, doi: https://doi.org/10.1016/j.jcp.2024.112918.

[237] E. Schiassi, M. De Florio, A. D'Ambrosio, D. Mortari, and R. Furfaro, "Physics-Informed Neural Networks and Functional Interpolation for Data-Driven Parameters Discovery of Epidemiological Compartmental Models," *Mathematics,* vol. 9, no. 17, p. 2069, 2021. [Online]. Available: https://www.mdpi.com/2227-7390/9/17/2069.

[238] Y. Ma, X. Xu, S. Yan, and Z. Ren, "A preliminary study on the resolution of electro-thermal multi-physics coupling problem using physics-informed neural network (PINN)," *Algorithms,* vol. 15, no. 2, p. 53, 2022.

[239] G. Coulaud and R. Duvigneau, "Physics-informed neural networks for multiphysics coupling: application to conjugate heat transfer," Université Côte d'Azur, Inria, CNRS, LJAD, 2023.

[240] D. Kim and J. Lee, "A review of physics informed neural networks for multiscale analysis and inverse problems," *Multiscale Science and Engineering,* vol. 6, no. 1, pp. 1-11, 2024.

[241] J. Parra, J. Navarro-Arenas, and P. Sanchis, "Silicon thermo-optic phase shifters: a review of configurations and optimization strategies," *Advanced Photonics Nexus,* vol. 3, no. 4, pp. 044001-044001, 2024.

[242] S. De, R. Das, R. K. Varshney, and T. Schneider, "Design and Simulation of Thermo-Optic Phase Shifters With Low Thermal Crosstalk for Dense Photonic Integration," *IEEE Access,* vol. 8, pp. 141632-141640, 2020, doi: 10.1109/ACCESS.2020.3013116.

[243] M. Moralis-Pegios *et al.*, *Optics-informed deep learning over silicon photonic hardware* (SPIE Photonex). SPIE, 2023.

[244] J. Xu and F. Wang, "Cardiac Mechano-Electrical-Fluid Interaction: A Brief Review of Recent Advances," *Eng,* vol. 6, no. 8, p. 168, 2025. [Online]. Available: https://www.mdpi.com/2673-4117/6/8/168.

[245] A. Satriano, E. J. Vigmond, D. S. Schwartzman, and E. S. Di Martino, "Mechano-electric finite element model of the left atrium," *Computers in Biology and Medicine,* vol. 96, pp. 24-31, 2018/05/01/ 2018, doi: https://doi.org/10.1016/j.compbiomed.2018.02.010.

[246] A. Zhao, D. Fattahi, and X. Hu, "Physics-informed neural networks for physiological signal processing and modeling: a narrative review," *Physiological Measurement,* vol. 46, no. 7, p. 07TR02, 2025/07/30 2025, doi: 10.1088/1361-6579/adf1d3.

[247] M. Movahhedi *et al.*, "Predicting 3D soft tissue dynamics from 2D imaging using physics informed neural networks," *Communications Biology,* vol. 6, no. 1, p. 541, 2023/05/18 2023, doi: 10.1038/s42003-023-04914-y.

[248] B. H. Nguyen, G. B. Torri, and V. Rochus, "Physics-informed neural networks with data-driven in modeling and characterizing piezoelectric micro-bender," *Journal of Micromechanics and Microengineering,* vol. 34, no. 11, p. 115004, 2024/10/18 2024, doi: 10.1088/1361-6439/ad809b.

[249] B. Moseley, A. Markham, and T. Nissen-Meyer, "Finite basis physics-informed neural networks (FBPINNs): a scalable domain decomposition approach for solving differential equations," *Advances in Computational Mathematics,* vol. 49, pp. 1-39, 2021.

[250] S. Anderson, V. Dolean, B. Moseley, and J. Pestana, "ELM-FBPINN: efficient finite-basis physics-informed neural networks," *ArXiv,* vol. abs/2409.01949, 2024.

[251] S. U. Rehman and W. Yousuf, "Extended Physics Informed Neural Network for Hyperbolic Two-Phase Flow in Porous Media," *ArXiv,* vol. abs/2511.13734, 2025.

[252] S. Nair, T. F. Walsh, G. Pickrell, and F. Semperlotti, "Multiple scattering simulation via physics-informed neural networks," *Engineering with Computers,* vol. 41, no. 1, pp. 31-50, 2025/02/01 2025, doi: 10.1007/s00366-024-02038-3.

[253] C. Shen, H. Zhao, and J. Jiao, "A multi-subnets physics-informed neural network (Ms-PINN) model for transient heat transfer analysis in materials with heterogeneous microstructures," *Engineering with Computers,* vol. 41, no. 5, pp. 3695-3717, 2025/10/01 2025, doi: 10.1007/s00366-025-02183-3.

[254] S. Khalid, M. H. Yazdani, M. M. Azad, M. U. Elahi, I. Raouf, and H. S. Kim, "Advancements in Physics-Informed Neural Networks for Laminated Composites: A Comprehensive Review," *Mathematics,* vol. 13, no. 1, p. 17, 2025. [Online]. Available: https://www.mdpi.com/2227-7390/13/1/17.

[255] X.-X. Chen, P. Zhang, and Z.-Y. Yin, "Physics-Informed neural network solver for numerical analysis in geoengineering," *Georisk: Assessment and Management of Risk for Engineered Systems and Geohazards,* vol. 18, no. 1, pp. 33-51, 2024/01/02 2024, doi: 10.1080/17499518.2024.2315301.

[256] X. Chen *et al.*, "Transfer learning for deep neural network-based partial differential equations solving," *Advances in Aerodynamics,* vol. 3, no. 1, p. 36, 2021/12/08 2021, doi: 10.1186/s42774-021-00094-7.

[257] Y. Shen, H. Zhang, W. Huang, C.-y. Liu, and Z.-g. Wang, "Geometric-perspective transfer learning for fast aerodynamic prediction in few-shot tasks," *Physical Review Fluids,* vol. 9, no. 10, p. 104101, 10/09/ 2024, doi: 10.1103/PhysRevFluids.9.104101.

[258] R. N. Asl, Y. Yamazaki, K. Taghikhani, M. Muramatsu, M. Apel, and S. Rezaei, "A Physics-Informed Meta-Learning Framework for the Continuous Solution of Parametric PDEs on Arbitrary Geometries," *ArXiv,* vol. abs/2504.02459, 2025.

[259] S. Wang, S. Sankaran, H. Wang, and P. Perdikaris, "An Expert's Guide to Training Physics-informed Neural Networks," *ArXiv,* vol. abs/2308.08468, 2023.
[260] S. A. Faroughi and F. Mostajeran, "Neural Tangent Kernel Analysis to Probe Convergence in Physics-informed Neural Solvers: PIKANs vs. PINNs," *ArXiv,* vol. abs/2506.07958, 2025.
[261] B. Gao, R. Yao, and Y. Li, "Physics-informed neural networks with adaptive loss weighting algorithm for solving partial differential equations," *Computers & Mathematics with Applications,* vol. 181, pp. 216-227, 2025/03/01/ 2025, doi: https://doi.org/10.1016/j.camwa.2025.01.007.
[262] O. V. Murashko and Y. V. Tkachov, "UNDERSTANDING PHYSICS-GUIDED MACHINE LEARNING: APPLICATIONS, TRENDS, AND CHALLENGES FOR AEROSPACE," *System design and analysis of aerospace technique characteristics,* vol. 36, no. №1, pp. 58-69, 06/06 2025, doi: 10.15421/472506.
[263] M. A. Nabian and H. Meidani, "Adaptive Physics-Informed Neural Networks for Markov-Chain Monte Carlo," *ArXiv,* vol. abs/2008.01604, 2020.
[264] H. Wang *et al.*, "Multi-scale Simulation of Complex Systems: A Perspective of Integrating Knowledge and Data," *ACM Comput. Surv.,* vol. 56, no. 12, p. Article 307, 2024, doi: 10.1145/3654662.
[265] Y.-H. Yeung, D. A. Barajas-Solano, and A. M. Tartakovsky, "Physics-Informed Machine Learning Method for Large-Scale Data Assimilation Problems," *Water Resources Research,* vol. 58, no. 5, p. e2021WR031023, 2022, doi: https://doi.org/10.1029/2021WR031023.
[266] L. Lu, X. Meng, Z. Mao, and G. E. Karniadakis, "DeepXDE: A Deep Learning Library for Solving Differential Equations," *SIAM Review,* vol. 63, no. 1, pp. 208-228, 2021, doi: 10.1137/19m1274067.
[267] T. Anandh, D. Ghose, H. Jain, and S. Ganesan, "FastVPINNs: Tensor-Driven Acceleration of VPINNs for Complex Geometries," *SIAM Journal on Scientific Computing,* vol. 47, no. 3, pp. C578-C600, 2025, doi: 10.1137/24m1658620.
[268] B. H. Theilman *et al.*, "Spiking Physics-Informed Neural Networks on Loihi 2," in *2024 Neuro Inspired Computational Elements Conference (NICE)*, 23-26 April 2024 2024, pp. 1-6, doi: 10.1109/NICE61972.2024.10548180.
[269] Z. Ren, S. Zhou, D. Liu, and Q. Liu, "Physics-Informed Neural Networks: A Review of Methodological Evolution, Theoretical Foundations, and Interdisciplinary Frontiers Toward Next-Generation Scientific Computing," *Applied Sciences,* vol. 15, no. 14, p. 8092, 2025. [Online]. Available: https://www.mdpi.com/2076-3417/15/14/8092.
[270] M. Zhou and G. Mei, "Transfer learning-based coupling of smoothed finite element method and physics-informed neural network for solving elastoplastic inverse problems," *Mathematics,* vol. 11, no. 11, p. 2529, 2023.
[271] P. Li, D. Grana, and M. Liu, "Bayesian neural network and Bayesian physics-informed neural network via variational inference for seismic petrophysical inversion," *Geophysics,* vol. 89, no. 6, pp. M185-M196, 2024.
[272] P. Dwivedi and B. Islam, "Balancing performance and energy efficiency: the method for sustainable deep learning: P. Dwivedi, B. Islam," *The Journal of Supercomputing,* vol. 81, no. 10, p. 1127, 2025.